\documentclass{MITcsail}
\usepackage{caption}
\usepackage{subcaption}
\usepackage{wrapfig}
\usepackage{hyperref}
\usepackage{comment}
\def\data{{\cal{D}}}

\def\Ndata{{{\rm{N}}_{\cal{D}}}}
\def\reference{{\cal{R}}}
\def\Nreference{{\rm{N}}_{\cal{R}}}
\def\Ndata{{\rm{N}}_{\cal{D}}}

\def\NRef{{\rm{N}}({\rm{R}})}

\newcommand\RH{{\rm{R}}}

\newcommand\w{{\rm{\bf{w}}}}
\newcommand\W{{\rm{\bf{W}}}}
\newcommand\what{{\widehat\w}}
\newcommand\HH{{\rm{H}}_{\rm{\bf{w}}}}
\newcommand\HHhat{{\rm{H}}_{\what}}
\usepackage[font=small,labelfont=bf]{caption}

\def\beq{\begin{equation}\displaystyle}
\def\eeq{\end{equation}}
\def\bea{\begin{eqnarray}\displaystyle} 
\def\eea{\end{eqnarray}}

\def\({\left(}
\def\){\right)}
\def\bry{\begin{array}}
\def\ery{\end{array}}

\usepackage{bbm}

\title{Anomaly-aware summary statistic from data batches\\
}

\author
{Gaia Grosso~$^{1,2.3}$\footnote{Correspondence E-mail: gaia.grosso@cern.ch}\\
\vspace{1em} 
\normalfont{\small $^{1}$NSF AI Institute for Artificial Intelligence and Fundamental Interactions}\\
\normalfont{\small $^{2}$MIT Laboratory for Nuclear Science, Cambridge, MA}\\
\normalfont{\small $^{3}$School of Engineering and Applied Sciences, Harvard University, Cambridge, MA}
}
\begin{document}

\maketitle
\thispagestyle{firstpagestyle} 

\begin{abstract}
%
Signal-agnostic data exploration based on machine learning could unveil very subtle statistical deviations of collider data from the expected Standard Model of particle physics.
The beneficial impact of a large training sample on machine learning solutions  motivates the exploration of increasingly large and inclusive samples of acquired data with resource efficient computational methods. 
In this work we consider the New Physics Learning Machine (NPLM), a multivariate goodness-of-fit test built on the Neyman--Pearson maximum-likelihood-ratio construction, and we address the problem of testing large size samples under computational and storage resource constraints. 
We propose to perform parallel NPLM routines over batches of the data, and to combine them
by locally aggregating over the data-to-reference density ratios learnt by each batch. 
The resulting data hypothesis defining the likelihood-ratio test is thus shared over the batches, and complies with the assumption that the expected rate of new physical processes is time invariant. 
We show that this method outperforms the simple sum of the independent tests run over the batches, and can recover, or even surpass, the sensitivity of the single test run over the full data. 
Beside the significant advantage for the offline application of NPLM to large size samples, the proposed approach offers new prospects toward the use of NPLM to construct anomaly-aware summary statistics in quasi-online data streaming scenarios.
\end{abstract}

\vspace{1cm}

\section{Introduction}
Numerous experimental tests have confirmed the Standard Model as the best description of the particle physics world, reaching levels of precision up to few parts per thousand. New physical processes, if detectable, should therefore manifest as very subtle statistical deviations of the data distributions with respect to the Standard Model expectations. Increasing the phase space acceptance and the statistics of the analysed datasets could unveil interesting data structures that were missed so far. Along with the regular signal-specific statistical analyses, signal-agnostic anomaly detection tools have grown of interest in the high energy physics community, and machine learning has proved to be a crucial ingredient to scale their discovery power. 

Anomaly detection broadly refers to the problem of identifying data patterns that do not align with their expected behaviours.
Events can be interpreted as anomalous for multiple reasons; either because their appearance in the dataset is rare, or because they manifest out of the expected data support. Moreover, anomalies can emerge as a collective behaviour affecting the statistical model describing the data occurrence in the experiment.
%
In this work we focus on collective anomalies, namely phenomena that rise as an unexpected deformation of the data probability distribution.
These kind of anomalies are tackled by means of statistical anomaly detection tools.
Statistical anomaly detection refers to the problem of recognising and quantifying distributional deviations of a dataset from its nominal expected behaviour.
Besides its use in new physics searches, statistical anomaly detection finds a wide spectrum of applications in experimental particle physics. The most striking examples are data quality monitoring of experimental setups at data taking time, and validation of simulated samples. 

When a tractable model of the data nominal behaviour is available, statistical anomaly detection reduces to a goodness-of-fit test: a test statistic is defined according to a notion of similarity between the data distribution and the nominal model, and the level of compatibility is quantified in terms of the test statistic $p$-value with respect to the test distribution in nominal conditions.
However, in most of the applications of statistical anomaly detection in high energy physics, a tractable model for the nominal data distribution is not available and needs to be replaced by a dataset of reference ( denoted as $\reference$ in this work), simulated according to the nominal model or built in a data-driven fashion. In these cases, the problem of statistical anomaly detection is solved as a two-sample-test, introducing a test statistic that measures the distance between two samples.

High energy physics, and in particular collider data, pose extremely challenging problems for statistical anomaly detection due to the level of precision required to test the Standard Model.
For instance, the accuracy of the statistical models employed in the process of scientific discovery is a particularly delicate matter that requires powerful testing tools. 
Similarly, the lack of new interesting tensions with the Standard Model hypothesis after the Higgs boson discovery seem to indicate that, if detectable, new physics should manifest at the LHC as a very rare process, whether in time or in phase space.
Anomaly detection algorithms must therefore be able to capture very subtle anomalous behaviours, and they should be run over the most inclusive set of data allowed by the computational and methodological constraints. 
An extensive review of goodness-of-fit algorithms routinely adopted in the high energy physics community can be found in  Ref.~\cite{Williams:2010vh}.

Machine learning (ML) has provided significant improvements to the statistical tools employed in data analysis at the LHC, allowing to deal with increasingly complex data structures, and exploiting a wider amount of the physics information produced by the experiments.
In the specific context of goodness-of-fit, machine learning based algorithms have opened the path to multivariate statistical analyses, able to capture mismodeling also in the correlations between variables Ref.~\cite{Weisser:2016cnc}.

In absence of constraining inductive biases, as it is the case in unknown anomalous signals searches, the power of machine learning strategies resides in their ability to learn from the training examples coming directly from the experimental data. However, the latter could be either scarce or completely lacking of anomalous events due to their rarity.
The impact of machine learning is therefore expected to become more and more evident by scaling the training sample size and,  as a consequence, the rate of collected anomalies. 
While beneficial to the training, larger sample sizes introduce computational and storage constraints, that need to be addressed.
In this work, we focus on the New Physics Learning Machine (NPLM) algorithm, an ML-empowered goodness-of-fit method based on the Neyman-Pearson hypothesis test (Ref.~\cite{DAgnolo:2018cun,DAgnolo:2019vbw, dAgnolo:2021aun,Letizia:2022xbe,Grosso:2023scl}), and we address the problem of how to test large datasets under computational and storage constraints.
The approach presented in this paper proposes parallel computing of the test over batches of the original sample and a final aggregation strategy that combines the local approximation of the data-to-reference ratios learnt by each instance of the algorithm. 
We outline three operational approaches for different resource constraints settings, aiming at maximally exploiting the available experimental data both in offline and quasi-online configurations.

We apply the split-aggregation approach to a one-dimensional toy model and two realistic LHC datasets of increasing complexity. The former is a dimuon final state characterised by 5 observables and sample size of $O(10^4)$ events, and a 24-dimensional problem defined by the tri-momenta of eight objects selected at the L1 trigger of the CMS experiment with sample size of $O(10^6)$ events.

Our numerical experiments for one-dimensional and five-dimensional problems show that the aggregation strategy can fully recover or, in same cases, even outperform the single batch implementation, thanks to the effect of regularization against statistical noise that the aggregation strategy introduces. 

Furthermore, the studies conducted on the 24-dimensional problem show the discovery potential of the various approaches in high-dimensional and large statistic scenarios.

The paper is organised as follows. In Section~\ref{sec:1} we describe the NPLM algorithm and the proposed batch solution for handling large sample sizes. In Section~\ref{sec:2} we study the performances of the split-aggregation approach on the one-dimensional and five-dimensional problems and we compare them with the original implementation of NPLM. In Section~\ref{sec:3}, we apply the split-aggregation strategy to a large size dataset simulating the CMS L1 trigger, showing the scalability of the algorithm to more complex scenarios. We conclude in Section~\ref{sec:4} discussing our main findings and future research directions. 
\section{Algorithm}
In this Section we briefly summarise the main features of the NPLM algorithm and we provide the statistical foundations for the new proposed split-aggregation solution.
\label{sec:1}
\subsection{New Physics Learning Machine (NPLM)}
New Physics Learning Machine (NPLM) is a ML-based signal-agnostic strategy to detect and quantify statistically significant deviations of the observed data $\data$ from a defined reference model $\RH$ Ref.~\cite{DAgnolo:2018cun}.
The NPLM method is inspired by the classical approach to hypothesis testing based on the log-likelihood-ratio Ref.~\cite{Neyman:1933wgr}
\begin{equation}\label{eq:t_np}
    t(\data) 
    = 2 \max\limits_{\w}\sum\limits_{x\in \data} \log\frac{{\cal L}(x|{\HH})}{{\cal L}(x|\RH)} \,,
\end{equation}
where the two compared hypotheses are the reference, $\RH$, and a composite alternative, $\HH$, whose nature is not known a priori.
A model $f_\w(x)$ with trainable parameters $\w$ is defined on the space of the data $x$ to parametrize  the density distribution of the data in the family of alternatives
\begin{equation}\label{eq:n_w}
    n(x|\HH)= n(x|\RH) e^{f_\w(x)}\,.
\end{equation}

Since the total number of events in $x$ is a Poissonian random variable whose expectation depends on the theory model, the test statistic in Eq.~\ref{eq:t_np} is computed as the extended log-likelihood-ratio, which in terms of $f_\w$ is written as
\begin{equation}\label{eq:t_np2}
t(\data)=2\max\limits_{\w}\left[ \NRef-{\rm{N}}(\HH) +\sum\limits_{x\in\data} f_\w(x)
\right]\,,
\end{equation}
with $\NRef$ and ${\rm{N}}(\HH)$ the expectation values of the total number of events in $\data$ under the reference and the alternative hypotheses respectively.
The maximisation problem described by Eq.~\ref{eq:t_np2} can be solved by a machine learning task
\begin{equation}\label{eq:t_np3}
t_{\w}(\data)=-2 \min\limits_{\w} L[f_{\w}]\,,
\end{equation}
where the loss function 
\begin{equation}\label{eq:loss}
L[f_{\w}]=\left[ \sum\limits_{x\in \reference}w_\reference(x)(e^{f_\w(x)}-1) -\sum\limits_{x\in\data} f_\w(x)
\right]\,.
\end{equation}
takes as input two training datasets: the dataset of interest $\data$, and a sample of reference $\reference$ drawn according to the reference hypothesis and weighted by $w_\reference(x)$ to match the experimental luminosity . 
Thus, the machine learning task learns the optimal values of the trainable parameters of $f$, $\what$, to approximate the log-density-ratio of the two samples
\begin{equation}
    f_\what(x) = \log\frac{n(x|\HHhat)}{n(x|\RH)}
\end{equation}

%
\begin{figure}[t]
    \centering
    \includegraphics[width=0.9\linewidth]{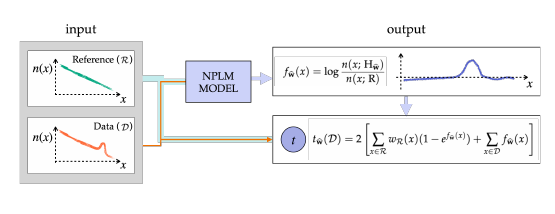}
    
    \caption{\textbf{NPLM training over a single dataset.} Schematic representation of a single run of the NPLM algorithm. The NPLM model, a kernel method in this work, takes as input two datasets, a reference sample ($\reference$) and a data sample $\data$, and it fit via maximum-likelihood the log-ratio of their densities $f(x;\,\what)$. The resulting model is evaluated in-sample on $\data$ and $\reference$ to compute the test statistic $t$.}
    \label{fig:scheme-1}
\end{figure}

As for any test statistic, the value of the test obtained for $\data$ has to be compared to the distribution of the test under the reference hypothesis, $p(t|\RH)$, and the $p$-value is finally used as a GoF metric.
We build $p(t|\RH)$ empirically, running pseudo-experiments on datasets that are generated according to the $\RH$ hypothesis.
The model $f_\w(x)$ could be a neural network as in Ref.~\cite{DAgnolo:2018cun, DAgnolo:2019vbw, dAgnolo:2021aun}, or it could be built with kernel methods Ref.~\cite{Letizia:2022xbe, Grosso:2023ltd}. 
In this work we employ kernel methods since they have been showed to be less time demanding than neural networks. Details about the NPLM training strategy based on kernel methods are postponed to Section~\ref{subsec:2}.

It should be mentioned that the test statistic in Eq.~\ref{eq:t_np3} has good statistical properties (i.e. its null distribution approximates a $\chi^2$ in the asymptotic regime) only if the number of events representing the reference hypothesis, ${\rm N}_{\reference}$ is large enough that 
the statistical fluctuations affecting the reference sample are negligible with respect to the one affecting the data. We refer the reader to  Ref.~\cite{DAgnolo:2019vbw} and  Ref.~\cite{Letizia:2022xbe} for a detailed discussion on the impact of the reference sample size on the NPLM test statistic properties.

Moreover, in realistic particle physics scenarios uncertainties affect the knowledge of the reference hypothesis requiring the introduction of nuisance parameters, defining families of transformations of the reference distribution that should be seen as not anomalous. A solution to extend the NPLM strategy to account for systematic errors has been introduced in  Ref.~\cite{dAgnolo:2021aun}. In our studies, we will consider systematic uncertainties as negligible and leave the treatment of systematic uncertainties within the batch approach for future studies. 
Finally, often times simulations are not accurate enough to provide a good reference model. In those circumstances, data driven techniques need to be implemented to build $\reference$. The problem of building an accurate reference model $\RH$ is a distinct problem and out of the scope of this work.
Comment on possible solutions to this problem in Section~\ref{sec:3}.
In the next Section we introduce the batching idea to address the problem of large sample size.

\subsection{Learning New Physics from batches}
Let's assume that the experimental data $\data$ are acquired in $N_{\rm b}$ batches $\data_i$
\beq
\data=\cup_i \data_i\,,
\eeq
and the size $N_i$ of each batch is a Poissonian random variable whose expectation is a function of the integrated data acquisition time.\footnote{more precisely, the expected number of observations for a given physics process ($H$) is a function of its physics cross section ($\sigma(H)$), the integrated luminosity ($L$), and the acceptance ($a$) and efficiency ($\epsilon$) of the experimental setup: $N(H)= \sigma(H) \times L \times a \times \epsilon$. }
In addition, let's assume that each batch $\data_i$ can be stored in a buffer long enough time to undergo the NPLM training task. 

Since the proton-proton collisions are independent events, the optimal test of the full dataset according to Neyman and Pearson would be the sum of the log-likelihood-ratio test performed over each single batch, with alternative hypothesis set to the True model ($\rm T$) underlying the data:

\begin{equation}\label{eq:t_comb_true}
    t_{id}(\data=\cup_i \data_i) 
    = 2 \sum\limits_{i=1}^{N_{\rm b}} \log\frac{{\cal L}(\data_i|{\rm T})}{{\cal L}(\data_i|\RH)}\,. 
\end{equation} 
Clearly, the True hypothesis T is not known a priori and we want to exploit the NPLM strategy to estimate it.
The solution we propose in this work is to run separate instances of the NPLM algorithm over different batches and recombine the information encoded in the log-density-ratio functions
\begin{equation}\label{eq:density-ratio-batch}
f_{\what,\, i}(x)= \log \frac{n(x;\, \HHhat^i)}{n(x;\,\RH)}
\end{equation}
learnt by the different instances.
Recalling the parametrization defined in Eq.~\ref{eq:n_w}, for each learnt model $f_{\w,\, i}$ we can write an estimate of the density ratio between the data batch and the reference $\reference$
\begin{equation}\label{eq:n_wi}
    n(x;\, \HHhat^i) = n(x;\,\RH) e^{f_{\what,i}(x)}\,,
\end{equation}
Each density $n(x;\, \HHhat^i)$ can be interpreted as an approximated view of the same True physical model. If the learning process outputs an accurate enough representation of T, then  
Eq.~\ref{eq:t_comb_true} can be straightforward approximated summing the likelihood-ratio-tests output by NPLM for the separate batches
\begin{equation}\label{eq:t-sum}
    t_{\rm SUM}^{N_{\rm b}}(\data)
    = \sum_{i=1}^{N_{\rm b}} t(\data_i)
    = 2 \sum_{i=1}^{N_{\rm b}} \log\frac{{\cal L}(\data_i|\HH^i)}{{\cal L}(\data_i|\RH)} 
    = 2 \sum_{i=1}^{N_{\rm b}} \left[ \sum\limits_{x\in \reference}w_\reference(x)(1-e^{f_{i,\what}(x)}) +\sum\limits_{x\in\data_i} f_{i,\what}(x)
\right]\,.
\end{equation}
This approach produces a powerful test if the exact shape and normalization of the signal is captured by all models. 
However, when the signal is very rare its impact in a small size data sample can be overlooked by the model. 
We can exploit two simple assumptions to provide a better solution. First, we assume that the new physics source is static, namely its occurence rate is uniform over time. Second, we assume that biases affecting single batch views are due to statistical noise whose impact over time is on average null.
Under these assumptions, averaging the density-ratio terms learnt in different batches provide a single alternative hypothesis $\HH^{N_{\rm b}}$ shared among batches, that helps to enhance systematic discrepancies, interpreted as signals, while suppressing random ones. 
The density of the data under $\HH^{N_{\rm b}}$ would then have the following form:
\begin{equation}
    n(x;\, \HH^{N_{\rm b}}) = \frac{1}{N_{\rm b}}\sum_{i=1}^{N_{\rm b}} n(x;\, \HH^i)\,.
\end{equation}
By expressing the model of the data density distribution in each batch in terms of the set of functions $\{f_{\w,i}\}_{i=1}^{N_{\rm b}}$, we can define an ``aggregated" model of the log-ratio of the densities as
\begin{equation}\label{eq:F}
    F^{N_{\rm b}}(x;\,\W)
    = \log \frac{n(x;\, \HH^{N_{\rm b}})}{n(x;\,\RH)} = \log\left[ \frac{1}{N_{\rm b}}\sum_{i=1}^{N_{\rm b}} e^{f_{i}(x;\, \w)}\right],
\end{equation}
and use it to compute the log-likelihood-ratio test statistics over the set of batches:
\begin{equation}\label{eq:t-avg}
    t_{\rm AGGR}^{N_{\rm b}}(\data)
    =2 \sum_{i=1}^{N_{\rm b}} \log\frac{{\cal L}(\data_i|\HH^{N_{\rm b}})}{{\cal L}(\data_i|\RH)} 
    = 2 \sum_{i=1}^{N_{\rm b}} \left[ \sum\limits_{x\in \reference}w_\reference(x)(1-e^{F^{N_{\rm b}}(x;\,\W)}) +\sum\limits_{x\in\data_i} F^{N_{\rm b}}(x;\,\W)
\right]\,.
\end{equation}

Batching solutions for computationally intensive calculations with kernel methods have already been proposed in the literature for supervised approaches (see for instance Ref.~\cite{JMLR:v16:zhang15d}). In  Ref.~\cite{JMLR:v16:zhang15d} the authors show theoretical guarantees for recovering the performances of the full sample training solution, provided that the number of splitting is not too large.

\subsection{Operational strategies in resource constrained settings}\label{subsec:strat}
Despite the advantage provided by splitting the dataset in batches, the computational requirements for the algorithm execution might still pose significant challenges to its application in realistic scenarios. 
In fact, computing $t_{\rm AGGR}^{N_{\rm b}}$ requires evaluating $F_{\W}^M(x)$ over each element $x$ in the total set of acquisitions $\data=\cup_i \data_i$. This means that all data batches have to be stored for the whole duration of the NPLM algorithm execution, and the memory locations at which the models are stored need to be accessible by all batches. 
Moreover, the execution time largely varies depending on several factors concerning the algorithm setup, such as the size of the batches, the choice of input variables, but also to the computational resources available and the way these are dispatched.
The possibility of running NPLM tasks in parallel on multiple GPUs provides, for instance, the potential to significantly reduce the computational time. 
%

On the other hand, if the storage resources are limited, it could happen that not all the data can be saved long enough for the NPLM routines to be all completed. In this situation, the test in Eq.~\ref{eq:t-avg} cannot be evaluated exactly.
Two alternatives can then be contemplated. 
The first one is to run NPLM over all the batches and save the final models $\{f_{\w,i}\}_{i+1}^{N_{\rm b}}$ to estimate $\HH^{N_{\rm b}}$, but running the test only on a subset of $N_{\rm test}$ batches that could be stored:
\begin{equation}\label{eq:t-avg_v2}
    t_{\rm AGGR}^{N_{\rm b}, N_{\rm test}}(\data)
    =2 \sum_{i=1}^{N_{\rm test}} \log\frac{{\cal L}(\data_i|\HH^{N_{\rm b}})}{{\cal L}(\data_i|\RH)} 
    = 2 \sum_{i=1}^{N_{\rm test}} \left[ \sum\limits_{x\in \reference}w_\reference(x)(1-e^{F_{\W}^{N_{\rm b}}(x)}) +\sum\limits_{x\in\data_i} F_{\W}^{N_{\rm b}}(x)
\right]\,.
\end{equation}
The second alternative is considering Eq.~\ref{eq:n_wi} as the limit of a smart multidimensional binning approach, where the bins centres are the elements of the reference sample $\reference$.
In this limit, each bin has counting equal to the unity under the reference hypothesis and value $e^{F_{\W}^{N_{\rm b}}(x)}$ under the alternative.
A ``saturated" binned likelihood-ratio test~\cite{Cousins:2018tiz} can then be computed: 

\begin{equation}\label{eq:t-sat}
    t_{\rm SAT}^{N_{\rm b}}(\reference)
    = 2 
    \sum\limits_{x\in \reference}w_\reference(x)\left[ (1-e^{F_{\W}^{N_{\rm b}}(x)})+e^{F_{\W}^{N_{\rm b}}(x)}\log F_{\W}^{N_{\rm b}}(x)
\right]\,.
\end{equation}
Notice that the test does not depend on the data $\data$ directly, but only via $F_{\W}^{N_{\rm b}}$, that has been learnt exploiting the dataset $\data$ in batches.
Moreover, the test does not compute the likelihood-ratio of the data, but it rather evaluates the distance between the density of the reference and that of the alternative hypothesis on the data points belonging to the reference sample $\reference$.
Eq.~\ref{eq:t-sat} is a powerful metric when the model $F_{\W}^{N_{\rm b}}$ is able to capture the signal shape well enough. Its performances deteriorates for signal events that lay in rare regions of the input space, or whose shape is too complex compared to the expressivity of the family of universal approximators used to define the models $\{f_{\w,i}\}$.
An alternative approach of online data summarization based on machine learning was proposed in Ref.~\cite{Butter:2022lkf}. There are two main conceptual differences between the approach presented in Ref.~\cite{Butter:2022lkf} and the one presented in the current paper. The first is that the authors of Ref.~\cite{Butter:2022lkf} propose to learn directly the probability distribution of the data, while via NPLM we propose to learn the ratio of the data density with respect to a reference. 
The second difference concerns the training strategy. The authors of Ref.~\cite{Butter:2022lkf} suggest to train a unique model (specifically, a normalizing flow) feeding the acquired data batches sequentially. Conversely, in the present work we propose to perform a complete and independent training for each batch, thus preventing catastrophic forgetting.\\ 


    
In the experiments reported in Section~\ref{sec:3} we consider three representative resource constrained scenarios:
\begin{enumerate}
    \item \textbf{No resource constraints (NPLM-ALL):}
    all batches are used both for training and for testing. The final test is the one in Eq.~\ref{eq:t-avg}.
    \item \textbf{Long term storage constraint (NPLM-ONE):} there are enough computational resources to train multiple instances of the algorithm in a quasi-online fashion but only one batch is available for testing. For this configuration, the final test is the one in Eq.~\ref{eq:t-avg_v2}, with $\rm N_{\rm test}=1$.
    \item \textbf{Saturated test (NPLM-SAT):} all data are temporarily available for a quasi-online training of the algorithm but they cannot be stored and tested. In this case, we only test the distance between the density of the data and that of the reference distribution by computing the saturated test 
    in Eg.~\ref{eq:t-sat}.
\end{enumerate}
%
\begin{figure}[H]
    \centering
\includegraphics[width=0.9\textwidth]{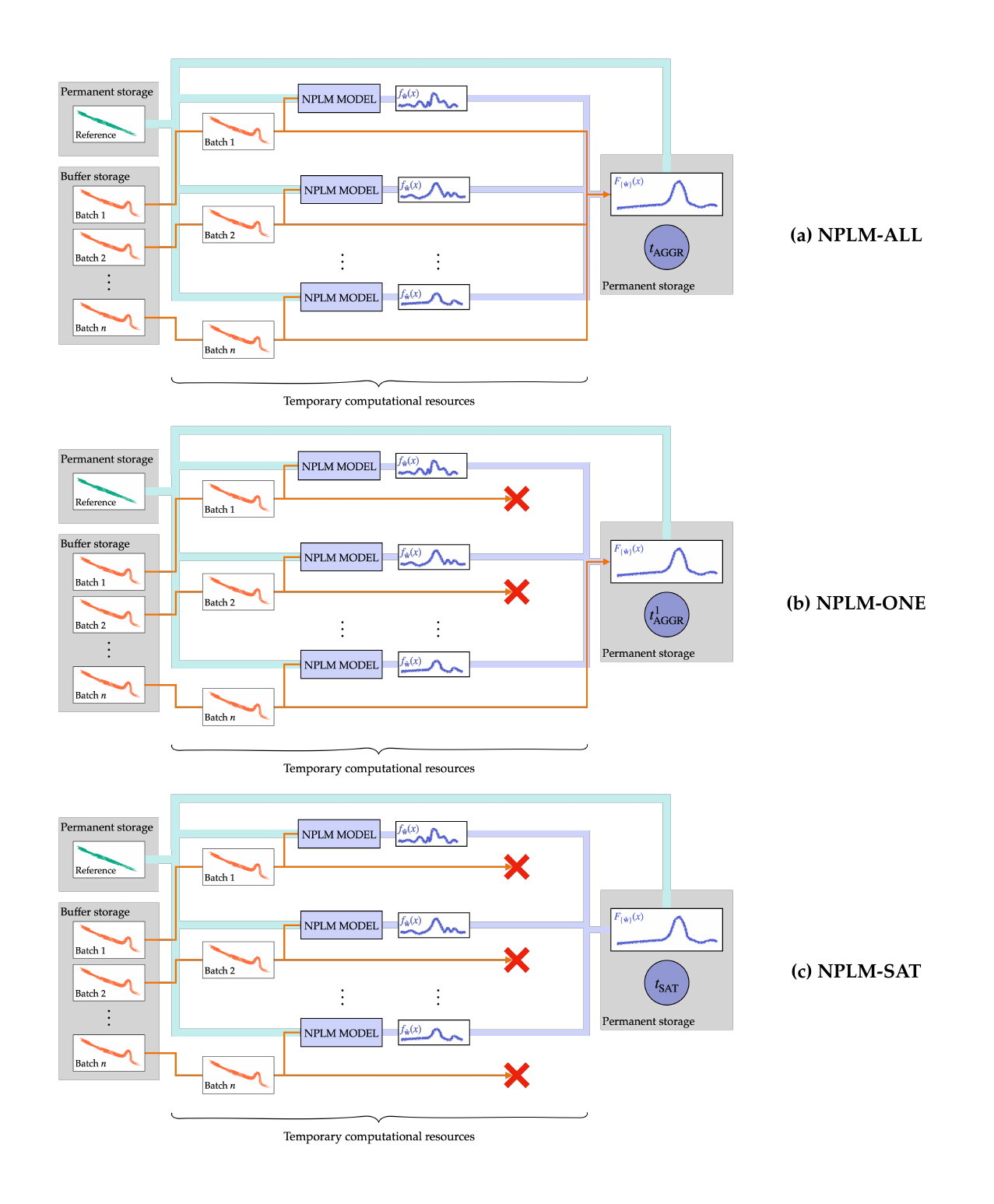}
    \caption{\textbf{Operational strategies in resource constrained scenarios.} Schematic representation of the three scenarios considered in this work. The experimental data are collected in a storage unit or buffer (orange box) and used to train multiple instance of the NPLM model (purple squares). In the NPLM-ALL scenario all data batches and trained models are sent downstream to compute the test statistic. In the NPLM-ONE scenario all the models but only one batch are sent downstream for testing. Finally, in the NPLM-SAT scenario only the models are stored and can be used to test for significant deviations. Due to the information reduction imposed by the resource constraints, the power of the test generally decreases from left to right.}
    \label{fig:scheme-2}

\end{figure}

\subsection{Time-efficient NPLM implementation with non-parametric models}\label{subsec:2}
To harness the execution time of a single instance of the NPLM algorithm we consider the kernel-based implementation of NPLM presented in Ref.~\cite{Letizia:2022xbe}, and applied in Ref.~\cite{Grosso:2023ltd}, which exploits the \textsc{Falkon} library~\cite{falkonlibrary2020}, a modern solver for kernel methods that leverages several algorithmic solutions developed in Ref. Ref.~\cite{rudi2017falkon,marteauferey2019globally,falkonlibrary2020} to allow for their use in large scale problems.
The model used in \textsc{Falkon} is a Nystrom approximated Kernel method defined as a finite weighted sum of kernel functions
\begin{equation}\label{kernel_mod}
f_\w(x)=\sum_{i=1}^{{M}} w_i k(x;\,x_i,\, \sigma_{\rm FLK})\,.
\end{equation}
The kernel function, $k(x;\,x_i,\, \sigma_{\rm FLK})$, adopted in this implementation is a Gaussian kernel
\beq
k(x;\,x_i,\, \sigma_{\rm FLK})=\exp\left(-\frac{\Vert x-x'\Vert^2}{2\sigma_{\rm FLK}^2}\right)\,,
\eeq
where the hyper parameter $\sigma_{\rm FLK}$ denotes the Gaussian width.
 
To approximate the log-ratio between the distribution of the $\data$ and $\reference$ datasets the training exploits a weighted binary cross-entropy loss
\begin{equation}\label{wbce}
L(f_\w) = \sum_{x\in\reference}w_\reference(x) \log \left[1+e^{f_\w(x)}\right]+\sum_{x\in\data}\log \left[1+e^{-f_\w(x)}\right],
\end{equation}
with a L2 regularisation term.
The NPLM method works under the assumption that the reference sample is large enough to be a good representation of the $\RH$ hypothesis, which 
typically leads to an imbalanced dataset with $\Nreference>\Ndata$. 
The weight $w_\reference(x)$, is thus introduced to balance the two classes' contributions.

\paragraph{Hyper parameters choice.} \label{subsubsec:hyperpar}
The hyper parameters choice for a given family of universal approximators defines the ``modes" of the data that the NPLM algorithm is able to capture.
For the kernel-methods considered in this work, there are three hyper parameters that need to be defined: the number of centres $M$, the Gaussian width $\sigma$, and the regularisation parameter $\lambda$.
In Ref.~\cite{Letizia:2022xbe} it has been showed that there exists a ``valley" of ($M$, $\sigma$, $\lambda$) configurations that guarantee the convexity of the loss function, arbitrarily small prediction error and reasonable convergence time.
However, no further indication can be provided to enhance the sensitivity to statistical anomalies. Different hyper parameters choices could be optimal to capture different types of data deviations from the reference distribution. For instance, small values of the Gaussian kernels width, $\sigma_{\rm FLK}$, would allow to better capture localised deviations, like narrow resonances, while larger values would be better suited to represent effects affecting the distribution on a wider range, like scale mismatches. 
If the nature of the anomaly is unknown and no good prior is available, then the most inclusive way of testing the data is combining multiple choices of the hyper-parameters configurations.
In this work we fix $M$ and $\lambda$
following the prescription proposed in Ref.~\cite{Letizia:2022xbe}. Namely, we chose $M=c\sqrt{N}$ with $c\sim(1)$ such that the training time is reasonable given the available computational resources, and we choose $\lambda$ as small as possible, provided no instabilities are encountered.

For the Gaussian width, $\sigma_{\rm FLK}$, we adopt the approach recently proposed in Ref.~\cite{NPLM-aggreg} rather than limiting to a single value we consider five values corresponding to the 5, 25, 50, 75 and 95 \% quantiles of the pairwise distance distribution between the elements in the reference set. 
We combine the results obtained with different values of $\sigma_{\rm FLK}$ by selecting the minimum:
\beq
P = \min\limits_{\{\sigma_{\rm FLK}\}} (p_{\sigma_{\rm FLK}})
\eeq

It is worth noticing that the authors of Ref.~\cite{JMLR:v16:zhang15d} comment on the impact of regularisation on the batched solutions, highlighting how the correct level of generalisation can be recovered if the trainings run on the batches is under-regularised. The NPLM algorithm is a signal-agnostic strategy and therefore an optimal regularisation level cannot be determined prior to testing the data. Nevertheless, we observe the benefit of the aggregation as a regularisation in our studies. Details are reported in Section~\ref{sec:2}.
\section{Numerical experiments}\label{sec:2}
In this Section we study the properties of the batch-based strategy outlined in Section~\ref{subsec:strat} over two proof-of-concept applications. We start considering a one-dimensional problem characterised by a smoothly falling reference distribution (\textbf{EXPO-1D}) for which the optimal test statistic according to Neyman and Pearson can be computed analytically and used as a target. We then move to the multidimensional problem of a dimuon final state observed at collider experiments (\textbf{DIMUON-5D}). 
A detailed description of the datasets' properties and signal benchmarks is given in Appendix~\ref{app:datasets}.
The details about the algorithm implementation, execution time and resource consumption are given in Section~\ref{subsec:resources}.

\subsection{Improving regularisation and sensitivity by aggregating over batches.}\label{subsec:all}
We start by studying the impact of the number of batches $N_{\rm b}$ on the power of the NPLM-ALL aggregated test statistic (Eq.~\ref{eq:t-avg}). 
We run the NPLM algorithm for different values of $N_{\rm b}$ 
and we compare the power curves of the aggregated test (Eq.~\ref{eq:t-avg}) and the simple sum of tests (Eq.~\ref{eq:t-sum}) resulting by combining them. 
We run the experiments on the various signal benchmarks reported in Table~\ref{tab:1d-dataset} for the \textbf{EXPO-1D} dataset and on Table~\ref{tab:5d-dataset} for the \textbf{DIMUON-5D} dataset.
 
\begin{figure}[h]
    \centering
    \includegraphics[width=\linewidth]{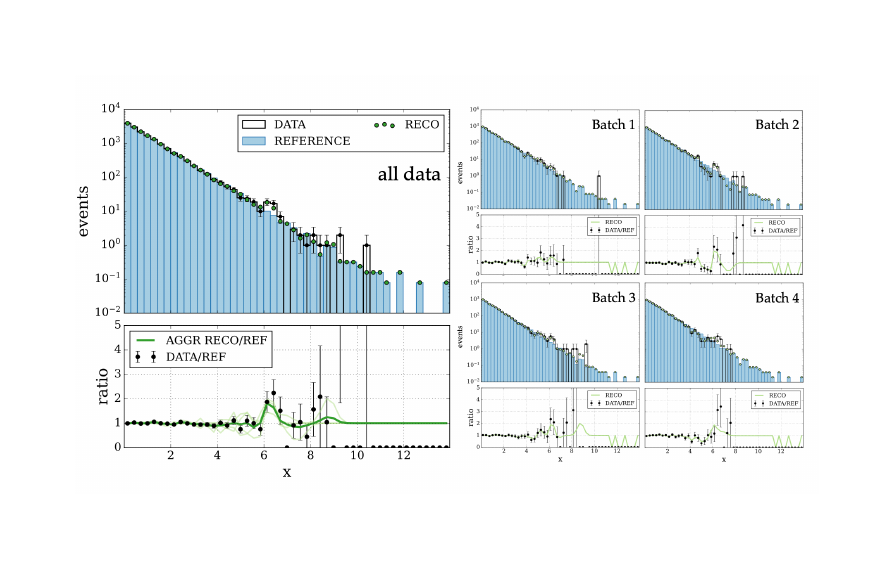}
    \vspace{-2cm}
    \caption{\textbf{Understanding the regularising effect of the batching strategy.} Visual representation of the densities functions reconstructed by the NPLM models trained on different batches (light green lines), and after combining them (dark green line in the main panel). The reconstruction after combination is more robust to statistical fluctuations and approximates the true signal model better than each single one.}
    \label{fig:1D-reco-narrow}
\end{figure}

Interestingly, the aggregated test shows similar or higher power as the number of batches increases. 
Conversely, we observe that the simple sum of tests suffers from significant power degradation as the number of batches increases.
We argue that the regularisation benefit introduced by averaging over the batches has higher impact than the sensitivity loss due to data splitting.
One possible explanation for such result can be found in the in-sample nature of the NPLM algorithm. Each trained model is by construction induced to (over)fit fluctuations in the data batch distribution even when those are not statistically significant in the single batch test. 
Combining the outcome of each batch training at the level of the functional forms of the model allow therefore to restore the suite of information carried along by each model, and exploit it to recover sensitivity to very rare signals.
An illustrative example of the dynamics mentioned above is given in Figure~\ref{fig:1D-reco-narrow}. The panels represent the data distribution (black histograms) and reference distribution (light blue histograms) for the one-dimensional \textbf{EXPO-1D} dataset, when the data are split in four batches. A rare Gaussian signal is injected in the tail of the exponentially falling distribution. The light green solid lines represent the models learnt in each training. On the right hand side of the Figure, we report four panels showing the individual batches and the relative trained models. The models follow the statistical fluctuations of the training sample producing an imprecise view of the true model of the data. On the left hand side, the four models are combined to produce the aggregated version, showed in dark green. It is easy to notice the improved stability of the aggregated model with respect to the ones trained on single batches.

\begin{figure}[h]
    \centering
\includegraphics[width=0.33\textwidth]{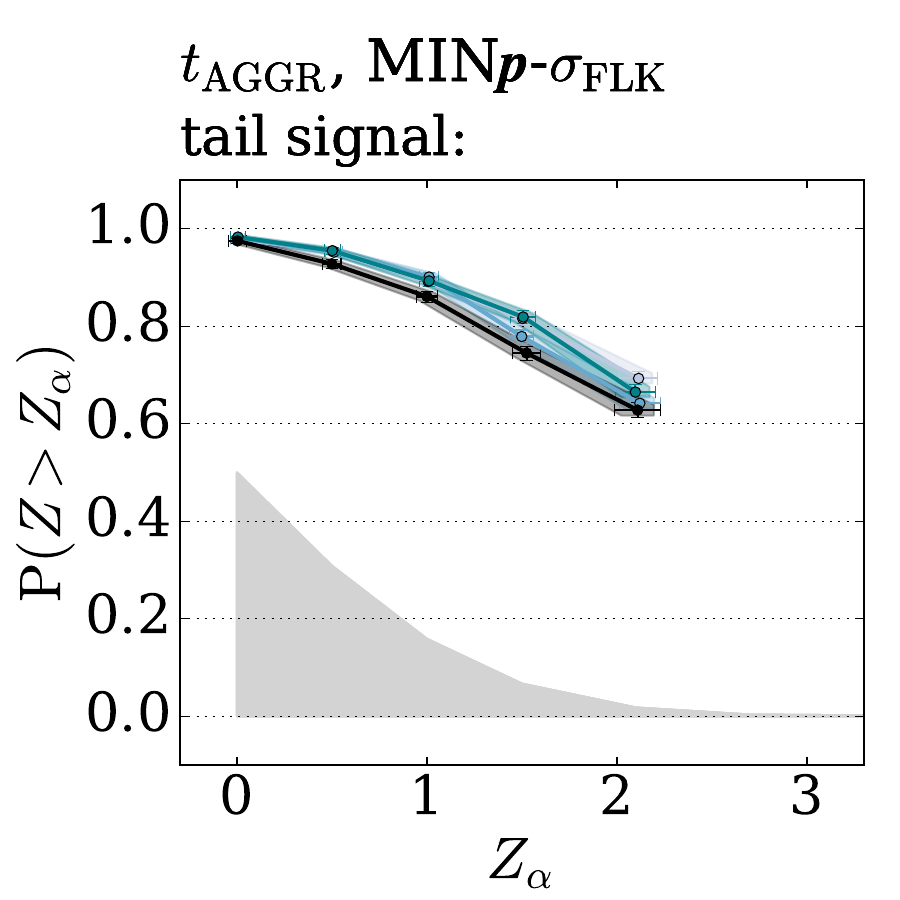}
\includegraphics[width=0.66\textwidth]{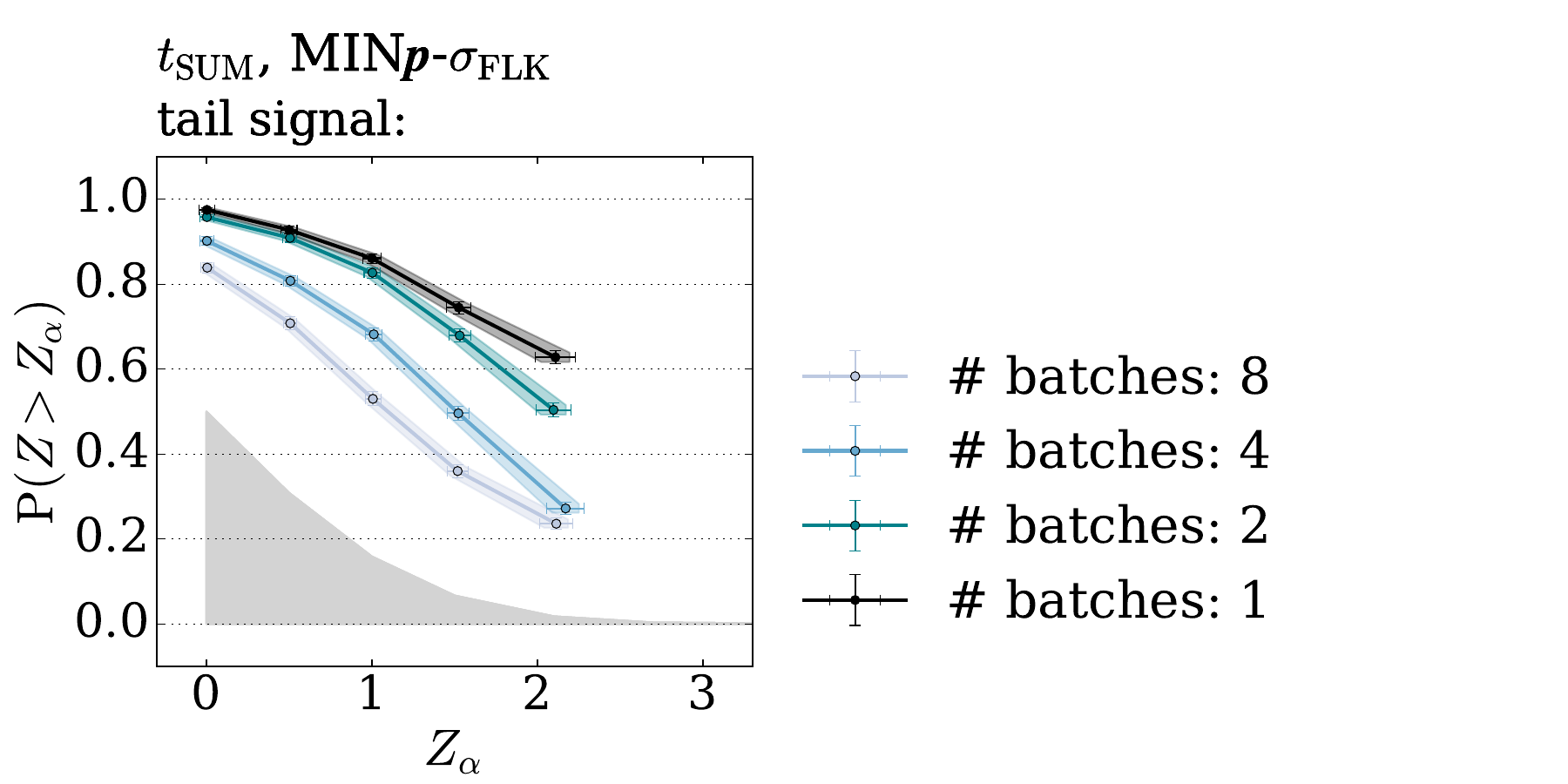}
\caption{\textbf{NPLM-ALL. EXPO-1D -- localised signal in the tail.} Power curves for the aggregated test $t_{\rm AGGR}$ (left panel) and the simple sum of tests $t_{\rm SUM}$ (right panel) at different number of batches. Data splitting improves the performances of $t_{\rm AGGR}$, while degrades $t_{\rm SUM}$.}\label{fig:1D-tail-1}
\end{figure}

\begin{figure}[h]
    \centering
\includegraphics[width=0.33\textwidth]{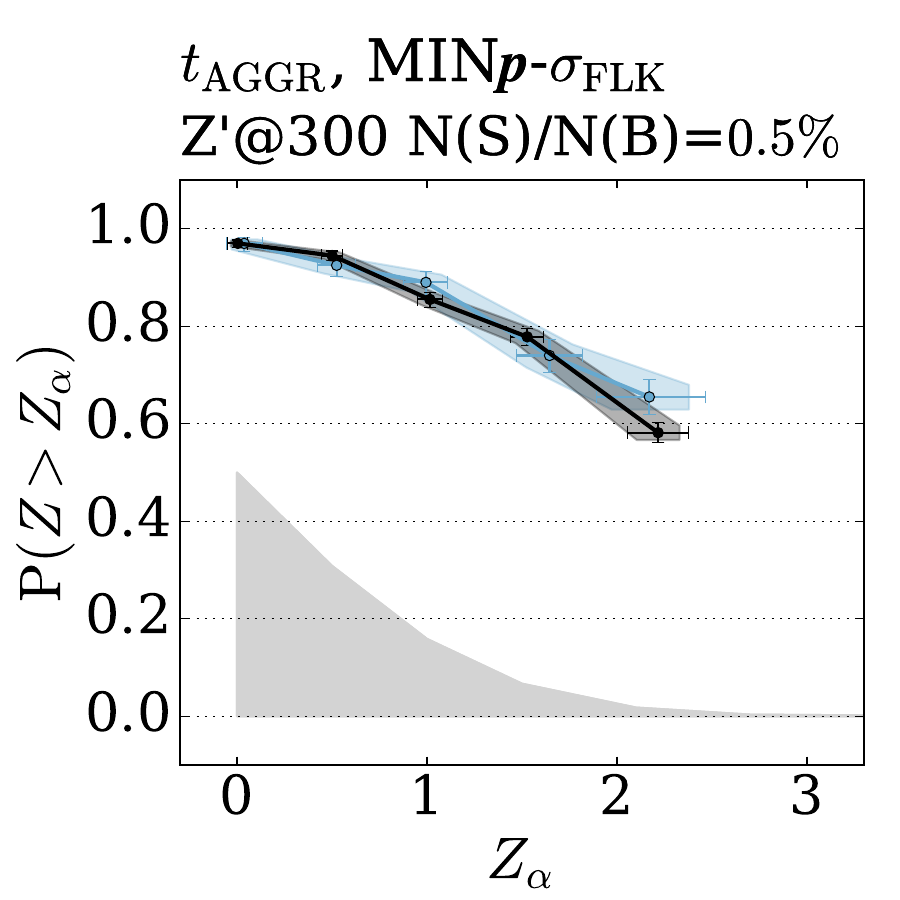}
\includegraphics[width=0.66\textwidth]{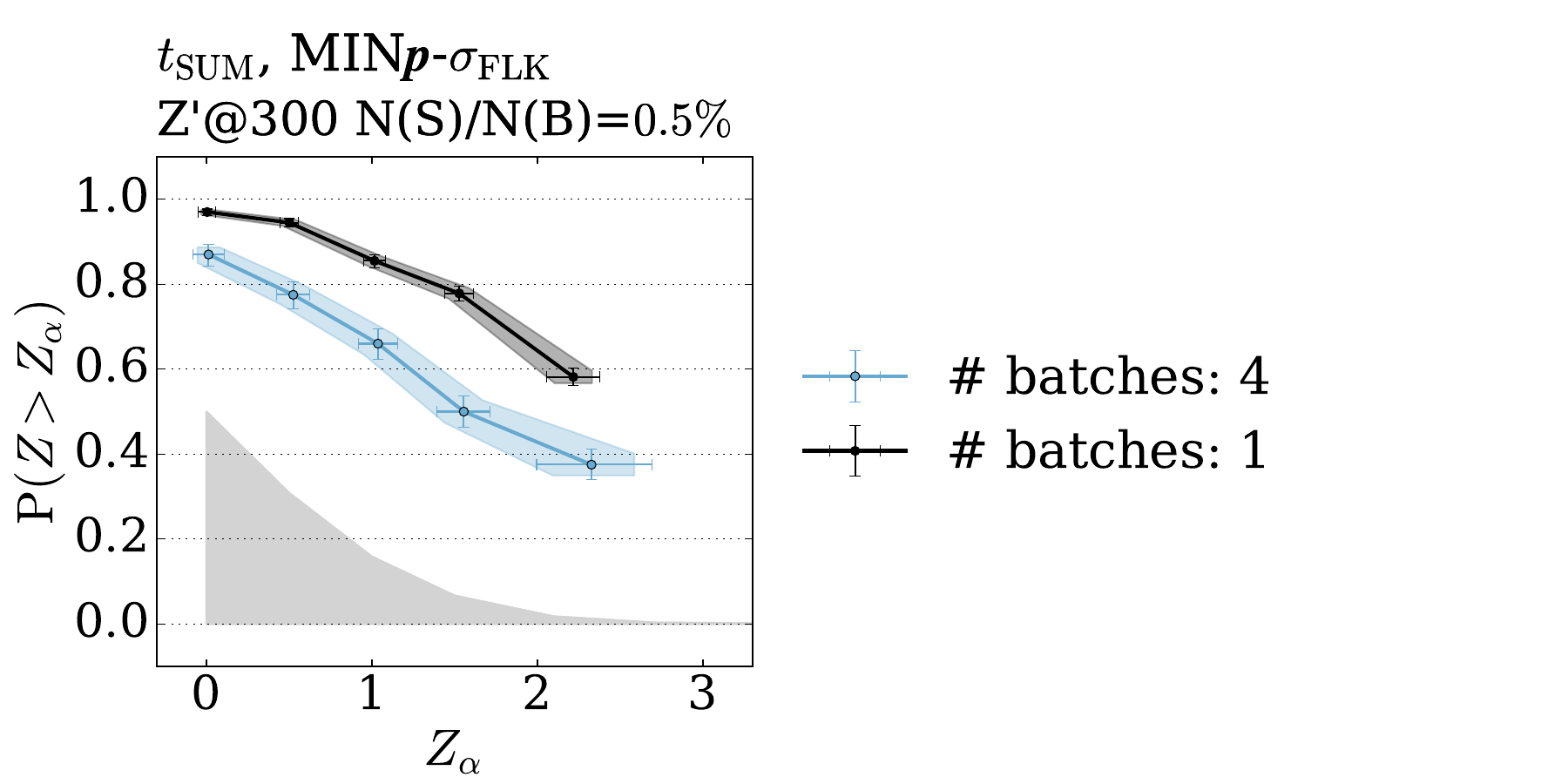}
    \caption{\textbf{NPLM-ALL. DIMUON-5D -- Z' resonance at 300 GeV.} Power curves for $t_{\rm AGGR}$ (left panel) and $t_{\rm SUM}$ (right panel) at different number of batches. Data splitting improves the performances of $t_{\rm AGGR}$, while degrades $t_{\rm SUM}$.}\label{fig:5D-zprime300-1}
\end{figure}

The observed behaviour is consistent across all the studied signal benchmarks for both datasets.
We report in Figure~\ref{fig:1D-tail-1} the power curves for one signal benchmark of the \textbf{EXPO-1D} dataset, and in Figure~\ref{fig:5D-zprime300-1} the power curves for one signal benchmark of the \textbf{DIMUON-5D} dataset.
Additional power plots for the aggregated and simple-sum tests for different values of $\sigma_{\rm FLK}$ on different datasets and signal benchmarks are reported in Appendix~\ref{app:1}.

\subsection{Improving anomaly detection on a single batch exploiting quasi-online learning.}\label{subsec:one}
\begin{figure}[h]
    \centering
\includegraphics[width=0.33\textwidth]{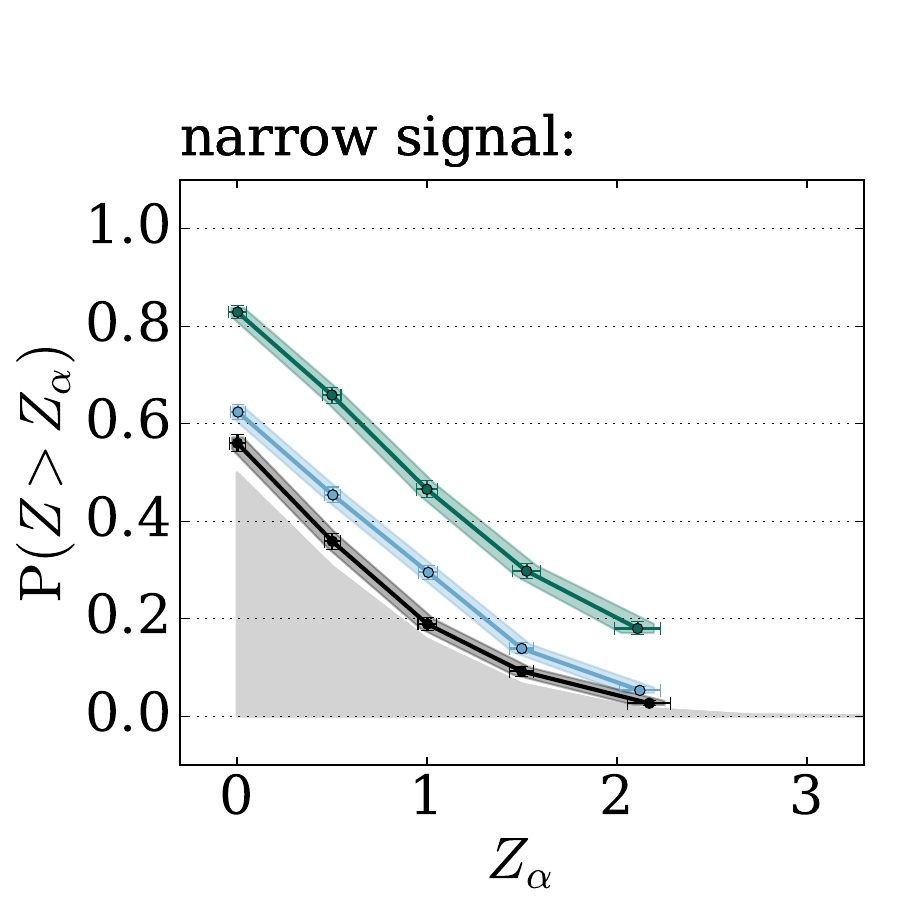}
\includegraphics[width=0.66\textwidth]{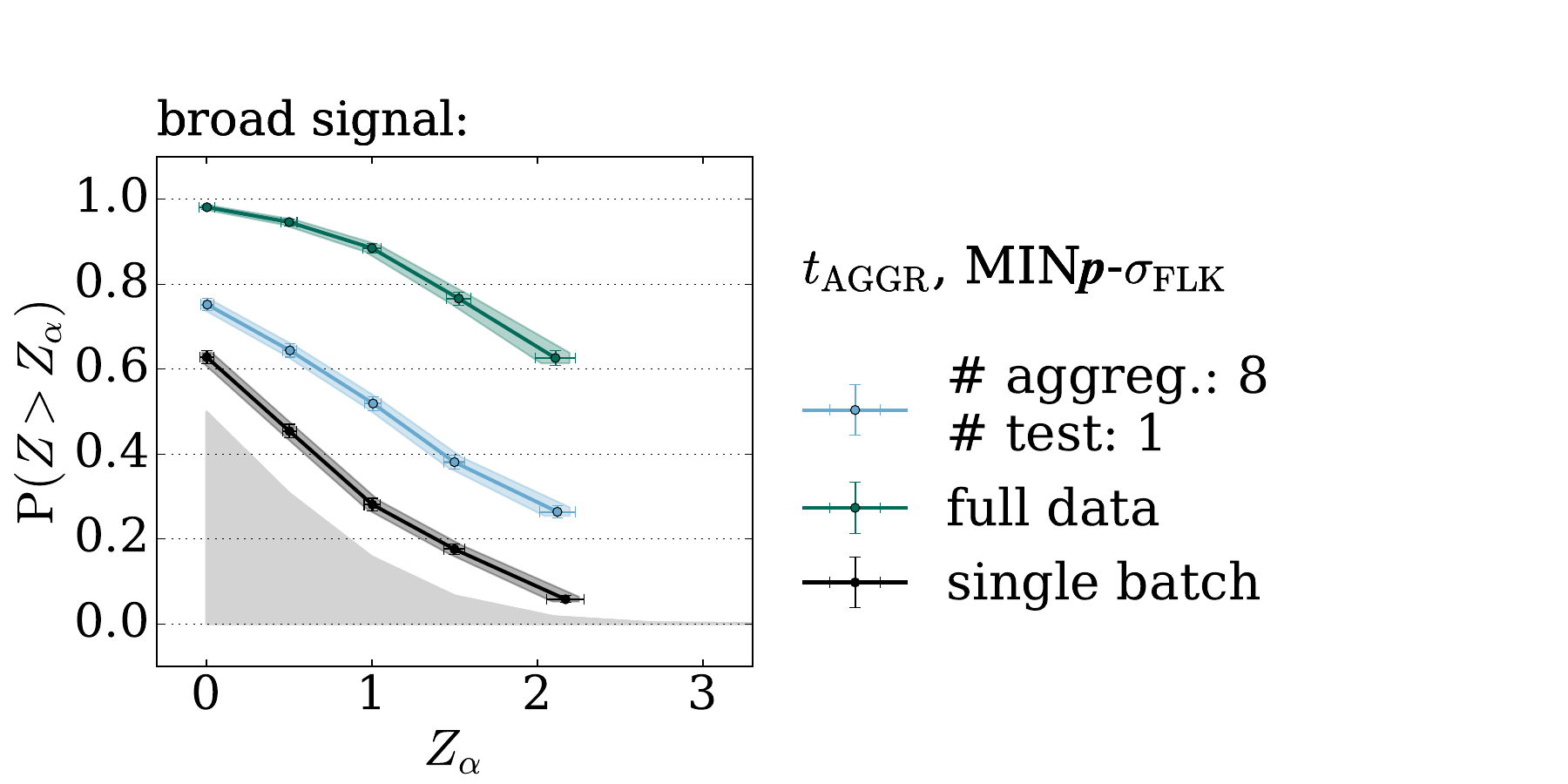}
    \caption{\textbf{NPLM-ONE. EXPO-1D.} Power curves for $t_{\rm AGGR}$ when evaluated over one single data batch (light blue line) and over the full dataset (green line), compared with the power of the test learnt and evaluated over one data batch (black line). The aggregation improves the accuracy over the learnt alternative model enhancing the sensitivity with respect to the single batch test. The two panels show the results of our tests for two signal benchmarks, the narrow peak (left panel) and the broad peak (right panel).}\label{fig:1D-nplm-one}
\end{figure}

\begin{figure}[h]
    \centering
\includegraphics[width=0.33\textwidth]{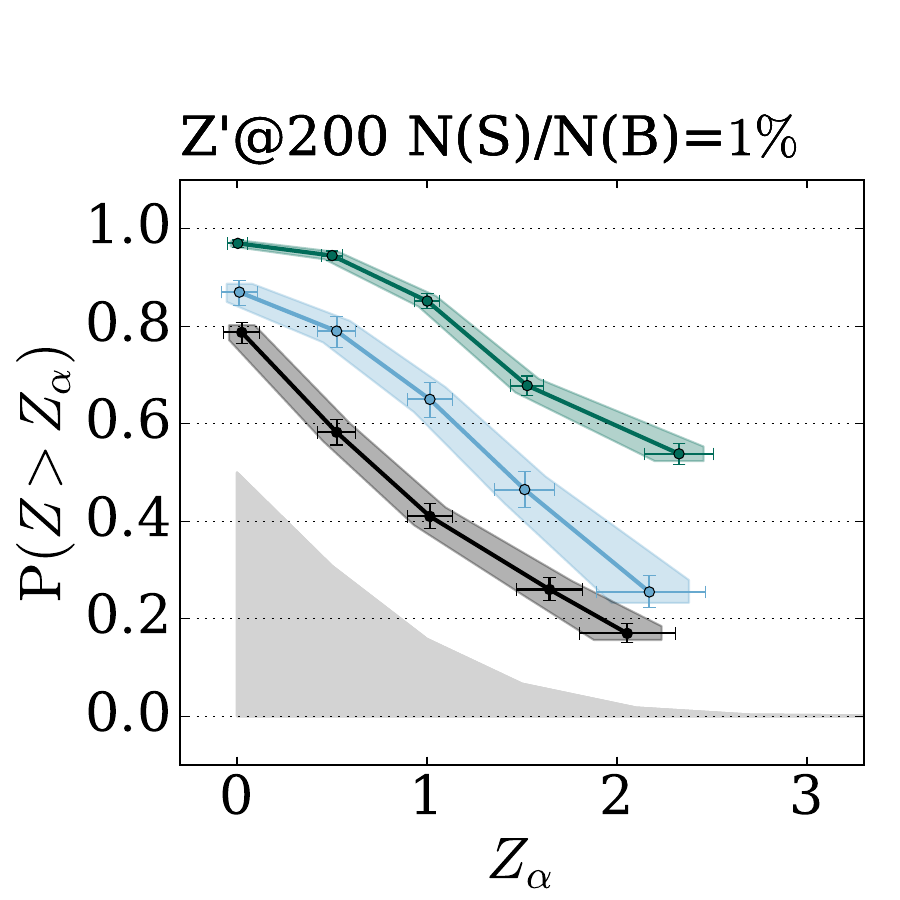}
\includegraphics[width=0.66\textwidth]{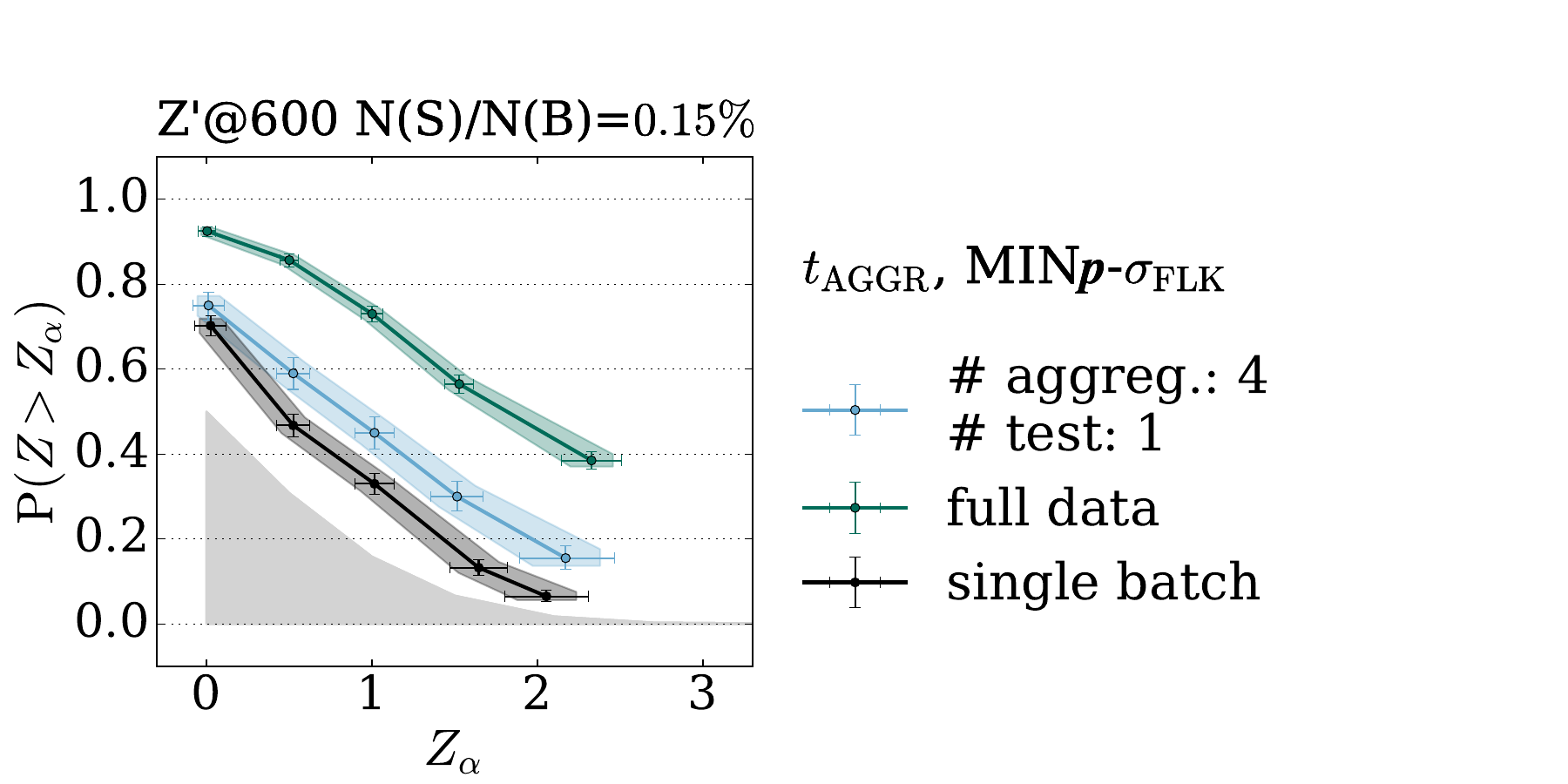}
    \caption{\textbf{NPLM-ONE. DIMUON-5D.} Power curves for $t_{\rm AGGR}$ when evaluated over one single data batch (light blue line) and over the full dataset (green line), compared with the power of the test learnt and evaluated over one data batch (black line). The aggregation improves the accuracy over the learnt alternative model enhancing the sensitivity with respect to the single batch test.The two panels show the results of our tests for two signal benchmarks, the a Z' resonance in the bulk of the mass spectrum (left panel) and and a Z' resonance in the tail of the mass spectrum (right panel).}\label{fig:5D-nplm-one}
\end{figure}

The results presented in Section~\ref{subsec:all} suggest that the aggregation improves the quality of the density-ratio modelling. 
This motivates the use of NPLM in a quasi-online setup to extract powerful information from data even when only a fraction of the latter can be eventually stored (storage constrained scenario). This is always the case at collider experiments, where trigger algorithms are designed to filter the data.
%
%
We study the impact of aggregating the modelling over multiple batches on the power of the NPLM test for a single data batch (NPLM-ONE scenario).
For the \textbf{EXPO-1D} dataset, the data are split in eight batches, all of them are analysed via NPLM but only one is saved for testing.
For the \textbf{DIMUON-5D} dataset, instead, the data are split in four batches.
Figure~\ref{fig:1D-nplm-one} shows the results of our study for the narrow (left panel) and broad (right panel) signal benchmarks in the \textbf{EXPO-1D} dataset, whereas Figure~\ref{fig:5D-nplm-one} shows the results for two different injections of a Z' resonant signal with invariant mass of 300 GeV, considered in the \textbf{DIMUON-5D} dataset. 
While not competitive with the ideal scenario of unlimited storage availability in which the full dataset can be tested (green line), the power curve of our experiment (lightblue line) shows significant improvement over the simple use of NPLM on a single batch (black line).
Similar behaviours are found for different signal benchmarks in both the univariate and multivariate datasets.
Additional plots 
are reported in Appendix~\ref{app:2}.

\subsection{Anomaly-preserving data compression via likelihood-ratio}\label{subsec:sat}
\begin{figure}[t]
    \centering
\includegraphics[width=0.33\textwidth]{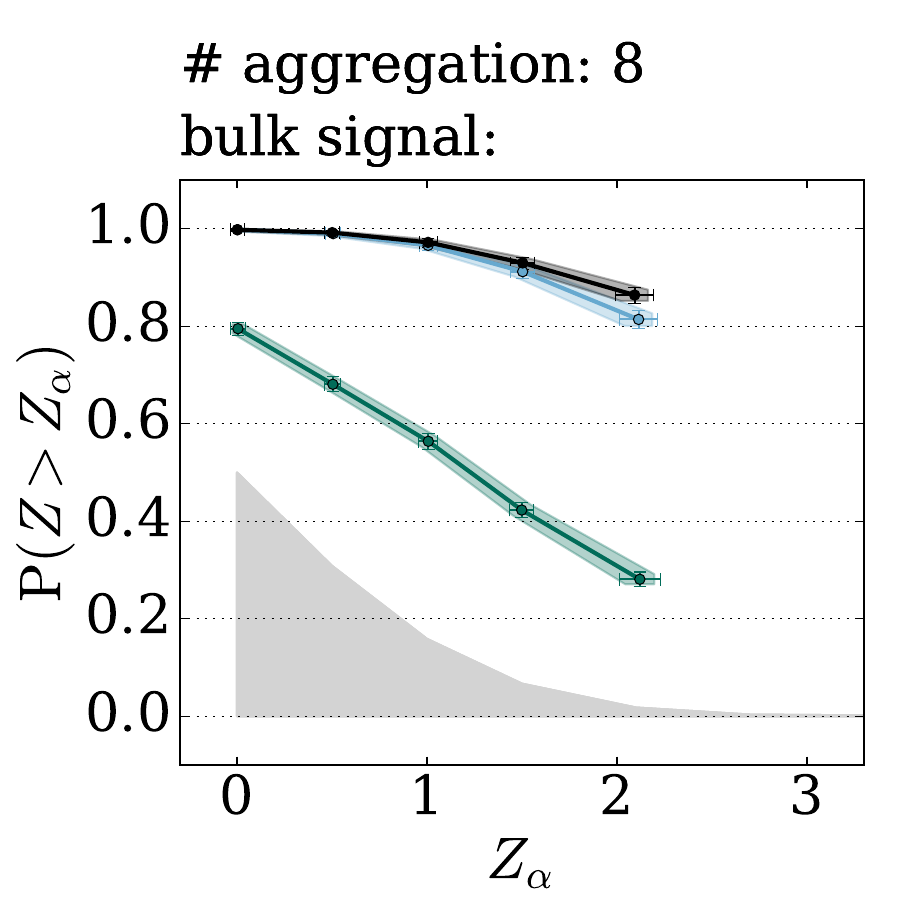}
\includegraphics[width=0.66\textwidth]{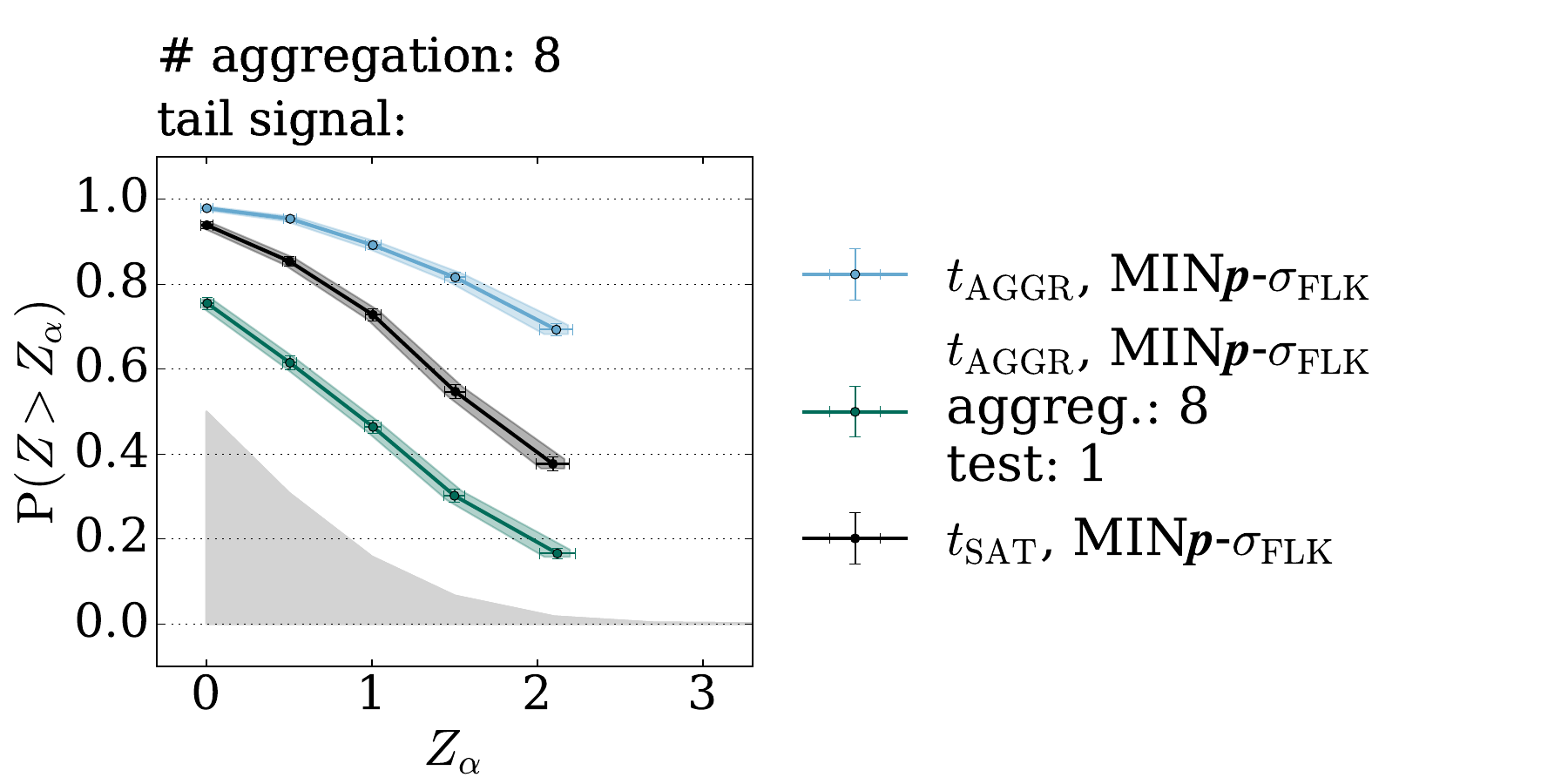}
    \caption{\textbf{NPLM-SAT. EXPO-1D.} Comparison of the aggregated and saturated test statistic power curves for a localized signal in the bulk (left panel) and in the tail (right panel). The experiments have been performed aggregating over 4 batches.}\label{fig:1D-nplm-sat}
\end{figure}
Finally, we consider the extreme scenario in which data cannot be permanently stored but are only available quasi-online for a limited time window. In this case, the proposed solution is storing the NPLM models $\{f_{\w,i}\}_{i+1}^{N_{\rm b}}$ as compressed representations of the data batches, and build the saturated likelihood-ratio test described in Eq.~\ref{eq:t-sat}.
The saturated test is not properly a likelihood-ratio test because the likelihood is not directly evaluated on the data points in $\data$. This in principle could lead to inefficiencies. A trivial failure example, for instance, is that of a signal localised in a low density region of the reference data distribution. Since the likelihood-ratio reconstructed by $F$ is only evaluated in the data points of the reference sample $\reference$, if the shape of the signal is underestimated then the test would not be sensitive to it. 
This intuition can be verified running numerical experiments of different nature. 
Figure~\ref{fig:1D-nplm-sat} shows two representative experiments run on the \textbf{EXPO-1D} dataset. In the Figure we compare the power curve of the saturated test (black line) with that of the aggregated test evaluated over the full dataset (light blue line), or on one batch only (green line). For the signal located in the bulk of the reference distribution (left panel), the saturated test has comparable power to the aggregated test evaluated over the full dataset. Instead, for the Gaussian signal in the tail of the reference distribution (right panel), the power of the saturated test deteriorates, as fewer reference data points are available in the signal region to properly quantify the discrepancy. 
Similar results are found for the \textbf{DIMUON-5D} dataset (see Figure~\ref{fig:5D-nplm-sat_aggr4}).
%
%
Additional Figures for the remaining signal benchmarks are reported in Appendix~\ref{app:3}.

%
%
\begin{figure}[t]
    \centering
\includegraphics[width=0.33\textwidth]{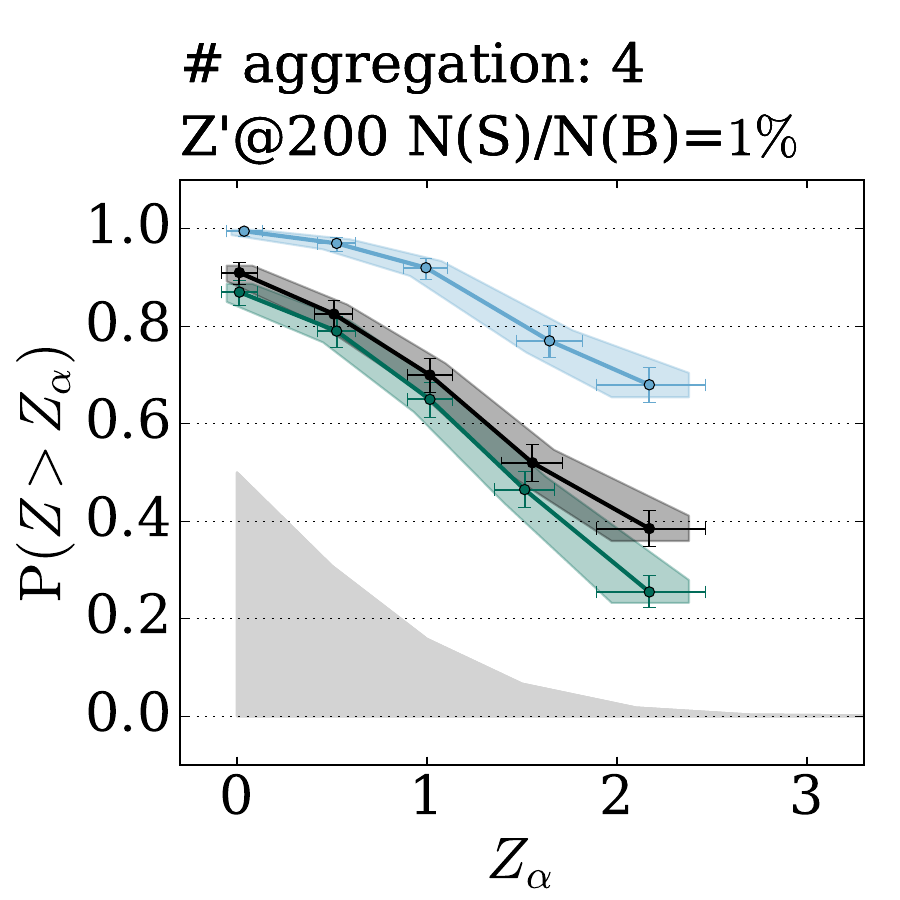}
\includegraphics[width=0.66\textwidth]{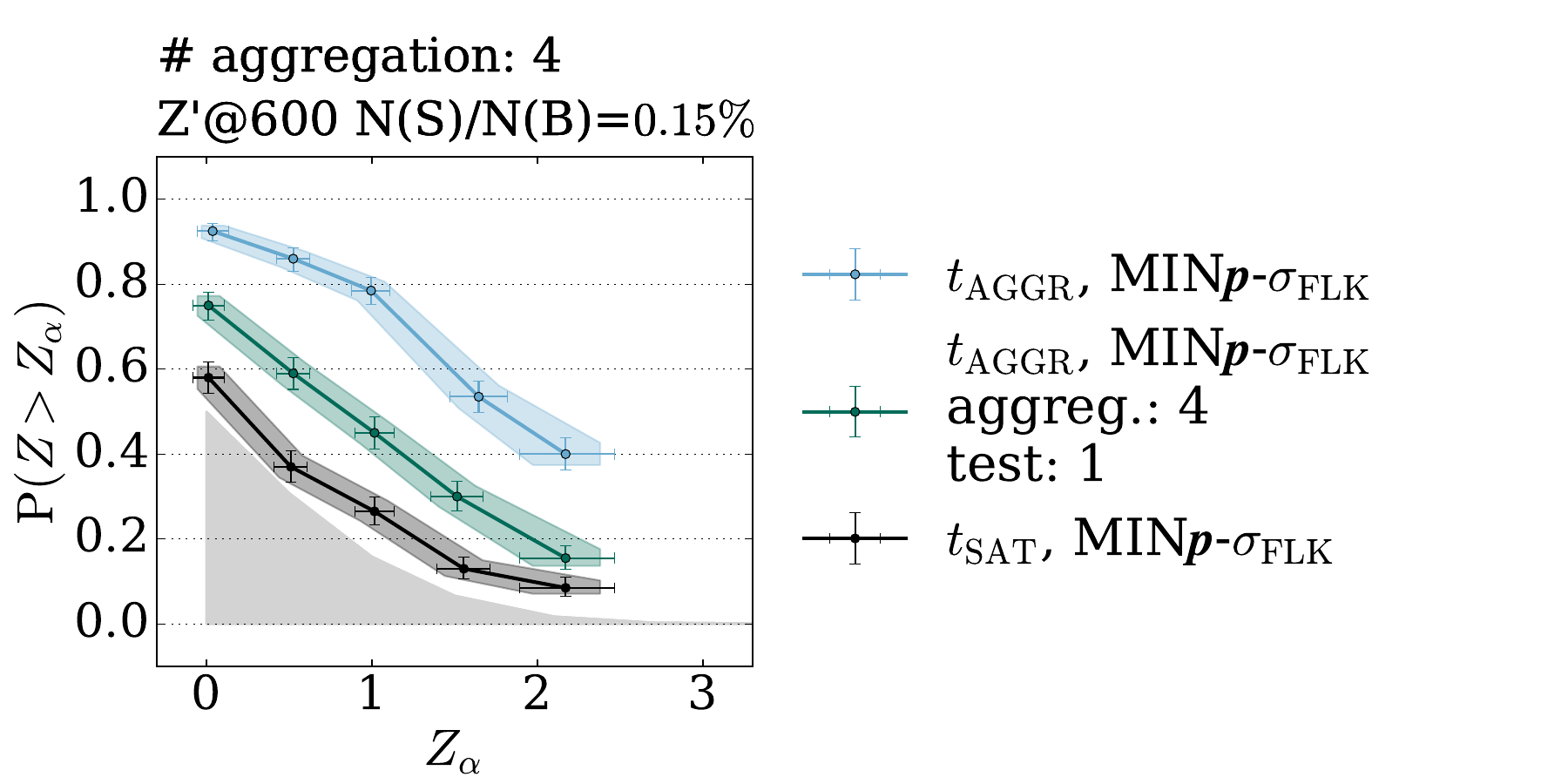}
    \caption{\textbf{NPLM-SAT. DIMUON-5D.} Comparison of the aggregated and saturated test statistic power curves for a Z' resonant signal in the bulk (left panel) and in the tail (right panel) of the mass spectrum considered in this work. The experiments have been performed aggregating over 4 batches.}\label{fig:5D-nplm-sat_aggr4}
\end{figure}
%

\section{Scaling up to big data LHC scenarios}\label{sec:3}
In this Section we apply the batch-aggregation approach to a larger scale dataset, emulating a typical data stream acquired at the CMS experiment with a one light lepton (electron or muon) requirement at the first level of data filtering (L1 trigger). The dataset, initially introduced in Refs.~\cite{Cerri:2018anq, Knapp:2020dde} and later adapted for anomaly detection~\cite{Govorkova:2021hqu} is publicly available on Zenodo~\cite{thea_aarrestad_2021_5046389,thea_aarrestad_2022_7152590,thea_aarrestad_2022_7152599,thea_aarrestad_2022_7152614,thea_aarrestad_2022_7152617}. It consists of simulated events for the Standard Model (SM) and four beyond the Standard Model signatures that we report below\footnote{Additional information about the dataset can be found in Ref.~\cite{Govorkova:2021hqu}.}: 
\begin{itemize}
    \item $A\rightarrow 4l$: a neutral scalar boson ($A$) decaying to two off-shell Z bosons, each forced to decay to two leptons;
    \item $h^{\pm}\rightarrow \tau^{\pm}\nu$: a charged scalar boson ($h^{\pm}$) decaying to a $\tau$ lepton and a neutrino;
    \item $h^0 \rightarrow \tau^+\tau^-$: a scalar boson ($h^0$) decaying to two $\tau$ leptons;
    \item ${\rm LQ}\rightarrow b\tau$: a leptoquark (LQ) decaying to a $b$ quark and a $\tau$ lepton.
\end{itemize}

The dataset contains a list of 19 physics objects (4 muons, 4 electrons, 10 jets and the event missing transverse energy) for each collision event. Each object is characterised by three kinematic variables and a class label, for a total of 76 features.
We run the batched version of NPLM over the 24 dimensional problem defined by considering the kinematic features of eight out of the nineteen objects: the missing transverse energy, the first two most energetic muons, the first two most energetic electrons, and first three most energetic jets.  We consider a data sample $\data$ of $1$ million SM events, on top of which signal events of the order of few parts per mill are injected to test the algorithm sensitivity.
The marginal distributions over the input features for the Standard Model background and the four signal benchmarks are reported in Figures~\ref{fig:cms-l1-input-panel1}
and~\ref{fig:cms-l1-input-panel2} in Appendix~\ref{app:datasets}. 

The reference sample $\reference$ is taken to be $10$ million SM events (ten times larger than the full $\data$ sample), and the $\data$ sample is split in 5 batches ($N_{\rm b}=5$).
As for the kernel methods hyper parameters, we set $M=10\,000$ and $\lambda=10^{-6}$. For sake of time, we study the performances of the model for a single value of the kernel width $\sigma_{\rm FLK}$, corresponding to the 90\% quantile of the distribution of pair-wise distance between SM data points.
We run the three versions of the algorithm presented in Section~\ref{sec:2}: train and test over all batches (NPLM-ALL), train over all and test only one batch (NPLM-ONE), and the saturated test on the model obtained averaging over all batches (NPLM-SAT).

\begin{figure}[H]
    \centering
    \includegraphics[width=\linewidth]{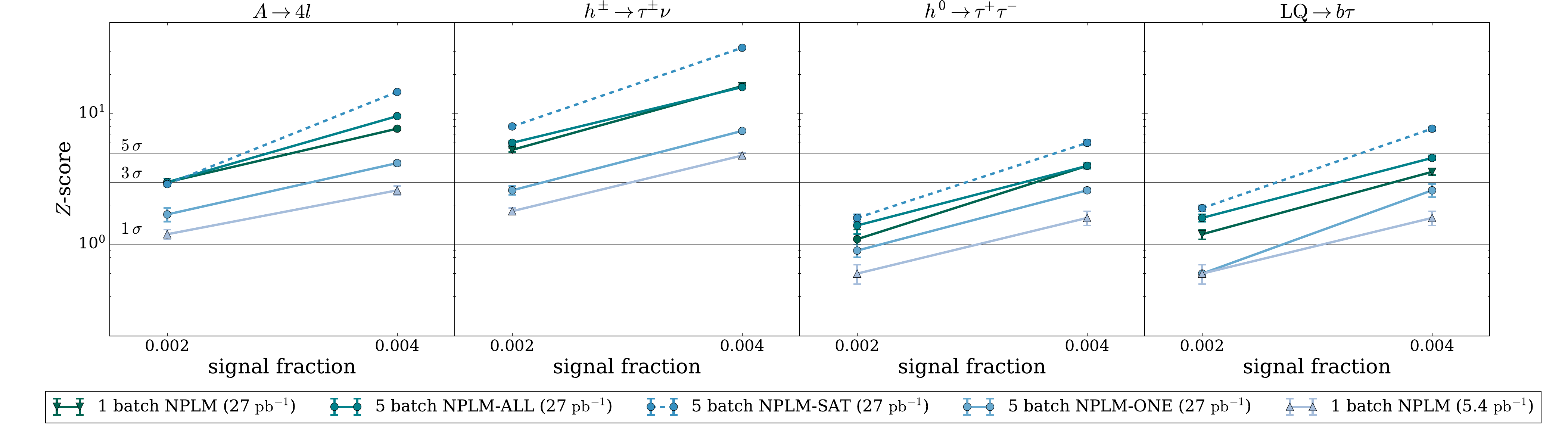}
    \caption{\textbf{CMS-L1-24D Summary plot. } Sensitivity reach in term of the median Z-scores as a function of the fraction of signal injection. The total number of background events sum to 1 million. The inspected signal fractions are of the order of few per mill. Each panel corresponds to a different signal benchmark.}
    \label{fig:summary-cms-l1}
\end{figure}

Figure~\ref{fig:summary-cms-l1} shows the median Z-score reached by the three approaches for a 0.2\% and 0.4\% signal injection of the four different signal benchmarks. For completeness, we also show the sensitivity of the algorithm if only a fifth of the luminosity, namely one batch, is exploited by NPLM.
The results of our tests confirm the main finding already outlined in Section~\ref{sec:2}. First, the NPLM-ONE approach exhibits increased performances with respect to the application of NPLM to a single batch, remarking that the aggregation via average proposed in this work allows to unveil sensitive information about the signal signature, even when the full set of data is not available at testing time.
Moreover, when all the data are available at testing time (NPLM-ALL), the NPLM algorithm can take advantage of the full sample statistics to make discovery possible. This is the case for the $A\rightarrow4l$ and $h^{\pm}\rightarrow\tau^{\pm}\nu$ signal benchmarks, where a 2 per mill signal injection (corresponding to a 2 sigma global normalization effect in our studies), can be raised to evidence (for $A\rightarrow4l$) or discovery (for $h^{\pm}\rightarrow\tau^{\pm}\nu$) level of significance
\footnote{The Z-score values reported in Figure~\ref{fig:summary-cms-l1} are computed empirically up to 3. For larger values of the default NPLM, NPLM-ALL and NPLM-ONE, we rely on the asymptotic $\chi^2$ distribution of the test statistic under the null hypothesis (the compatibility of the NPLM test with a $\chi^2$ for kernel methods was previously studied in Refs.~\cite{Letizia:2022xbe, Grosso:2023ltd}). For NPLM-SAT we don't observe the emergence of an asymptotic $\chi^2$. However the null distribution has a good agreement with a normal distribution. We therefore use the normal asymptotic to extrapolate an estimate of Z-scores above 3. Points above 3 in the Figure should therefore be taken more as a qualitative trend than an accurate estimate.}. 
For the $A\rightarrow4l$ and $h^{\pm}\rightarrow\tau^{\pm}\nu$ signal benchmarks, we observe impressive reaches for the NPLM-SAT approach as well. This result aligns with what observed in the one-dimensional and five-dimensional experiments, namely that NPLM-SAT is as successful as NPLM-ALL when the signal support is mainly in the bulk of the reference sample support, as it is the case for the signal benchmarks considered here (see marginal distributions reported in Figures~\ref{fig:cms-l1-input-panel1} and \ref{fig:cms-l1-input-panel2}). Additionally, this result points out the potential of the NPLM split-aggregation method as a 
tool to compute \textit{anomaly-preserving summary statistics}, allowing to save information about data that are only temporarily available and eventually discarded by the data acquisition system.
We conclude this Section with commenting on the performances of NPLM on the $h^{\pm}\rightarrow \tau^{\pm}\nu$ and ${\rm LQ}\rightarrow b\tau$ signals, that are found to be poorer. We believe that the reason could reside in the choice of the features adopted in these studies -- only a subset of the original nineteen physics objects is retained for this analysis.
Moreover, the algorithm performances are sensitive to the size of the reference sample $\reference$ -- our studies suggest that the larger the sample the more accurate the result becomes, though more computationally demanding. Finally, one peculiar feature of this dataset is the so called ``zero padding", namely filling with zeros the entries corresponding to specific objects in the event that are not observed. The zero padding gives rise to sharp features that project great part of the data on lower dimensional surfaces of the input manifold. The problem of variable size events could be solved choosing a different data representations, model architecture, or by mapping the data to a lower dimensional embedding. These interesting directions are out of the scope of this paper and are left for future work.
\section{Computational resources and training time}\label{subsec:resources}
In this Section, we report the details of the algorithm implementation, highlighting the typical execution time and computational resources needed for the numerical experiments studied in this work.
\paragraph{Execution time.}
The execution time of a single instance of the NPLM algorithm based on kernel methods depends on several factors: the total number of data points, the number of kernels $M$, the kernel width $\sigma_{\rm FLK}$, the smoothness parameter $\lambda$, and on the hardware resources employed. Our implementation relies on the Falkon library~\cite{falkonlibrary2020}, that allows to speed up the fitting time by optimally exploiting GPUs~\cite{Letizia:2022xbe}.
\begin{table}[H]
    \centering
    \scalebox{0.99}{
    \begin{tabular}{c|ccccc|c}
    dataset & $M$ & $\sigma_{\rm FLK}$ &$\lambda$ & $\Nreference$ & $\Ndata$ (all) & Time (single batch)\\
    \toprule
    EXPO-1D 
        & 1k
        & [0.1, 0.3, 0.7, 1.4, 3.0]
        & $10^{-6}$
        & 200k
        & 16k
        & 0.5--1.5 s\\
    DIMUON-5D
        & 10k
        & [0.8, 1.9, 3.0, 4.4, 6.6]
        & $10^{-6}$
        & 40k
        & 8k
        & 3--12 s\\
    L1CMS-24D
        & 10k
        & [10.5]
        & $10^{-6}$
        & 10M
        & 1M
        & 40-70 s\\
    \end{tabular}}
    \caption{Model hyper parameters and execution time. Unless stated otherwise, all instances of the NPLM algorithm for each dataset follow the settings described in this table.}
    \label{tab:hyper}
\end{table}
We report in Table~\ref{tab:hyper} the typical time range for a single batch execution for the hyper parameters choice adopted in our experiments. 
The variance within experiments on the same dataset is mainly due to the $\sigma_{\rm FLK}$ value, smaller values require a higher number of iterations of the Falkon algorithm to reach convergence.
The main difference in execution time between the univariate and multivariate cases is due to the model size, being the number of kernels used in the five-dimensional problems a factor 10 larger than in the one-dimensional case.
The training time is not significantly affected by the batch size, as the reference sample size is dominating the overall size of the training sample. For large samples and complex models the execution time increases exponentially with the dataset size (see studies reported in Ref.~\cite{Letizia:2022xbe}), and the benefit of batching the dataset in terms of execution time become more striking. This has been observed, for instance, when studying the L1CMS-24D.

\paragraph{Storage resources.}
Aggregating over batches requires both temporary and permanents computational and storage resources. The temporary resources are used to train each single instance of the algorithm and should be sufficient to store the data batch input to the model, and the model itself. Importantly, the model location should be accessible by all data batches to allow estimating $F^{N_{\rm b}}$ according to Eq.~\ref{eq:F}.
The reference sample $\reference$ and the values that the aggregated model $F^{N_{\rm b}}$ takes in the test points should instead be stored in a permanent location.
The reference sample is a static input to the NPLM algorithm and is common to all training instances.\footnote{As previously stated, this work doesn't cover the case of an imperfect reference, namely the case in which the reference model is affected by systematic uncertainties and its optimal configuration given the experimental condition is not constant. We leave this case for future work.} Together with the test data, it is used to asses the value of the test in all three versions described in Section~\ref{sec:2}.
The amount of memory resources to be allocated for the aggregated model depends on the number of data points required by the test. For the test in Eq.~\ref{eq:t-avg}, the aggregated function $F^{N_{\rm b}}$ is evaluated over all data points and reference points, requiring a vector of length $\Nreference+\Ndata$. For the test in Eq.~\ref{eq:t-avg_v2}, only a subset $N_{\rm test}<N_{\rm b}$ of batches is used for evaluation and therefore the vector length reduces to $\Nreference+ \frac{N_{\rm test}}{N_{\rm b}}\Ndata$. Finally for the saturated test in Eq.~\ref{eq:t-sat}, the aggregated function is only evaluated over the reference set, requiring a $\Nreference$ length vector only.
For the first two cases, the storage resources needed scale linearly with the data taking time, while in the last case they are constant. The implications of information loss on the sensitivity of the different summary statistics are discussed in Section~\ref{subsec:one} and~\ref{subsec:sat}.

\section{Conclusions}\label{sec:4}
In this work we explore the possibility of scaling the use of NPLM to test large size data samples in presence of variable computational and storage resource constraints.
We propose to split the dataset in batches and run parallel instances of the NPLM algorithm on each of them. The final results are combined averaging the density-ratio functions learnt by each instance of the NPLM algorithm.
We apply this strategy to both univariate and multivariate benchmarks representative of typical data distributions in particle physics experiments.
Our experiments show that the split-aggregation strategy preserves or even surpass the sensitivity power of the NPLM test applied to the full sample (original implementation). 
The reason for the success is mainly related to the uniqueness of its training strategy, aiming at finding the maximum likelihood fit of the model to the input data, a representation of the true model underneath the data that over-fits the specific sample used for training.
While over-fitting is commonly evaded in machine learning algorithms, here it assumes a crucial role in preserving data structures that do not generalise, either because they resemble statistical fluctuations or because they are especially rare traits of the data.
Preserving rare structures is what makes the NPLM batching successful in recombining subtle hints of anomalous events into a statistically significant signal, also suggesting that the density-ratio functions learnt by the NPLM algorithm could be good candidates to build \textit{anomaly-preserving summary statistics}.\\

Moreover, the new NPLM split-aggregation strategy offers a solution to offline signal-agnostic analyses of collider data, often characterised by large sample sizes. It is beneficial both in terms of training time and sensitivity performances. thanks to its effect of \textit{regularization} against statistical fluctuations, able in some cases to increase the significance of the observed discrepancy.
It should be mentioned that rigorously treating systematic uncertainties affecting the reference sample in this context is crucial. In this work we don't address the problem of systematic uncertainties. A way to address systematic uncertainties in the original NPLM implementation was proposed in Ref.~\cite{dAgnolo:2021aun} for neural network based models. In the split-aggregation approach additional challenges arise from the fact that each batch has access only to ``local" information about uncertainties. An accurate global assessment of the systematic effects via nuisance parameters estimation needs therefore the design of a successful aggregation strategy over the batches (see for instance Ref.~\cite{Heinrich:2023bmt}). The extension  of the approach proposed in Ref.~\cite{dAgnolo:2021aun} to handle systematic uncertainties in the split-aggregation NPLM strategy is subject of ongoing study and left for future work. \\
%
%

The potential application of the new batch-based NPLM strategy extends beyond the offline analysis, to quasi-online signal-agnostic analyses of streamed data that are only temporarily available. 
Online approaches to data analyses allow to inspect collider data at the experiments prior the stage of filtering applied by the triggers, opening the way to the exploration of phase space regions never analysed before, at significantly high rates. Efforts in this direction are ongoing at the LHC experiments.
The CMS experiment is investigating new approaches to analyse the experimental data at the collision rate of 40 MHz with focus on the muon detector chambers  and the electromagnetic calorimeters~\cite{CMS:2023ukn, Migliorini:2024htd}. 
Moreover, the LHCb experiment has recently upgraded to a triggerless data readout at a rate of roughly 30 MHz, with partial event reconstruction that allows to reduce the rate down to 1 MHz prior the final trigger stage~\cite{Aaij:2019zbu}.
In this work we proposed two possible solutions to run the NPLM algorithm over a continuous stream of data under storage and computational resources constraints.
In case of limited storage, we propose to train NPLM over multiple batches and evaluate the test only on one batch available for long term storage (NPLM-ONE). In complete absence of permanent storage, we propose to train NPLM over multiple batches and perform a goodness-of-fit test of the aggregated density model following the saturated likelihood-ratio approach (NPLM-SAT).
Experiments run under constrained storage scenarios (NPLM-ONE) show that performing the aggregation over batches improves the modelling of the signal increasing the power of the NPLM test applied to a single batch.
As for the saturated test, we observe sensitivity powers comparable to those of the NPLM algorithm applied to the full statistics when the signal lays on high density regions, while the power is only partially recovered for signals localised in the tails. In both cases, the sensitivity is expected to scale as the statistics of the processed data increases. 
%
%
Online analysis is a promising avenue for anomaly detection. The smart use of machine learning algorithms can leverage large statistics to efficiently navigate unexplored data and recognise novel rare processes of unexpected nature or location.
The studies presented in this work are a first step in the direction of investigating online or quasi-online solutions for statistical anomaly detection with the NPLM algorithm. Detailed studies on scalability and uncertainties quantification in typical LHC data streaming scenarios are left for future work.

\section*{Acknowledgments} 
This work is supported by the National Science Foundation under Cooperative Agreement PHY-2019786 (The NSF AI Institute for Artificial Intelligence and Fundamental Interactions, http://iaifi.org/). 
 Computations
in this paper were run on the FASRC Cannon cluster supported by the FAS Division of Science
Research Computing Group at Harvard University. 
The author would like to thank Siddharth Mishra-Sharma, Phil Harris, Marco Zanetti and Marco Letizia for the useful conversations and constructive feedback on the manuscript.

\bibliographystyle{ieeetr}
\bibliography{bibliography}

\beginsupplement

\section{Supplementary Materials}
\subsection{Datasets details}\label{app:datasets}
\paragraph{EXPO 1D.} This simple univariate setup introduced in Ref.~\cite{DAgnolo:2018cun} and further studied in  Ref.~\cite{dAgnolo:2021aun, Grosso:2023scl} represents an energy or transverse-momentum spectrum that falls exponentially. Such type of distributions are fairly common in collider physics experiments. Studying GoF techniques in this setup is thus illustrative of some of the challenges associated with the search for new physics at these experiments.
The density distribution under the reference model is defined as
\beq
n(x|{\RH}) = {\rm N(R)}\, e^{-x}\, ,
\eeq
where ${\rm N(R)}$ denotes the number of expected events in the dataset. 
In our studies we set ${\rm N(R)}$ at $16{\rm k}$ events for the full dataset $\data$, and we consider batches of size $8{\rm k}$ (2 batches), $4{\rm k}$ (4 batches) or $2{\rm k}$ (8 batches) events.
To test the model ability to detect features of various narrowness and in different locations, we consider four Gaussian signals
\beq
n(x|{\rm G}_{\mu,\sigma}) = {\rm N(S)}\frac{1}{\sqrt{2\pi}\sigma_{\rm NP,i}}\,e^{-\frac{(x-\mu)^2}{2\sigma^2}}\,,
\eeq
and an excess in the tail defined by
\beq
n(x|{\rm E}) = \frac{\rm N(S)}{2}x^2e^{-x}\,.
\eeq

The data distribution under the alternative hypotheses has therefore the following form
\beq
n(x|{\rm H_i})= \,n(x|{\rm R}) + n(x|{\rm S})\,,
\eeq
with ${\rm S}=\{{\rm G}_{\mu,\sigma}, {\rm E}\}$.
For this toy model, the ideal test statistic according to Neyman--Pearson can be computed analytically as
\begin{equation}\label{eq:t_id_1D}
t_{id}^{\rm S}({\data})= 2\log\left[e^{\rm -N(S)}\prod_{x\in {\data}}\frac{n(x|{\rm S})+n(x|{\rm R})}{n(x|{\rm R})}\right]\,.    
\end{equation}

The fraction of signal injection over reference events, $\rm N(S)/N(R)$, the Z-score\footnote{The Z-score is defined as the quantile of a standard normal distribution whose survival function matches the $p$-value $p$:
$$Z = \Gamma^{-1}(1-p)$$} for the ideal test ($Z_{id}$),  as well as the location $\mu$ and scale $\sigma$ of the Gaussian signals are reported in Table~\ref{tab:1d-dataset}.
\begin{table}[h]
    \centering
    \begin{tabular}{llcccc}
         &&\textbf{$\mu$}&  \textbf{$\sigma$}& \textbf{N(S)/N(R)} & \textbf{$Z_{id}$ (full batch)}\\
         \toprule
         \textbf{Gaussian peaks}
         &bulk   &  1.6  &  0.16& $1.5\cdot 0^{-2}$&4.9\\
         &broad  &  4    &  0.64& $6.5\cdot 0^{-3}$&4.2\\
         &narrow &  4    &  0.01& $1.5\cdot 0^{-3}$&4.8\\
         &tail   &  6.4  &  0.16& $1.5\cdot 0^{-3}$&4.0\\
         \midrule
         \textbf{Tail excess}&excess &       &      & $1.5\cdot 0^{-2}$&4.5\\
         \bottomrule
    \end{tabular}
    \caption{Summary of the 1D signal benchmarks. }
    \label{tab:1d-dataset}
\end{table}
\begin{figure}[t]
    \centering
    \includegraphics[width=1\linewidth]{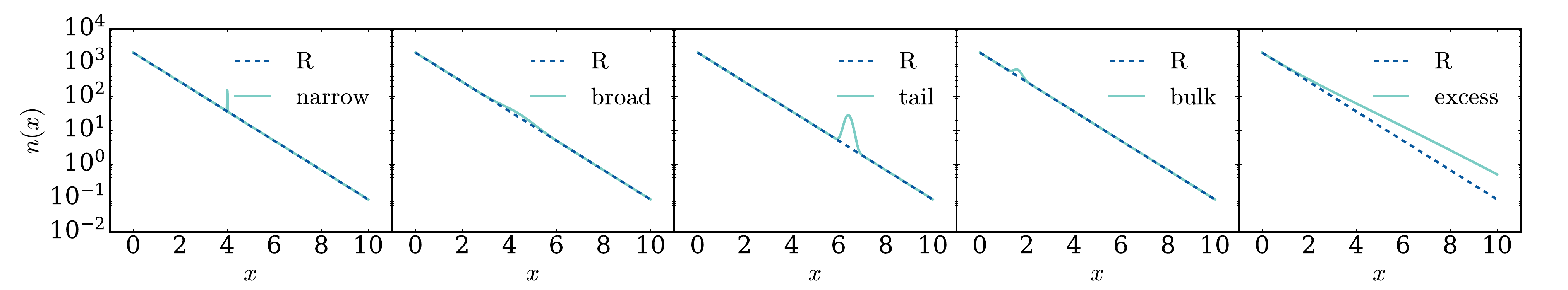}
    \caption{Expo-1D signal benchmarks}
    \label{fig:1d-data}
\end{figure}

Running the NPLM algorithm requires a reference sample ($\reference$) to compare the data $\data$ with. Our $\reference$ sample is composed of $\Nreference=200k$ events, sampled according to the $\RH$ hypothesis.
It should be noticed that this studied is conducted under the assumption that we can sample synthetic data from the $\RH$ hypothesis, which is perfectly known, at will.

\paragraph{Dimuon 5D.}
The second benchmark considered in this work consists of a set of Monte Carlo simulated proton-proton collisions happening at 13 TeV at the LHC, with two opposite charged muons in the final state Ref.~\cite{grosso_gaia_2021_4442665}.
In this dataset, each event is represented by five variables describing the kinematics of the dimuon system: the transverse momentum of each muon ($p_{T,1(2)}$), their pseudorapidities ($\eta_{1(2)}$), and the relative azimuthal angle between the two objects ($\Delta\phi_{12}=\phi_1-\phi_2$). We focus on final states with transverse momenta greater than 20 GeV, pseudorapidities lower than 2.1 in absolute value, and with invariant mass of the dimuon system larger than 60 GeV. The dominant background process in this configuration is the Drell-Yan, that for sake of simplicity we consider as the only source of background.
In our experiments the total number of expected events in the SM background hypothesis is around $\NRef=8000$, after the acceptance selections are applied. We then study the impact of splitting the data in four batches and combining them via the batch-based strategy proposed in Section~\ref{sec:2}.
As signal benchmarks, we considered a set of $Z'$ bosons with variable mass and width and a set of EFT scenarios described by the dimension-6 4-fermions Lagrangian $\frac{c_W}{\Lambda}J_{L\mu}^aJ_{La}^{\mu}$, with variable Wilson coefficients. Details on the signal benchmarks are given in Table~\ref{tab:5d-dataset}. 
\begin{table}[H]
    \centering
    \begin{tabular}{ccccc}
    
         &  \textbf{Mass (GeV)}&  \textbf{$\boldsymbol{\Sigma}$ (GeV)}& \textbf{N(S)/N(R)}&\textbf{$Z_{id}$ (full batch)}\\
         \toprule
         \textbf{Z' scenarios}
         &  \multirow{2}{*}{600}&  16& $6\cdot 10^{-4}$& 3\\
         &  &  16& $1.5\cdot 10^{-3}$& 7.1\\
         &  \multirow{2}{*}{300}&  8& $2.5\cdot 10^{-3}$  & 5\\
         &  &  8& $5\cdot 10^{-3}$ & 8.8\\
         &  \multirow{2}{*}{200}&  5& $5\cdot 10^{-3}$ & 5.1\\
         &  &  5& $1\cdot 10^{-2}$ & 9.1\\
         &  &  &  & \\
         &  \multicolumn{2}{c}{\textbf{$\boldsymbol{c_w ({\rm TeV^{-2}})}$}}& \textbf{N(D)/N(R)}\\
         \toprule
         \textbf{EFT scenarios}
         & \multicolumn{2}{c}{1.0}  & 1.002 & 3.7\\
    \end{tabular}
    \caption{Summary of the MUMU signal benchmarks. The signal yields, N(S), and data yileds, N(D), are the ones expected for the luminosity of one batch in our experiments configuration.}
    \label{tab:5d-dataset}
\end{table}

\begin{figure}[H]
    \centering
    \begin{subfigure}[h]{\textwidth}
    \includegraphics[width=1\linewidth]{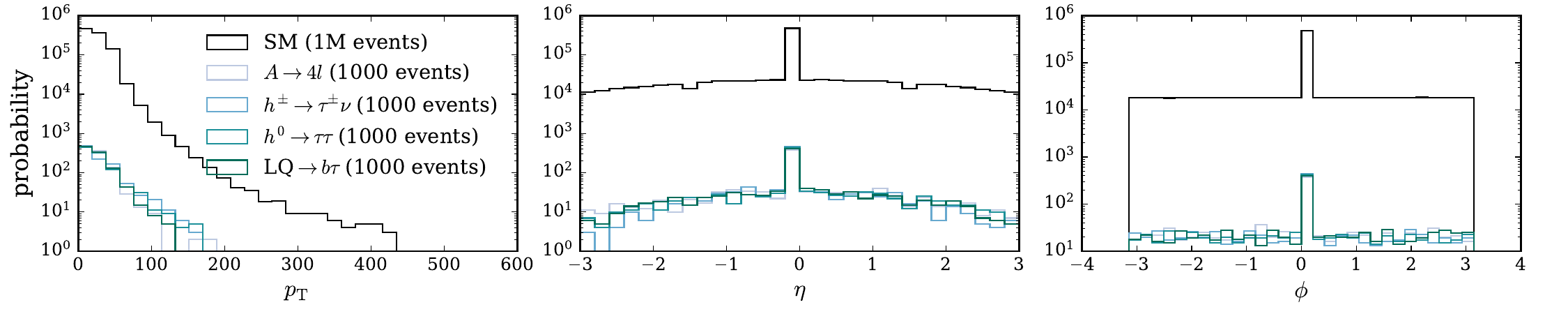}
    \caption{Leading muon.}
    \end{subfigure}\\
    \begin{subfigure}[h]{\textwidth}
    \includegraphics[width=1\linewidth]{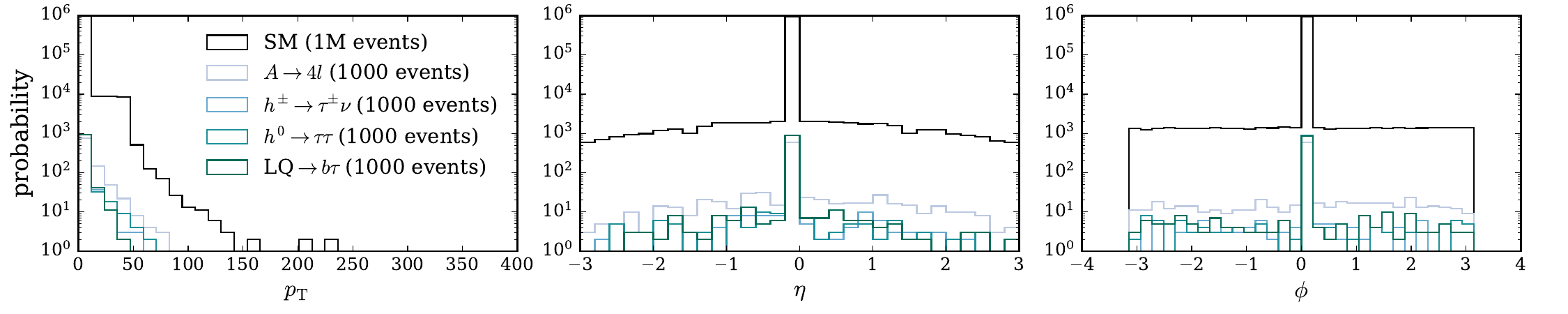}
    \caption{Subleading muon.}
    \end{subfigure}\\
    \begin{subfigure}[h]{\textwidth}
    \includegraphics[width=1\linewidth]{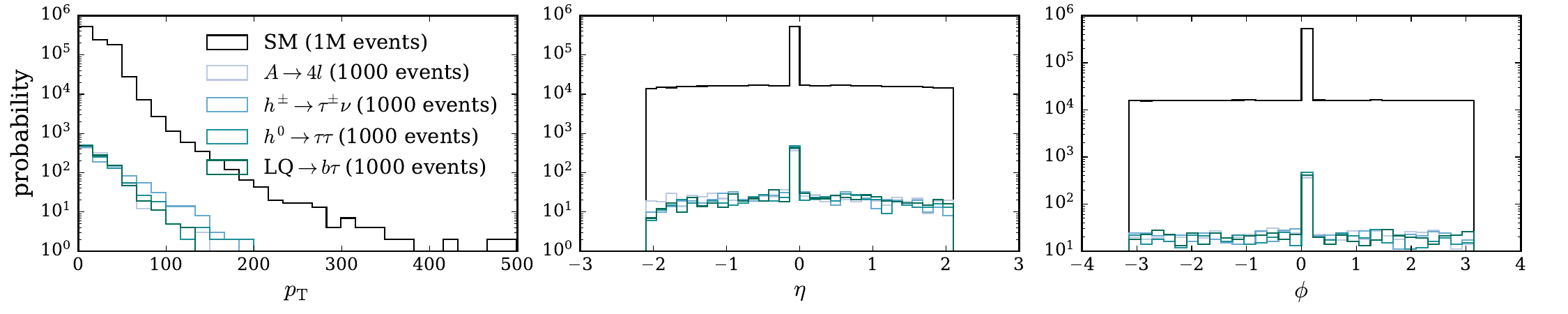}
    \caption{Leading electron.}
    \end{subfigure}\\
    \begin{subfigure}[h]{\textwidth}
    \includegraphics[width=1\linewidth]{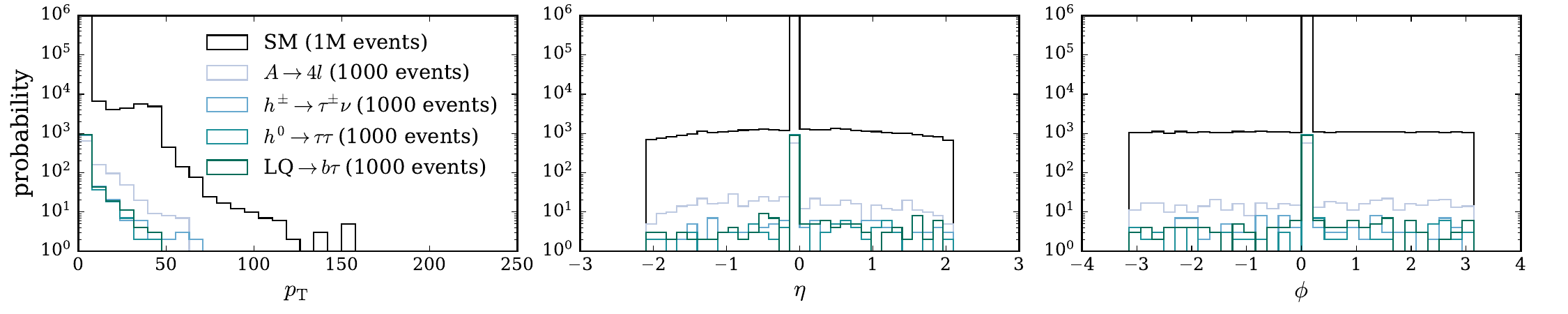}
    \caption{Subeading electron.}
    \end{subfigure}\\
    \caption{\textbf{Input features for the CMS-L1-24D dataset (panel 1).} In each row we show the transverse momentum, pseudorapidity and azymuthal angle of an object. }
    \label{fig:cms-l1-input-panel1}
\end{figure}

\begin{figure}[H]
    \centering
    \begin{subfigure}[h]{\textwidth}
    \includegraphics[width=1\linewidth]{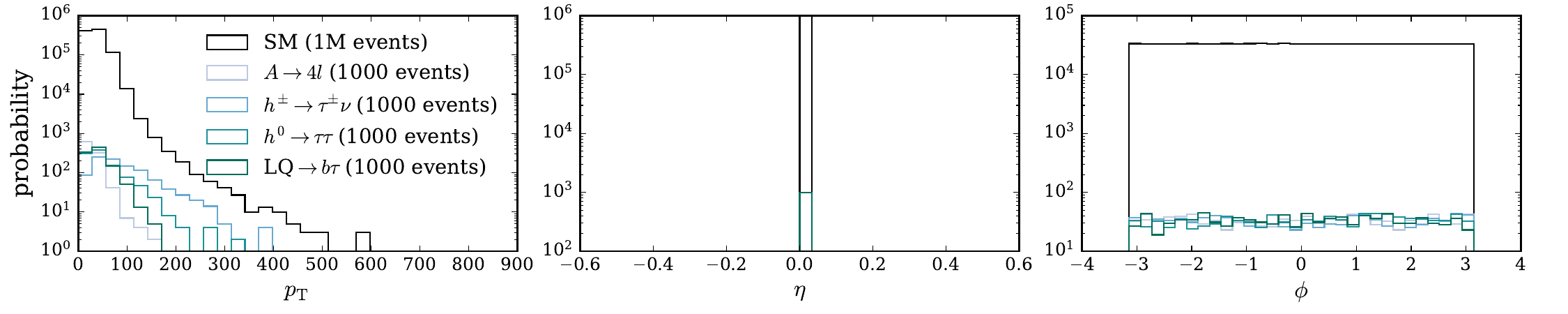}
    \caption{Missing transverse energy.}
    \end{subfigure}\\
    \begin{subfigure}[h]{\textwidth}
    \includegraphics[width=1\linewidth]{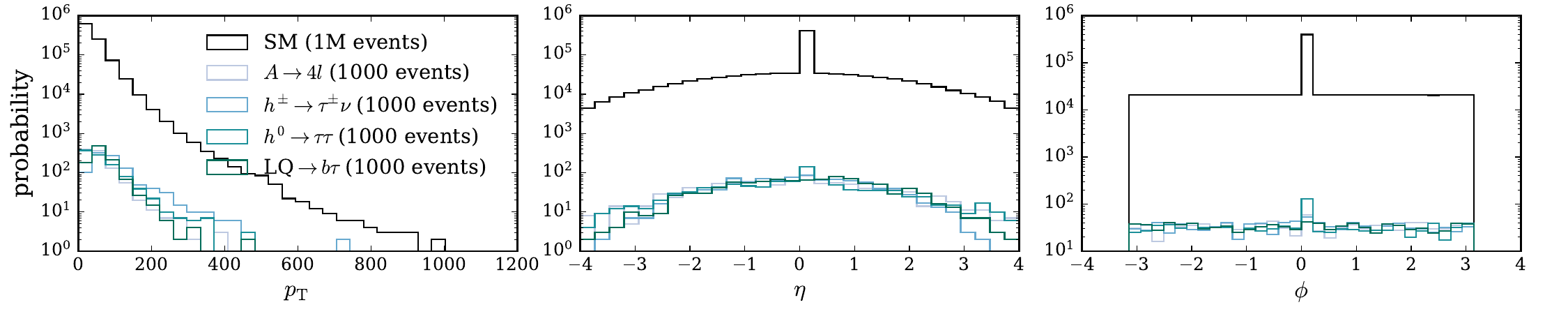}
    \caption{Leading jet.}
    \end{subfigure}\\
    \begin{subfigure}[h]{\textwidth}
    \includegraphics[width=1\linewidth]{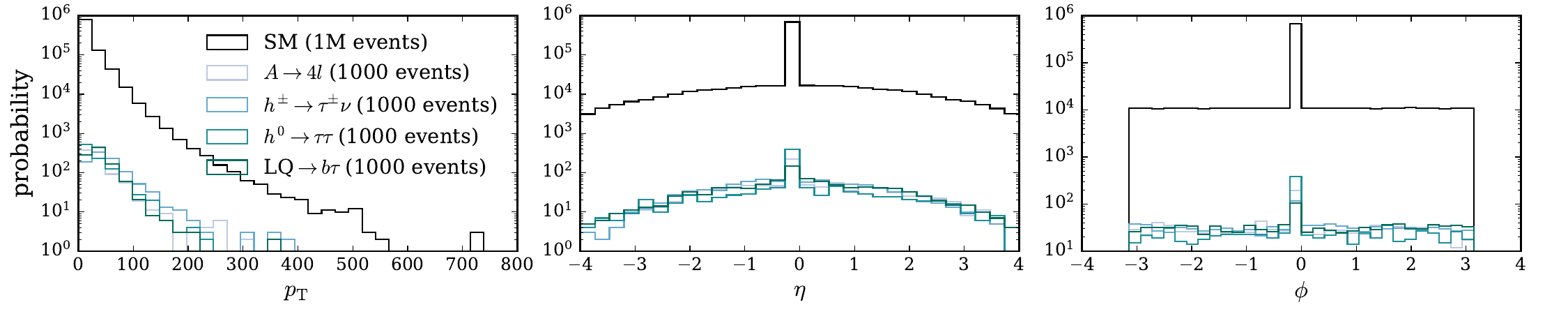}
    \caption{Second leading jet.}
    \end{subfigure}\\
    \begin{subfigure}[h]{\textwidth}
    \includegraphics[width=1\linewidth]{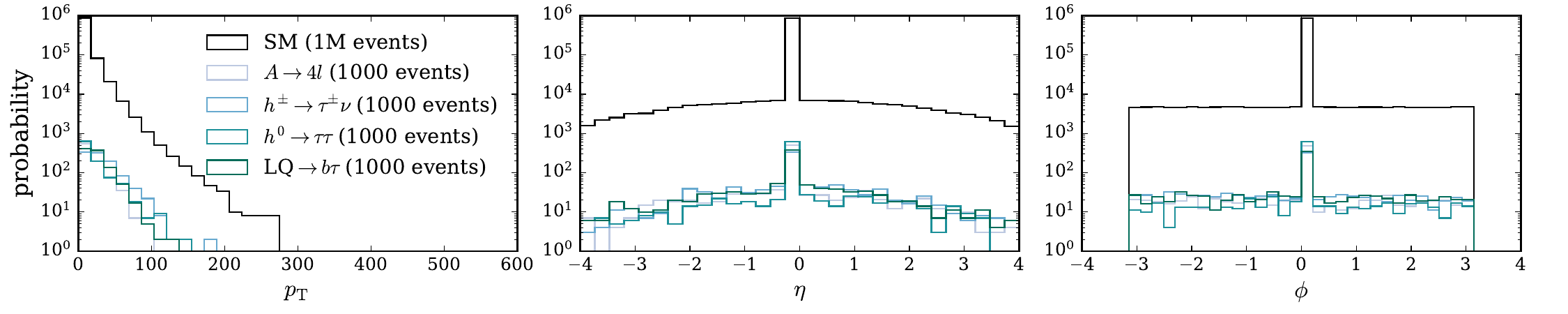}
    \caption{Third leading jet.}
    \end{subfigure}\\
    \caption{\textbf{Input features for the CMS-L1-24D dataset (panel 2).} In each row we show the transverse momentum, pseudorapidity and azymuthal angle of an object. }
    \label{fig:cms-l1-input-panel2}
\end{figure}

\clearpage
\paragraph{CMS-L1 24D.} As supplementary material we report the marginal distribution of the SM dataset and the four signal benchmarks over the 24 input variables considered in this work (Figures~\ref{fig:cms-l1-input-panel1},~\ref{fig:cms-l1-input-panel2}). The narrow peak at zero present in all plots is the effect of ``zero padding" (see main text for more comments).

\clearpage
\subsection{Improving regularization and sensitivity by aggregating over batches.}\label{app:1}
\begin{figure}[H]
    \centering
\includegraphics[width=0.19\textwidth]{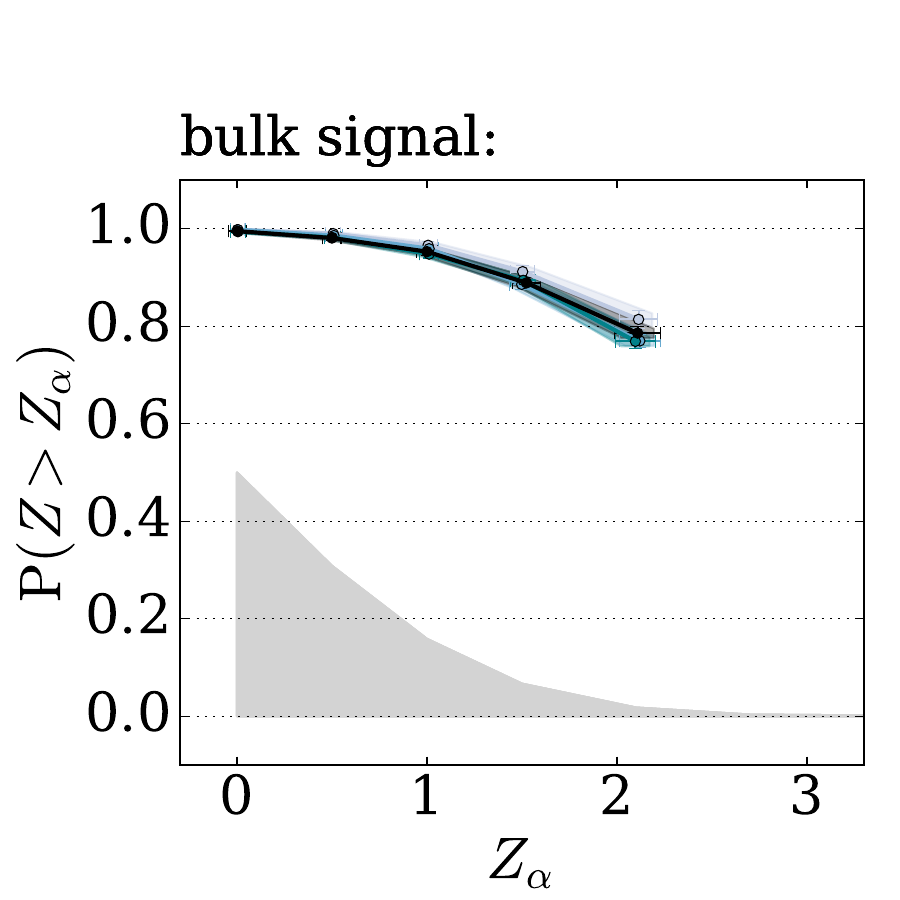}
\includegraphics[width=0.19\textwidth]{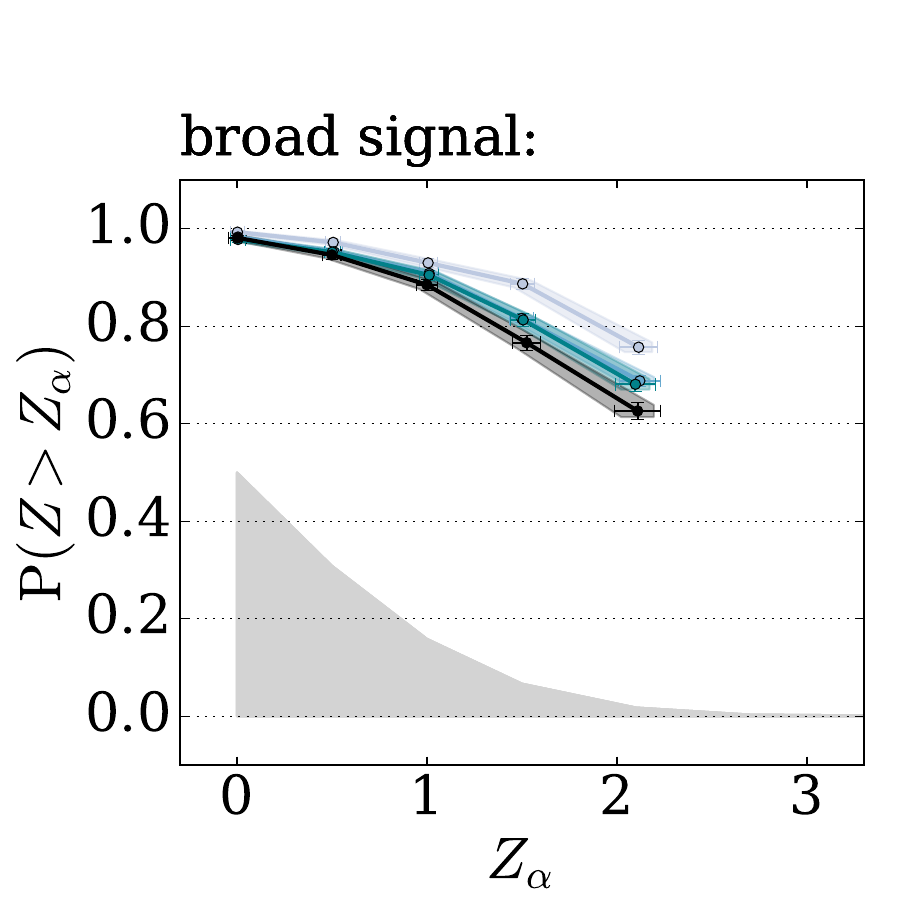}
\includegraphics[width=0.19\textwidth]{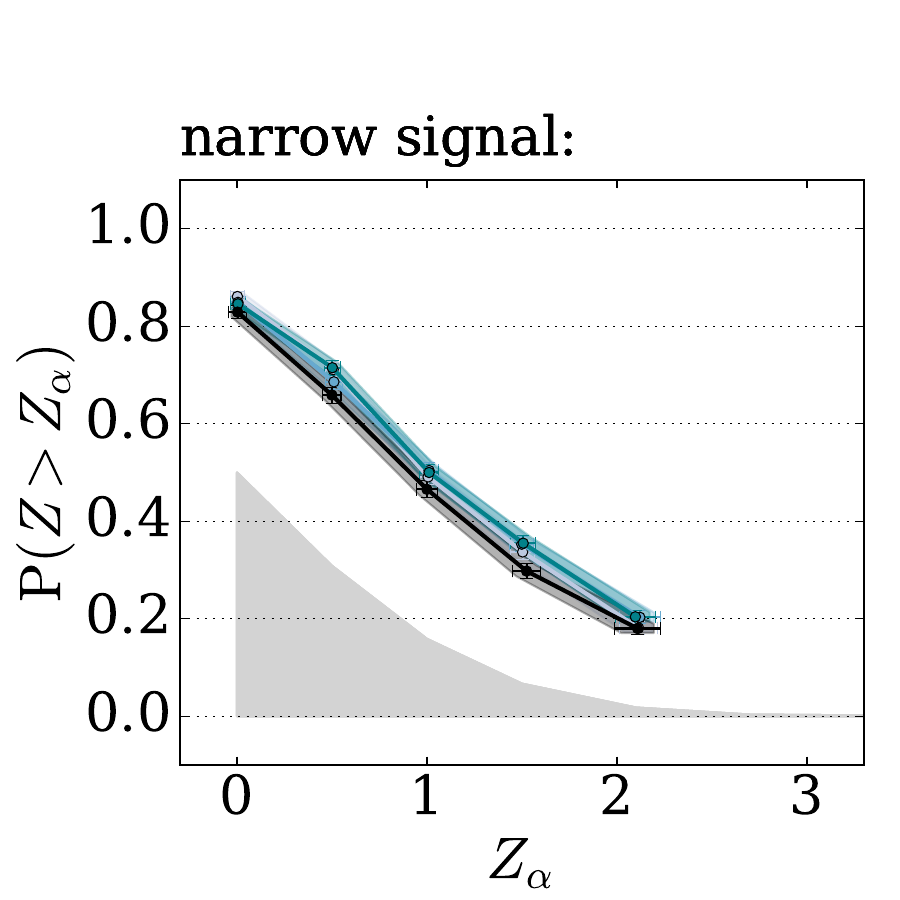}
\includegraphics[width=0.38\textwidth]{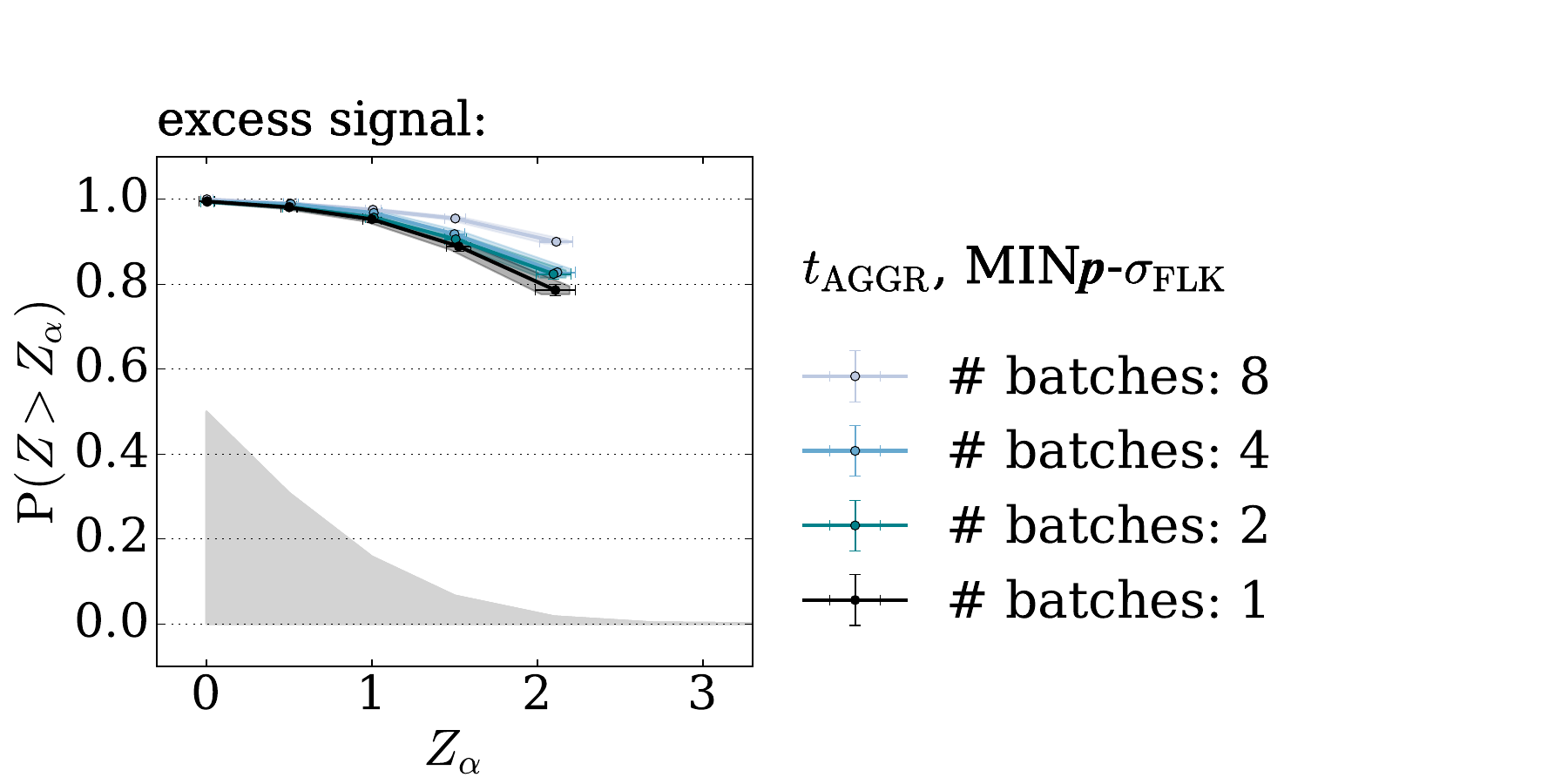}\\

\includegraphics[width=0.19\textwidth]{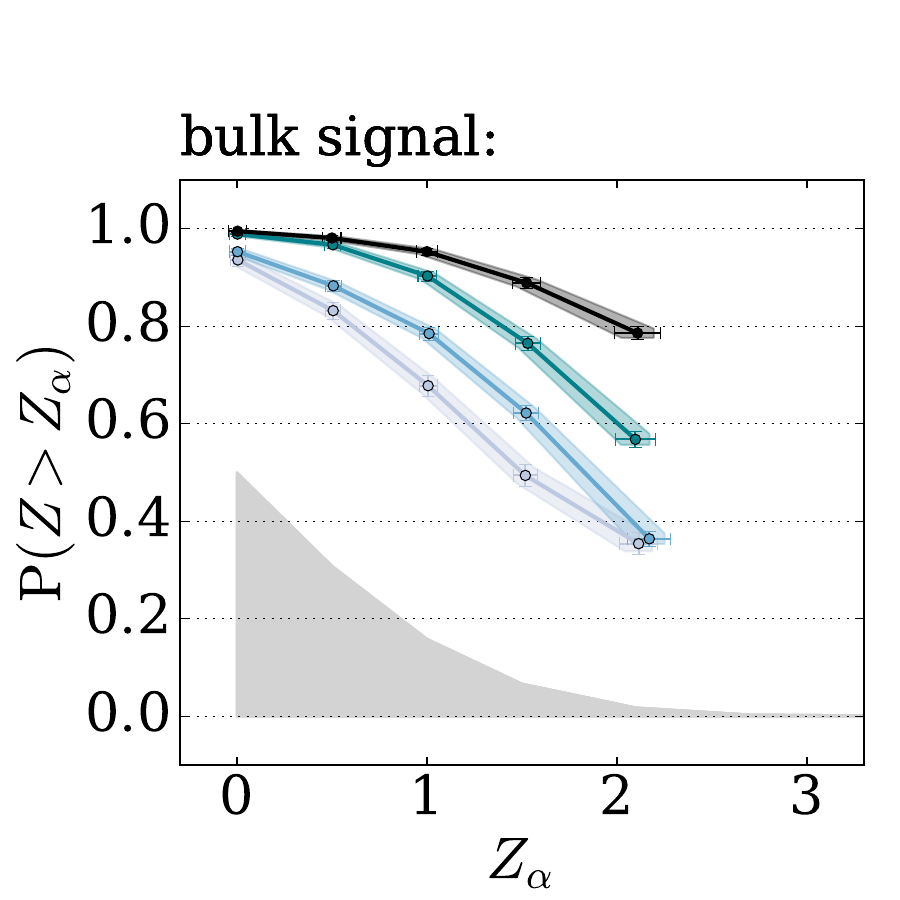}
\includegraphics[width=0.19\textwidth]{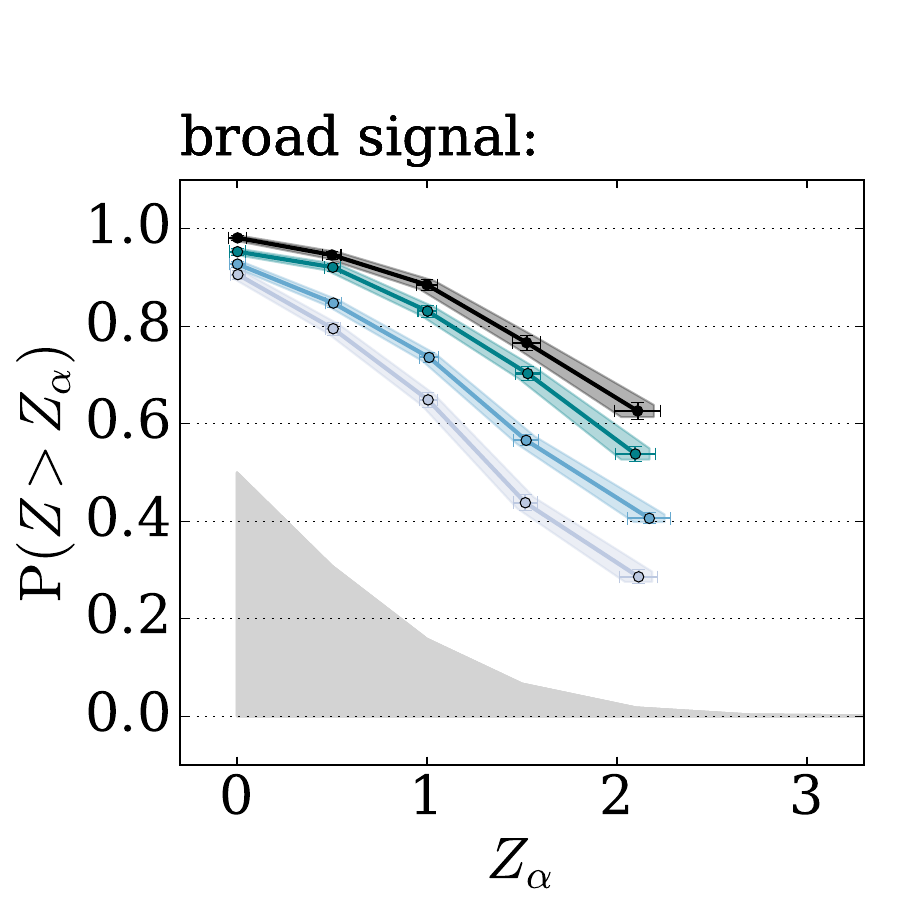}
\includegraphics[width=0.19\textwidth]{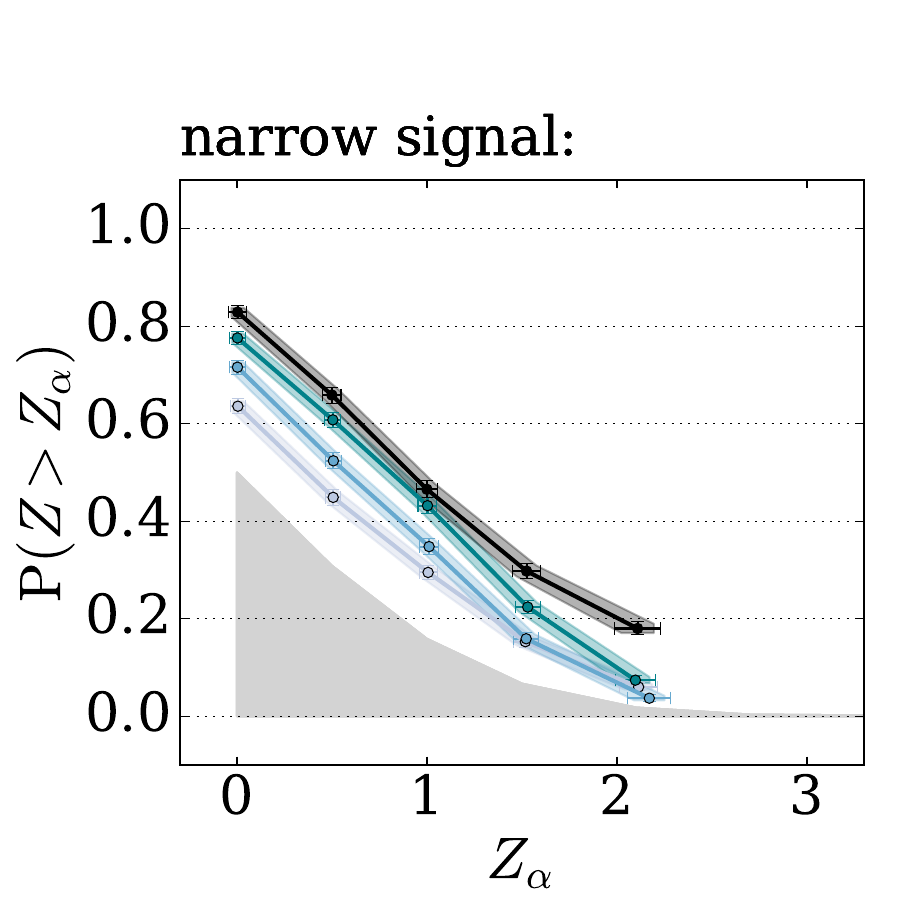}
\includegraphics[width=0.38\textwidth]{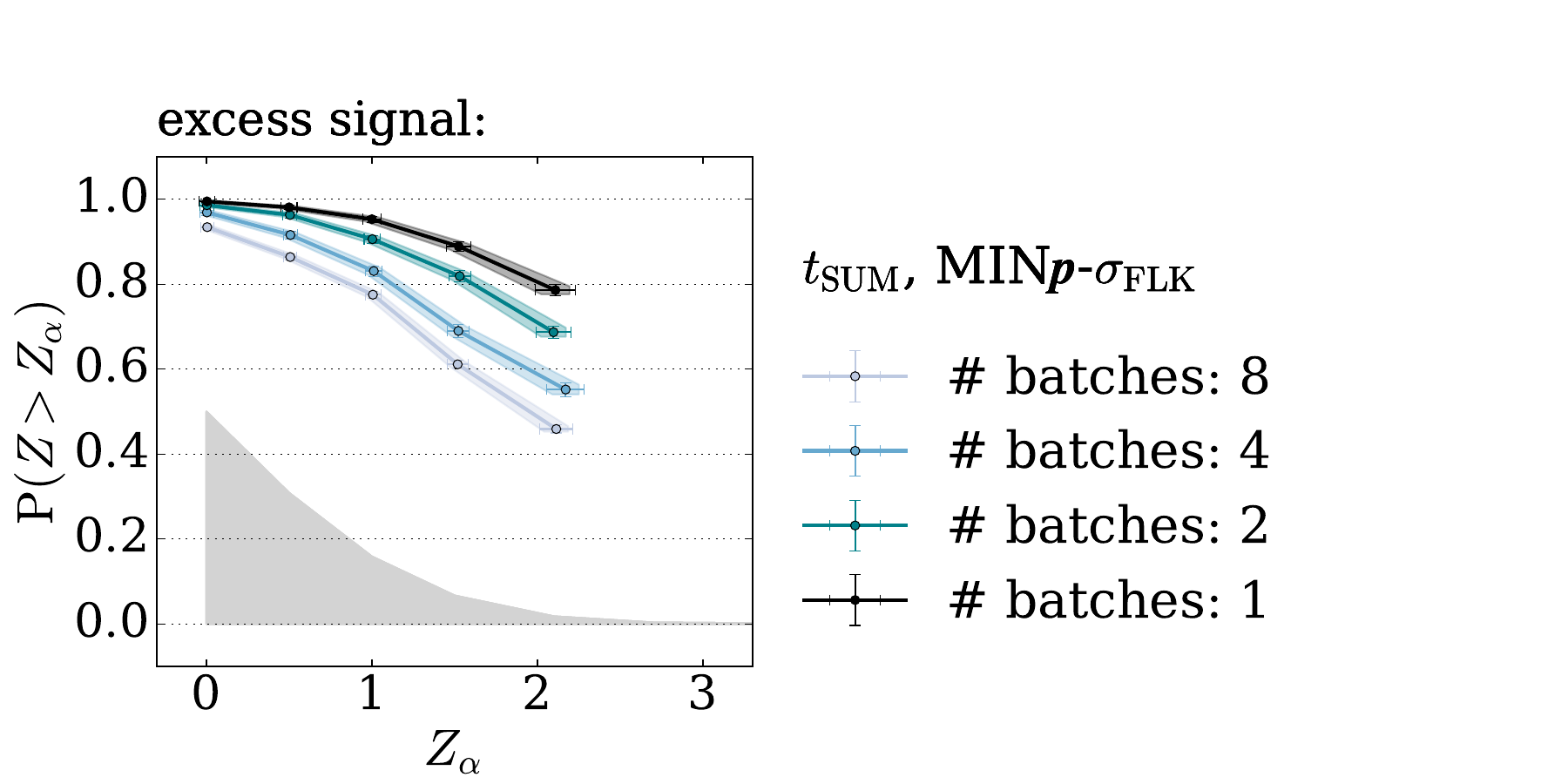}
\caption{\textbf{EXPO-1D, NPLM-ALL.} Power curves for $t_{\rm AGGR}$ (top row) and $t_{\rm SUM}$ (bottom row) at different number of batches. Each column shows a different signal benchmark. Data splitting improves the performances of $t_{\rm AGGR}$, while degrades $t_{\rm SUM}$.}\label{fig:1D-nplm-all-app}
\end{figure}

\begin{figure}[H]
    \centering
\includegraphics[width=0.24\textwidth]{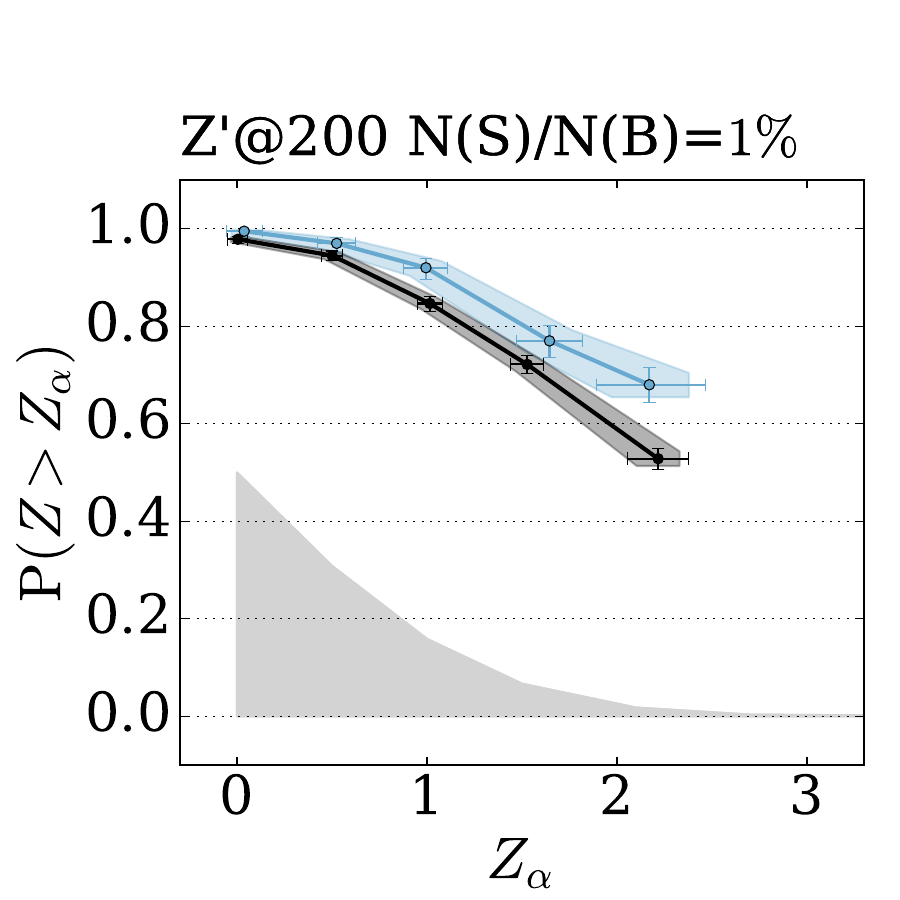}
\includegraphics[width=0.24\textwidth]{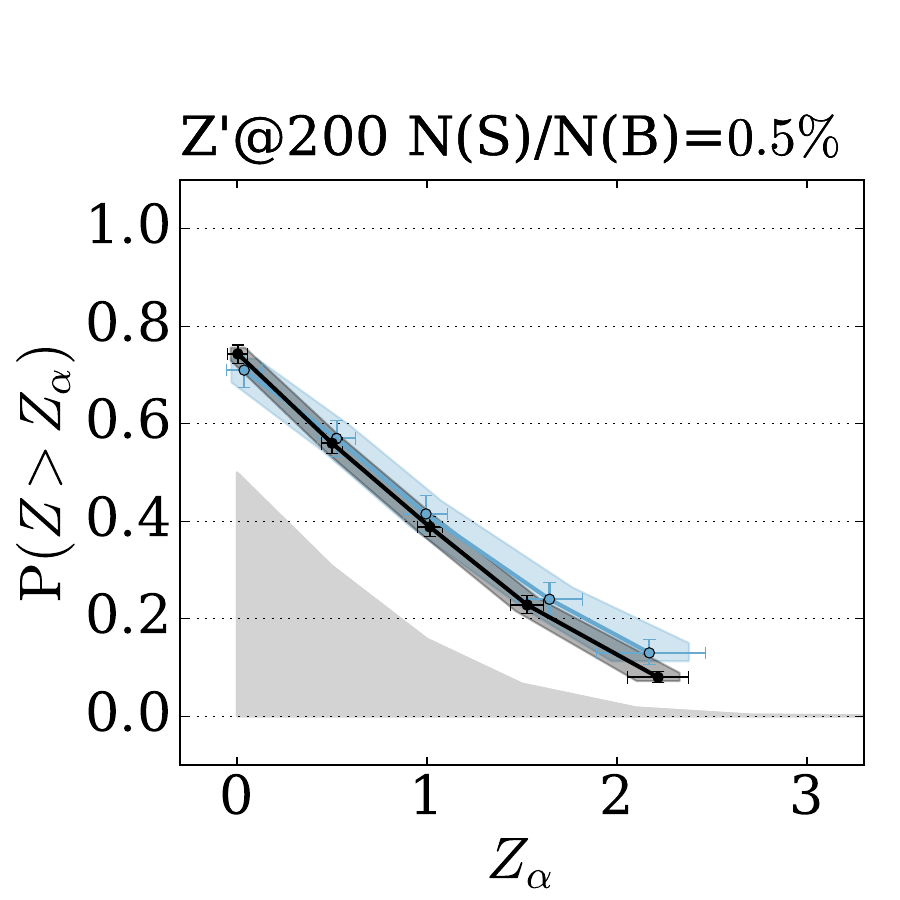}
\includegraphics[width=0.48\textwidth]{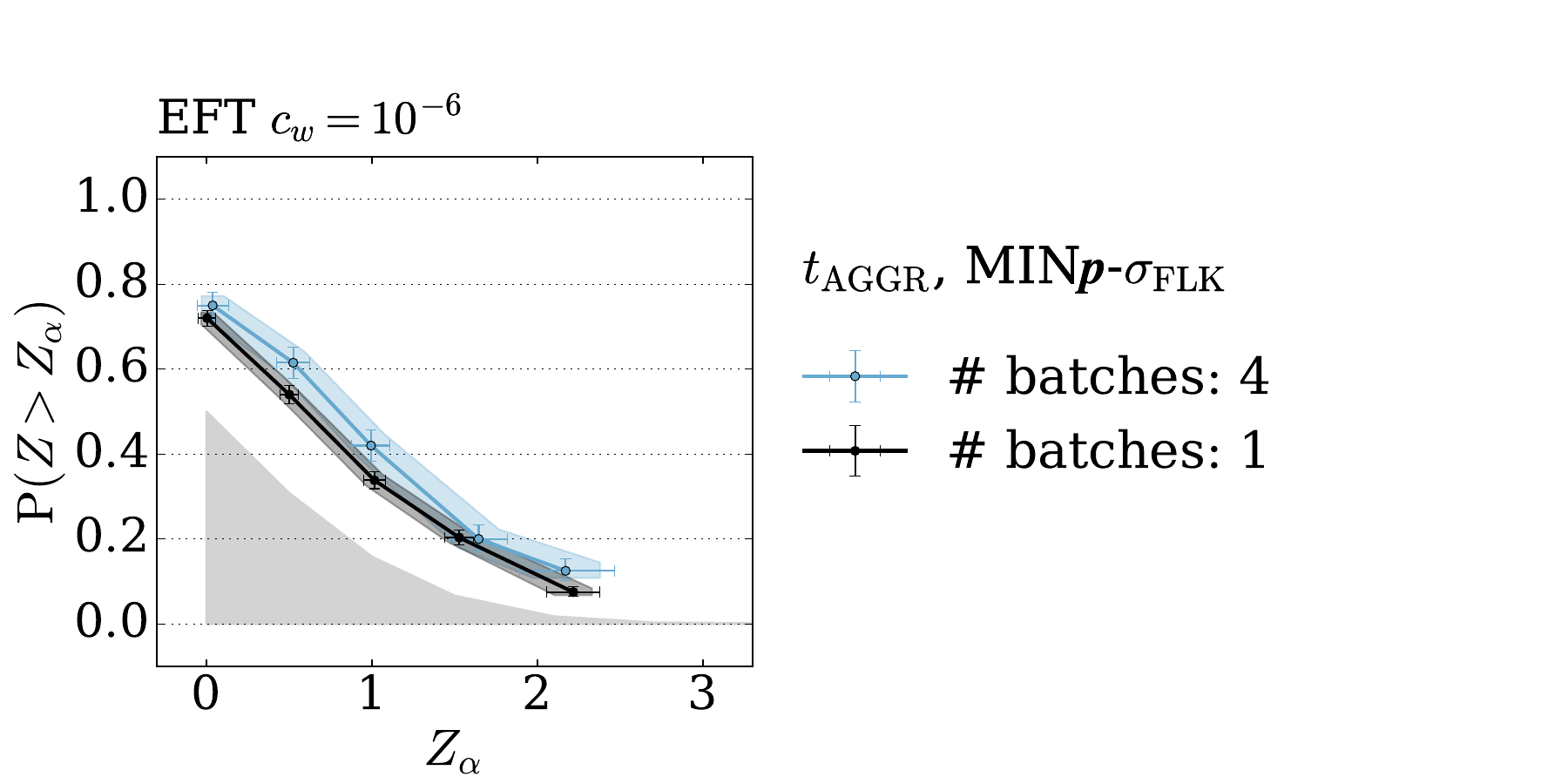}\\

\includegraphics[width=0.24\textwidth]{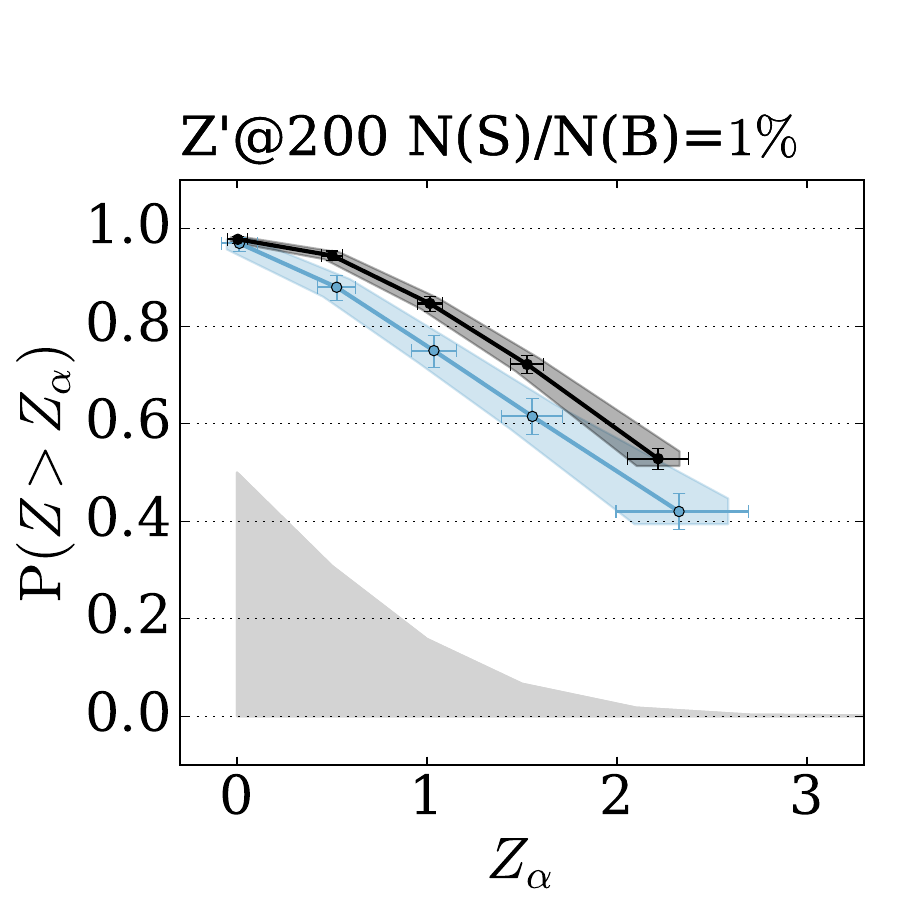}
\includegraphics[width=0.24\textwidth]{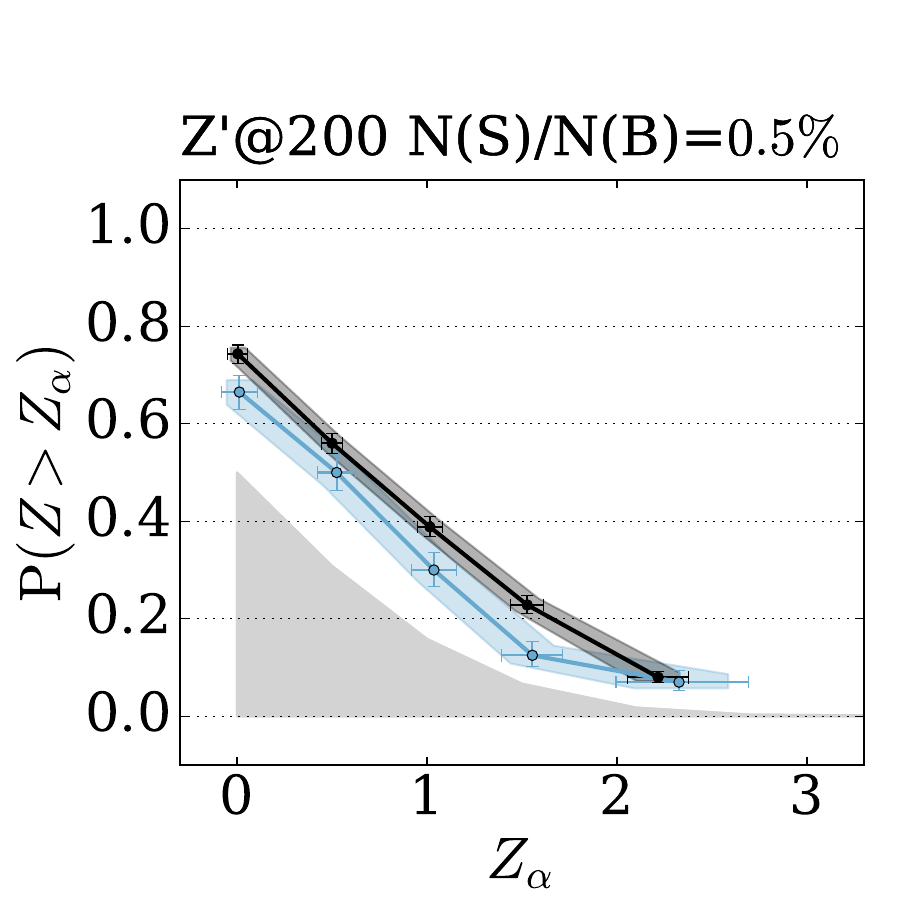}
\includegraphics[width=0.48\textwidth]{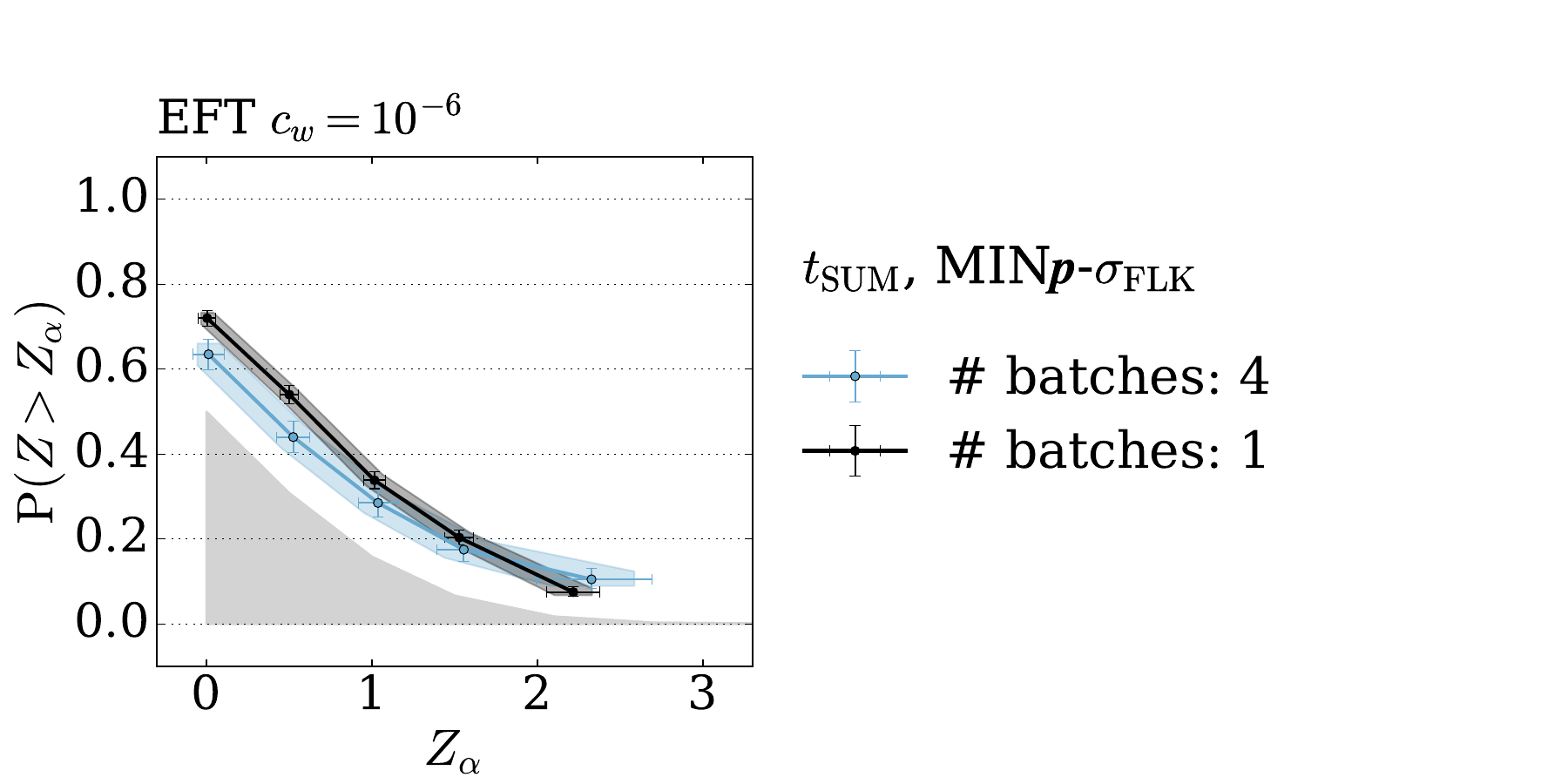}\\
\caption{\textbf{MUMU-5D, NPLM-ALL (panel 1).} Power curves for $t_{\rm AGGR}$ (top row) and $t_{\rm SUM}$ (bottom row) at different number of batches. Each column shows a different signal benchmark. Data splitting maintains or improves the performances of $t_{\rm AGGR}$, while degrades $t_{\rm SUM}$.}\label{fig:5D-nplm-all-1}
\end{figure}

\begin{figure}[H]
    \centering
\includegraphics[width=0.24\textwidth]{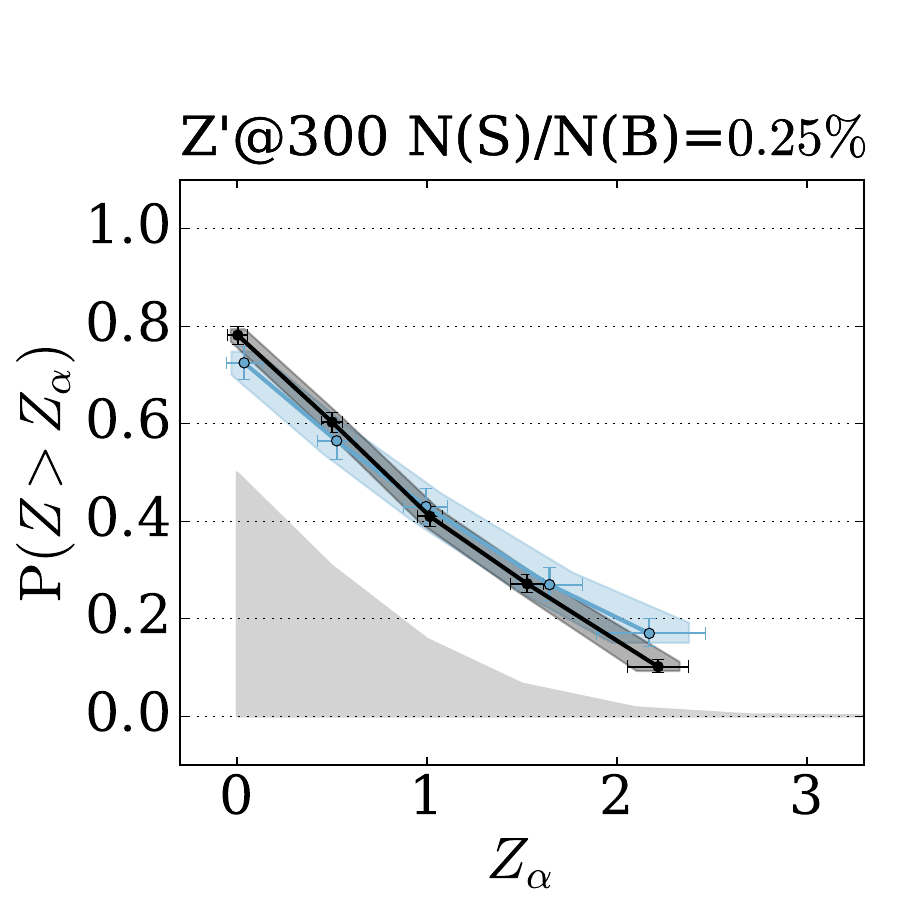}
\includegraphics[width=0.24\textwidth]{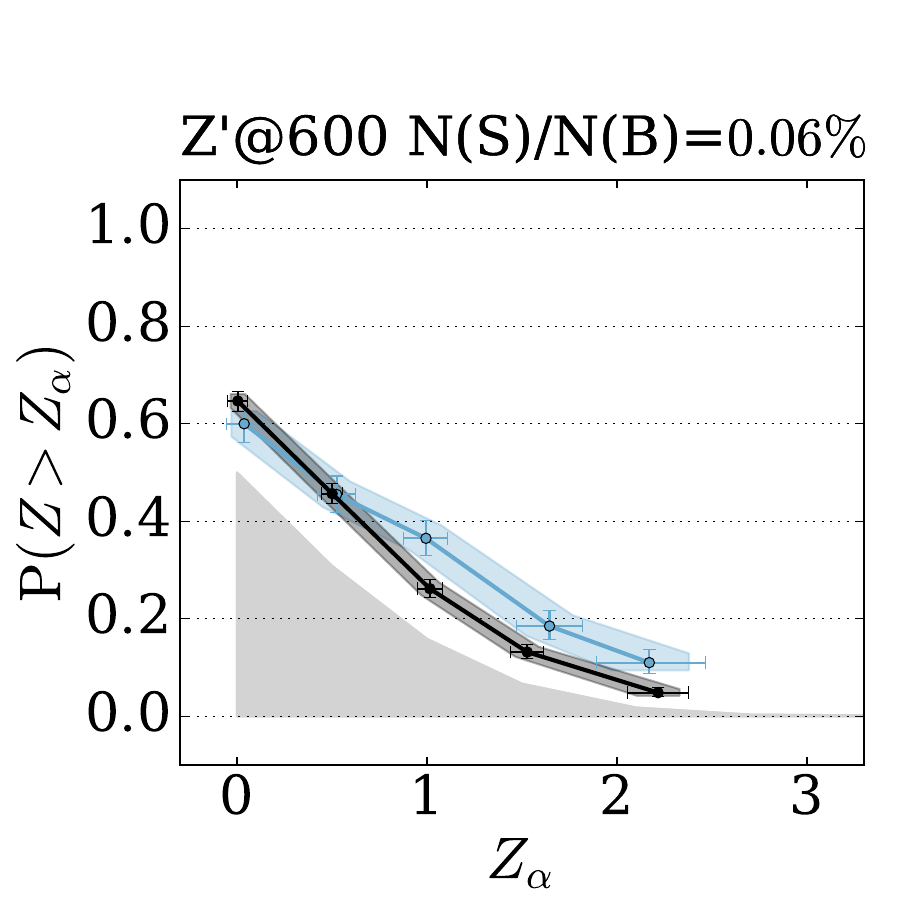}
\includegraphics[width=0.48\textwidth]{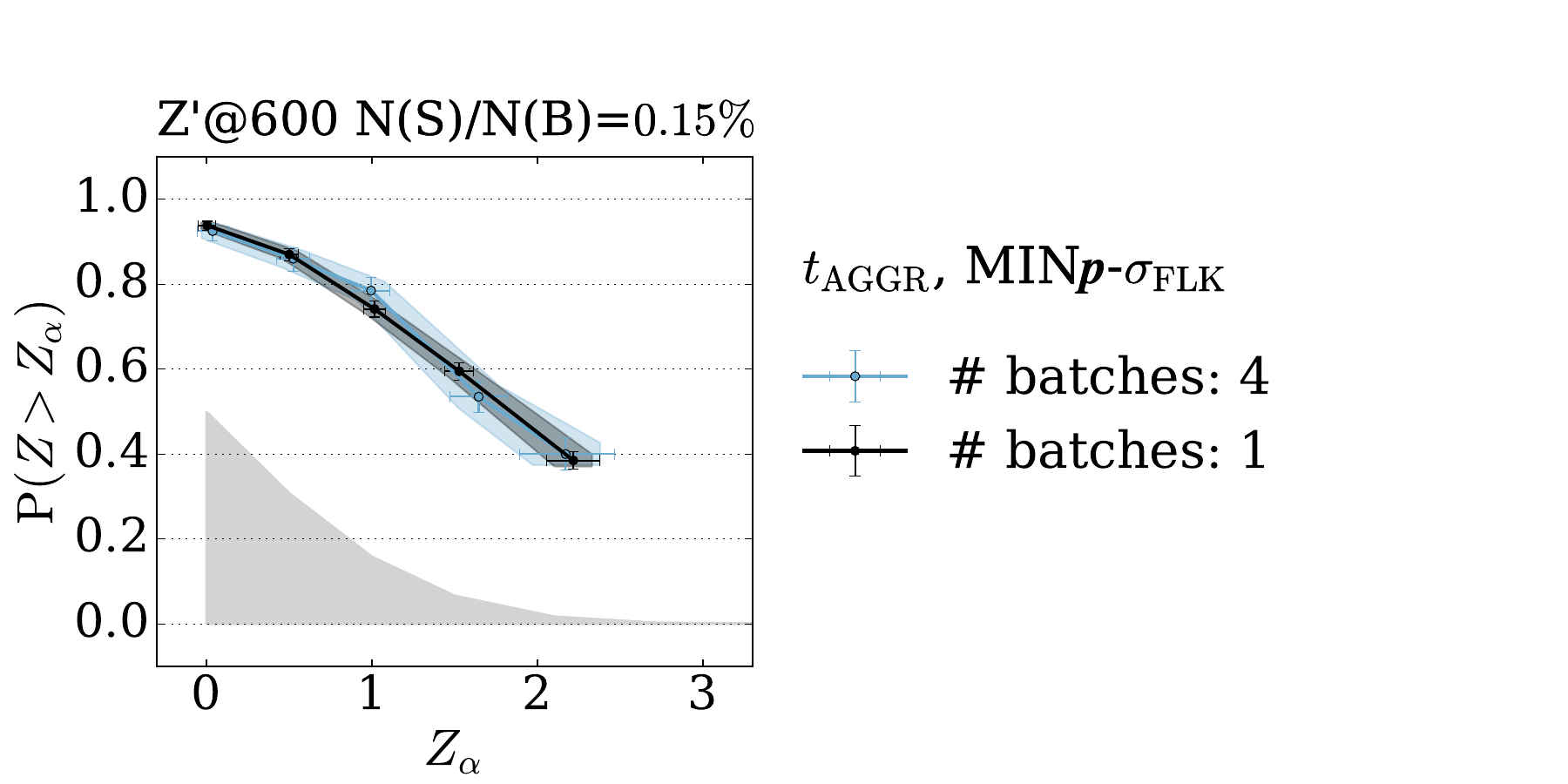}
\includegraphics[width=0.24\textwidth]{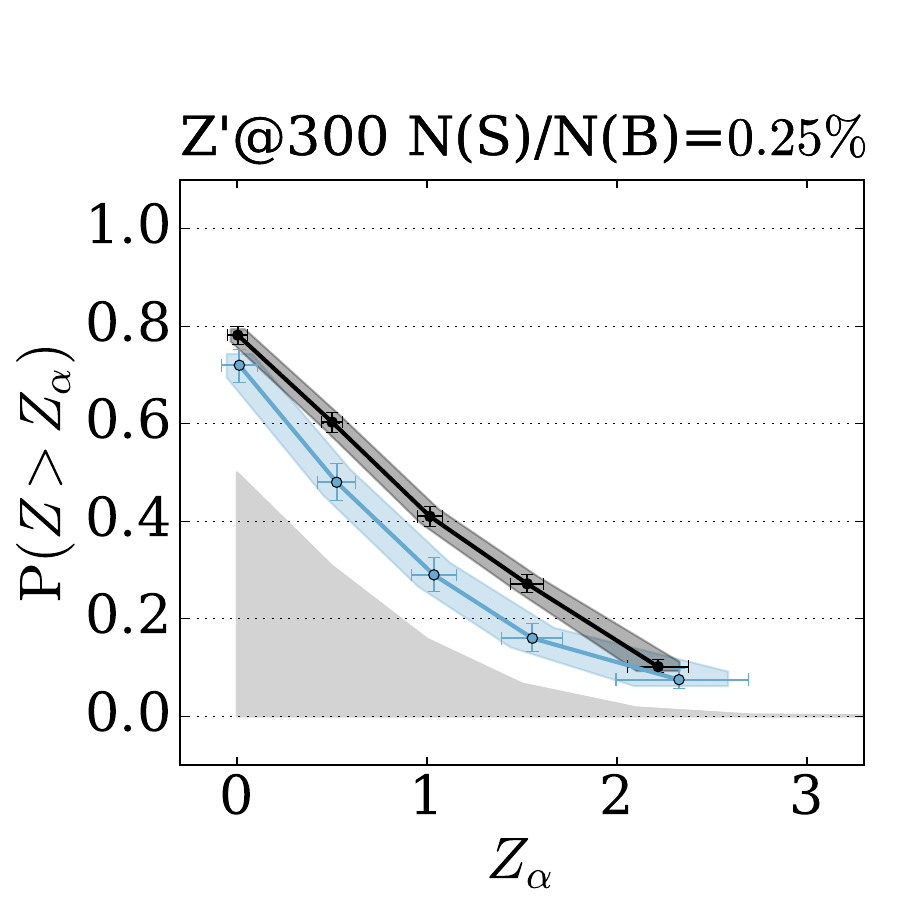}
\includegraphics[width=0.24\textwidth]{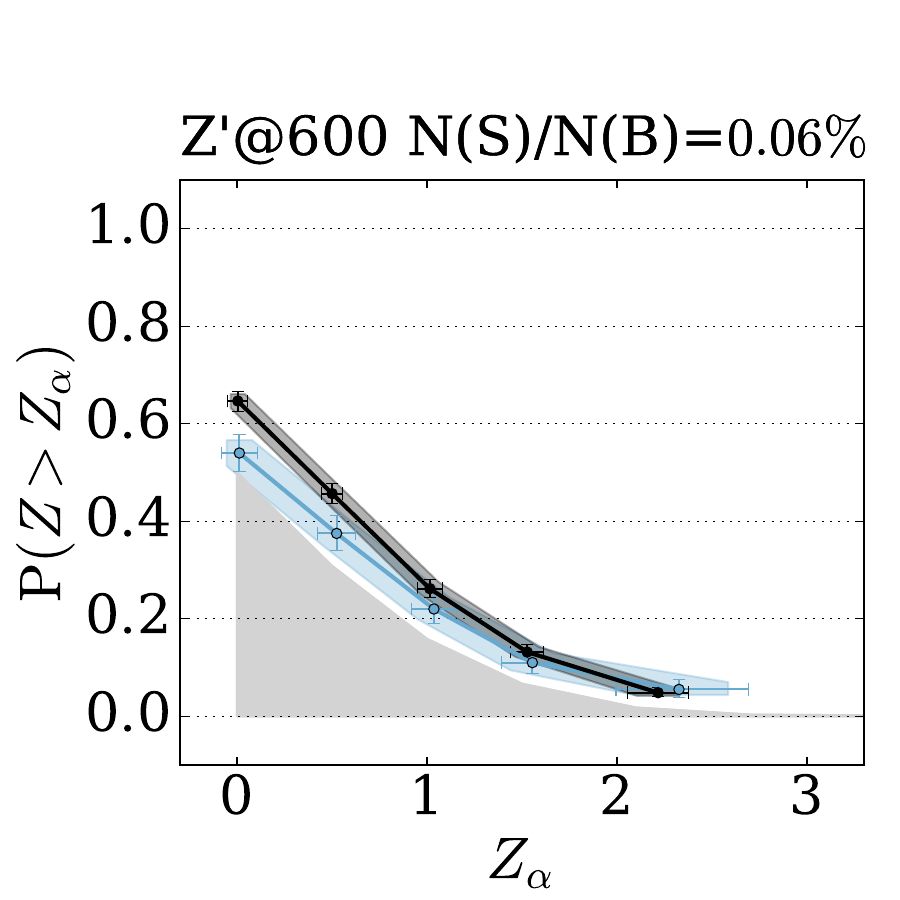}
\includegraphics[width=0.48\textwidth]{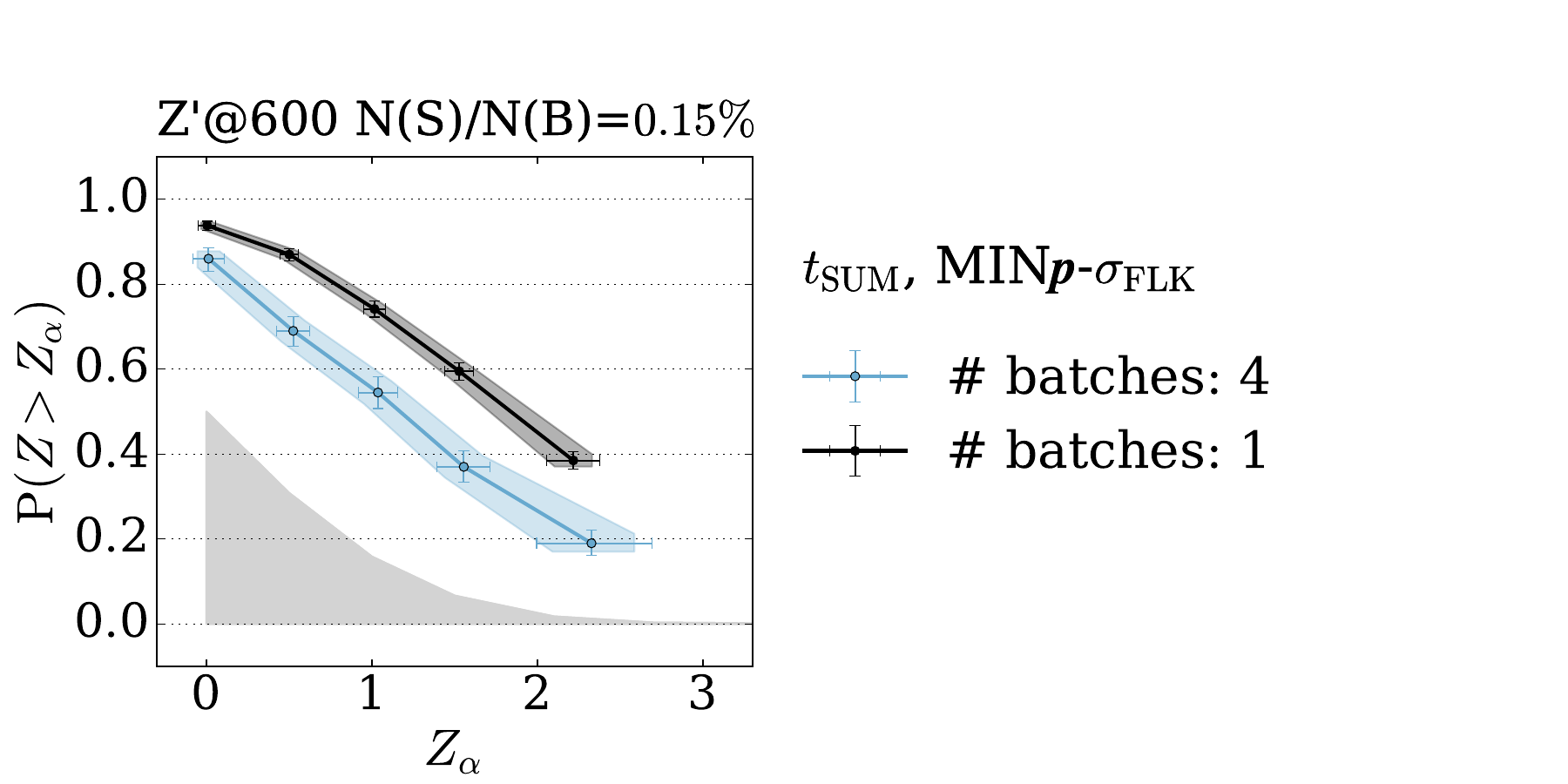}
\caption{\textbf{MUMU-5D, NPLM-ALL (panel 2).} Power curves for $t_{\rm AGGR}$ (top row) and $t_{\rm SUM}$ (bottom row) at different number of batches. Each column shows a different signal benchmark. Data splitting maintains or improves the performances of $t_{\rm AGGR}$, while degrades $t_{\rm SUM}$.}\label{fig:5D-nplm-all-2}
\end{figure}

In this Appendix we provide additional plots showing the power curves of the aggregation method NPLM-ALL proposed in this work, both for the univariate \textbf{EXPO-1D} dataset (Figure~\ref{fig:1D-nplm-all-app}) and for the multivariate \textbf{MUMU-5D} one (Figures~\ref{fig:5D-nplm-all-1} and~\ref{fig:5D-nplm-all-2}). The power of the method is compared with a simple sum of tests (details in Sections~\ref{sec:2} and~\ref{subsec:all}).


\subsection{Impact of aggregation on a single batch test}\label{app:2}
\begin{figure}[h]
    \centering
\includegraphics[width=0.32\textwidth]{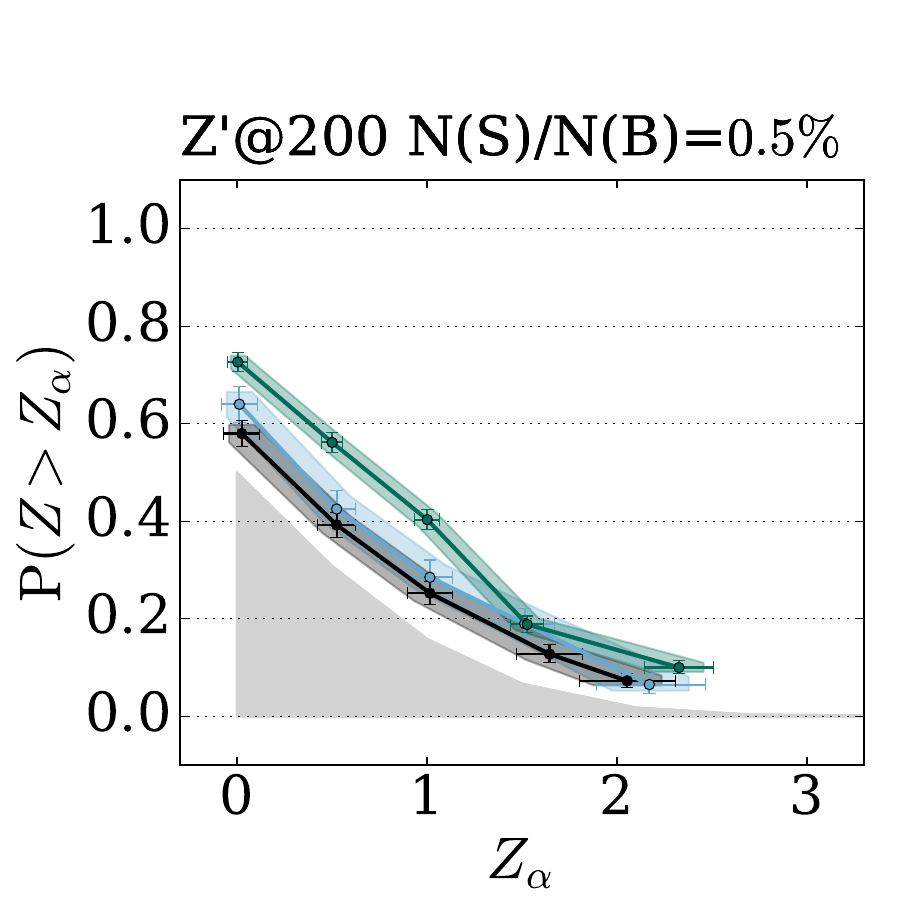}
\includegraphics[width=0.32\textwidth]{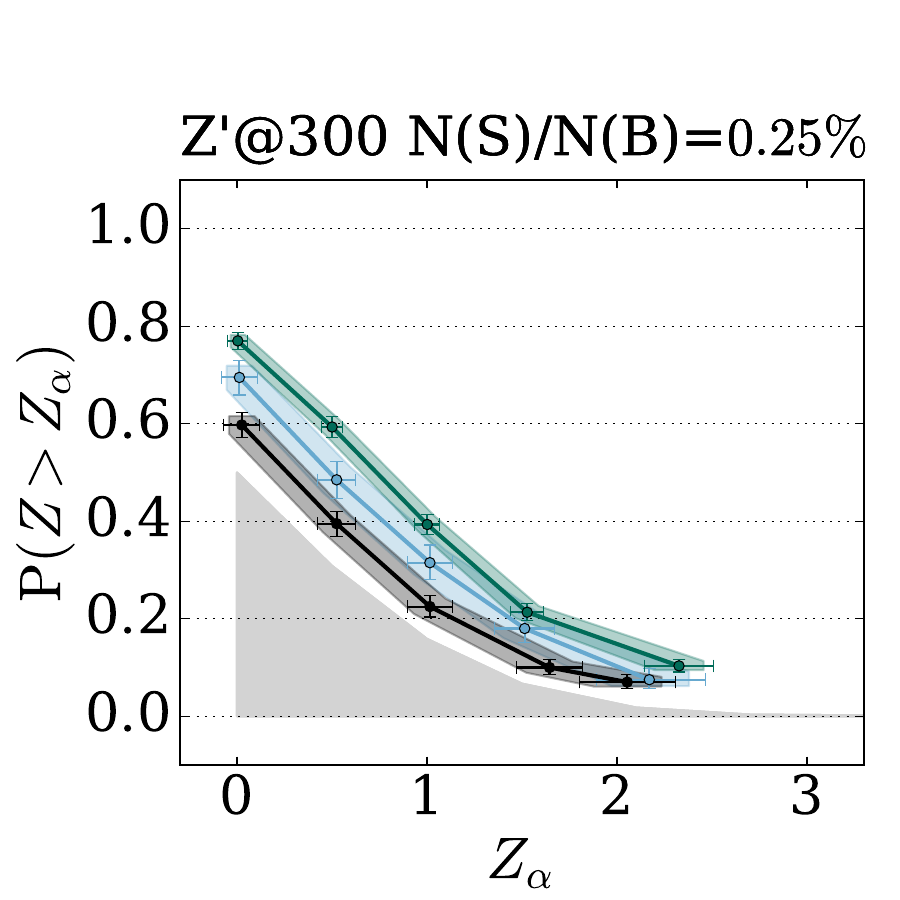}
\includegraphics[width=0.32\textwidth]{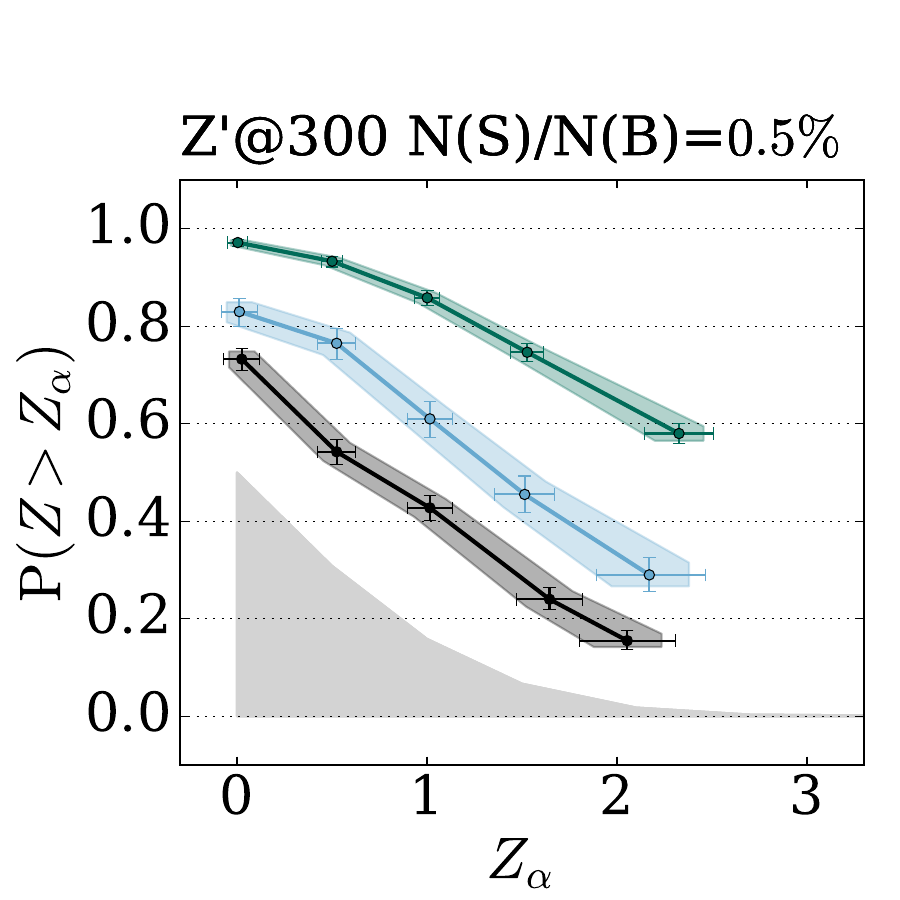}\\
\includegraphics[width=0.32\textwidth]{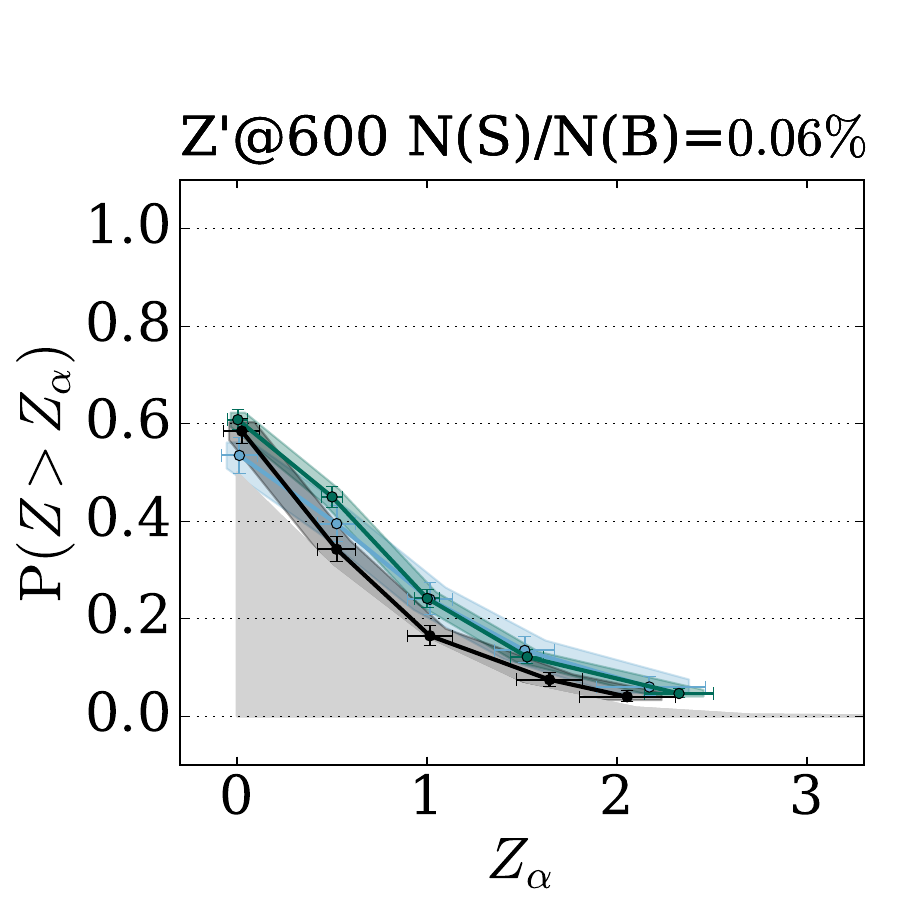}
\includegraphics[width=0.64\textwidth]{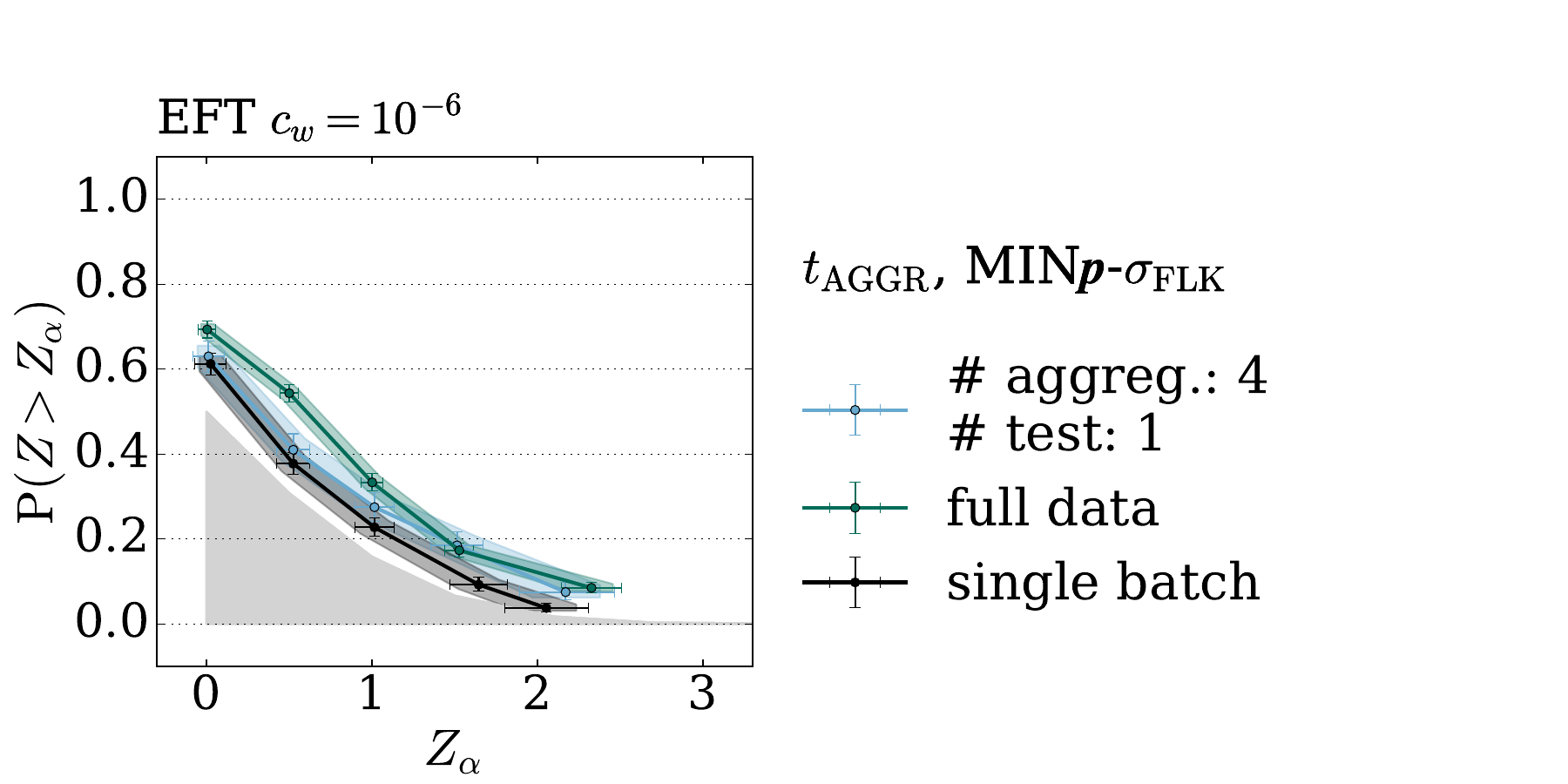}
    \caption{\textbf{NPLM-ONE. DIMUON-5D. } Power curves for the aggregated test $t_{\rm AGGR}$ when evaluated over one single data batch (light blue line) and over the full dataset (green line), compared with the power of the test learnt and evaluated over one data batch (black line). The aggregation improves the accuracy over the learnt alternative model enhancing the sensitivity with respect to the single batch test.}\label{fig:5D-nplm-one-app}
\end{figure}

We provide in Figure~\ref{fig:5D-nplm-one-app} additional plots showing the power curves of the aggregation method NPLM-ONE proposed in this work. The power of the method is compared with a simple sum of tests (details in Sections~\ref{sec:2} and~\ref{subsec:one}).

\subsection{Saturated test statistic}\label{app:3}
\begin{figure}[H]
    \centering
\includegraphics[width=0.24\textwidth]{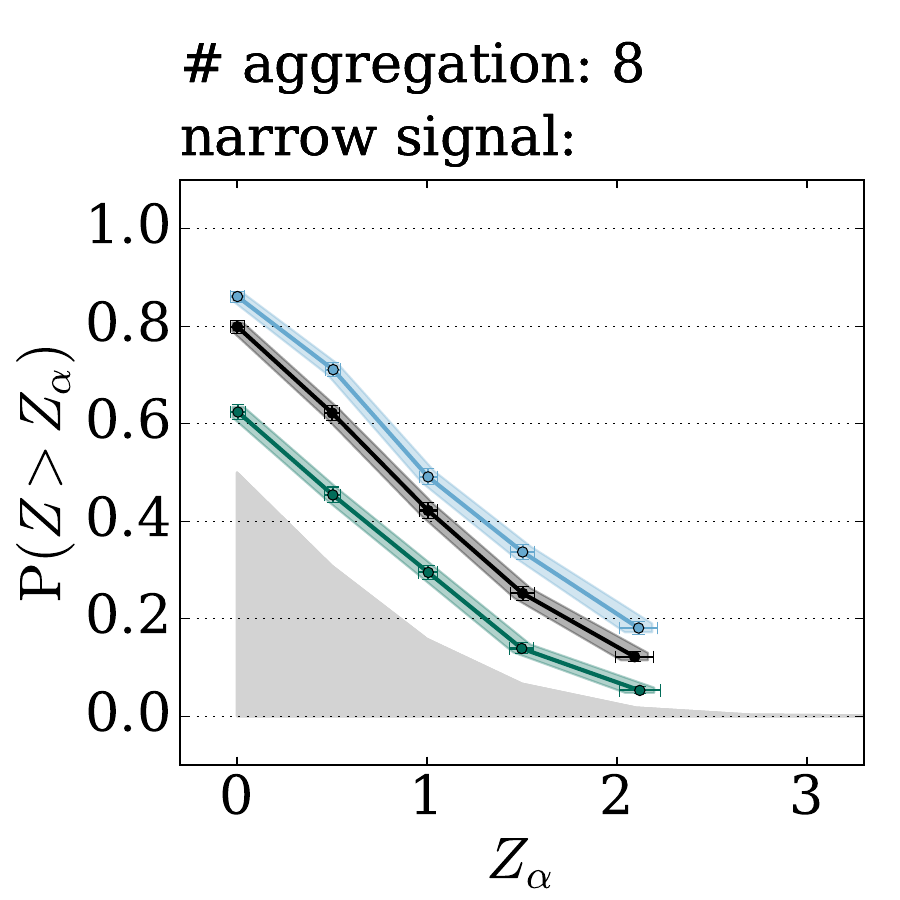}
\includegraphics[width=0.24\textwidth]{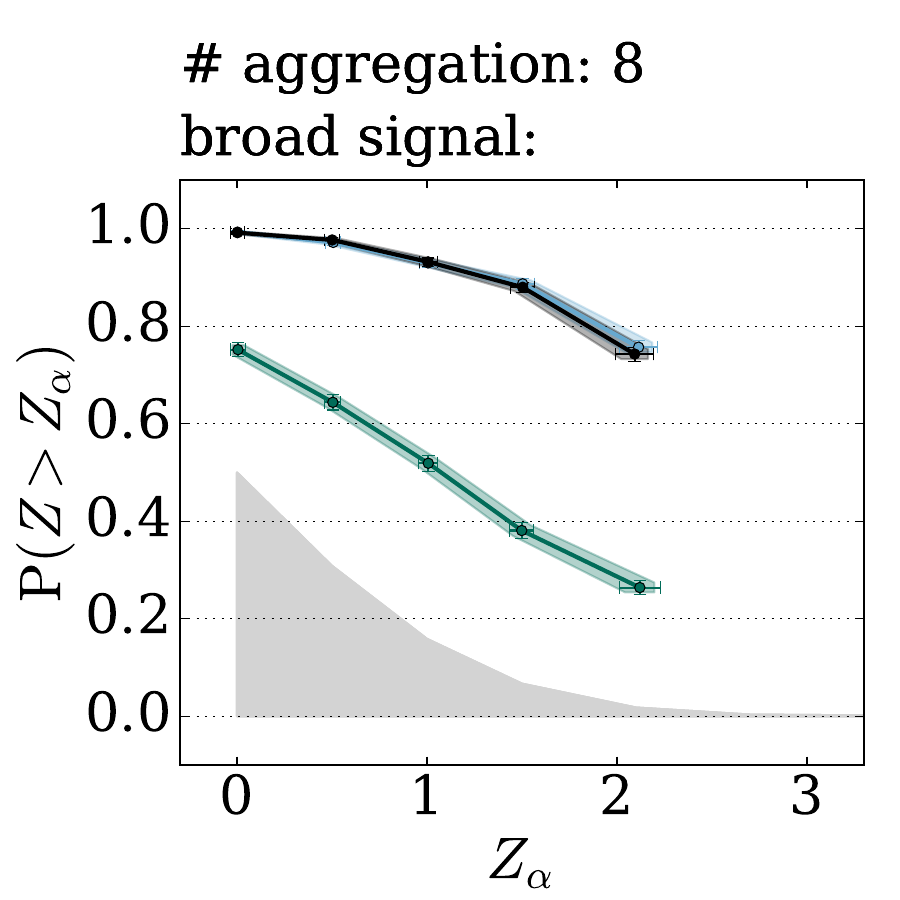}
\includegraphics[width=0.48\textwidth]{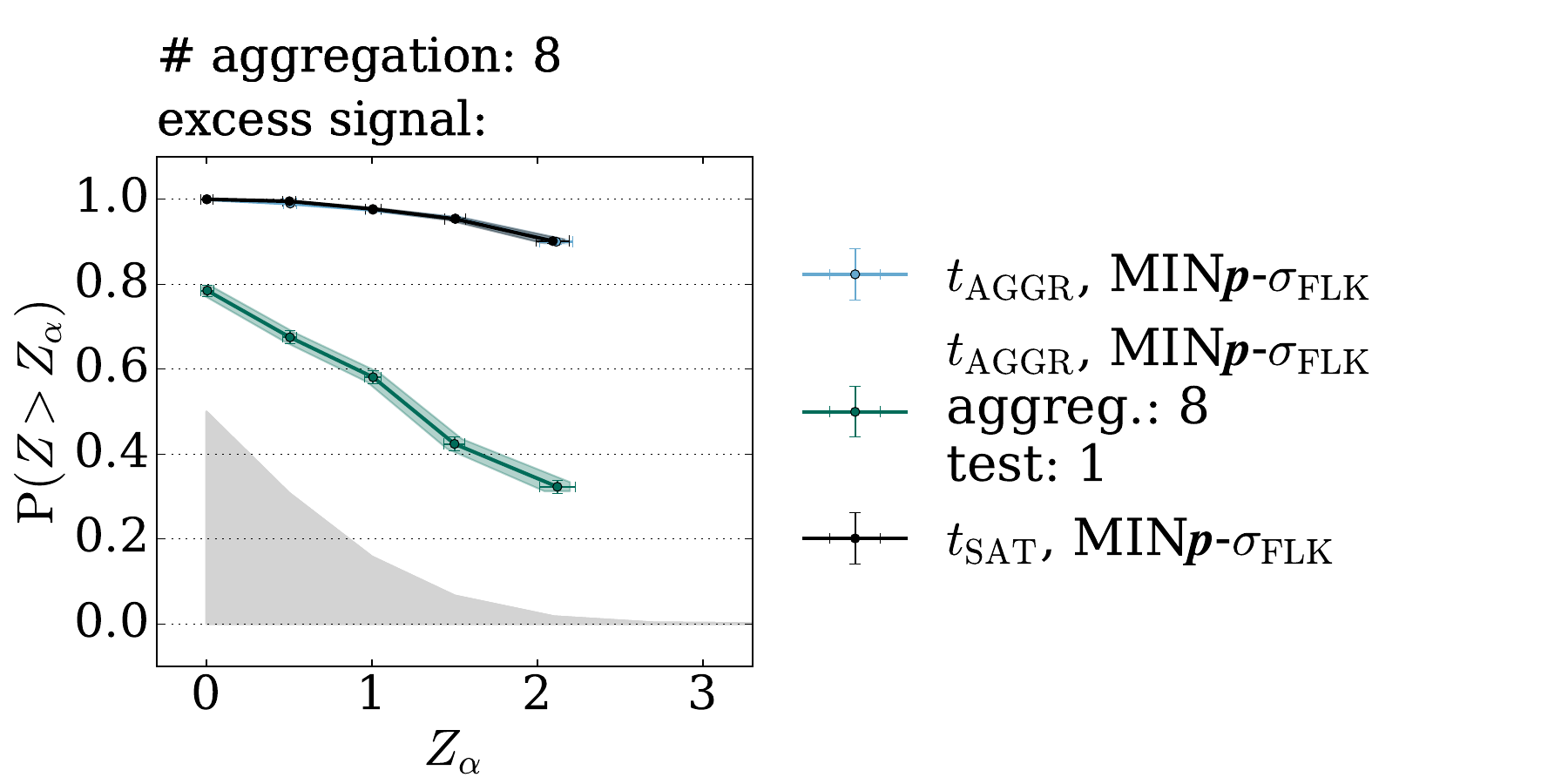}
    \caption{\textbf{NPLM-SAT. EXPO-1D.} Comparison of the aggregated and saturated test statistic power curves for a narrow gaussian signal (left panel), a broad gaussian signal (central panel), and a non resonant excess in the tail (right panel) of the exponential falling reference distribution. The experiments have been performed splitting the dataset in 8 batches. We show in black the power curve of the saturated test (NPLM-SAT), in lightblue the power curve of the aggregated test evaluated over the full dataset (NPLM-ALL), and in green the one of the aggregated test evaluated over one batch (NPLM-ONE).}\label{fig:1D-nplm-sat_ap}
\end{figure}

\begin{figure}[H]
    \centering
\includegraphics[width=0.32\textwidth]{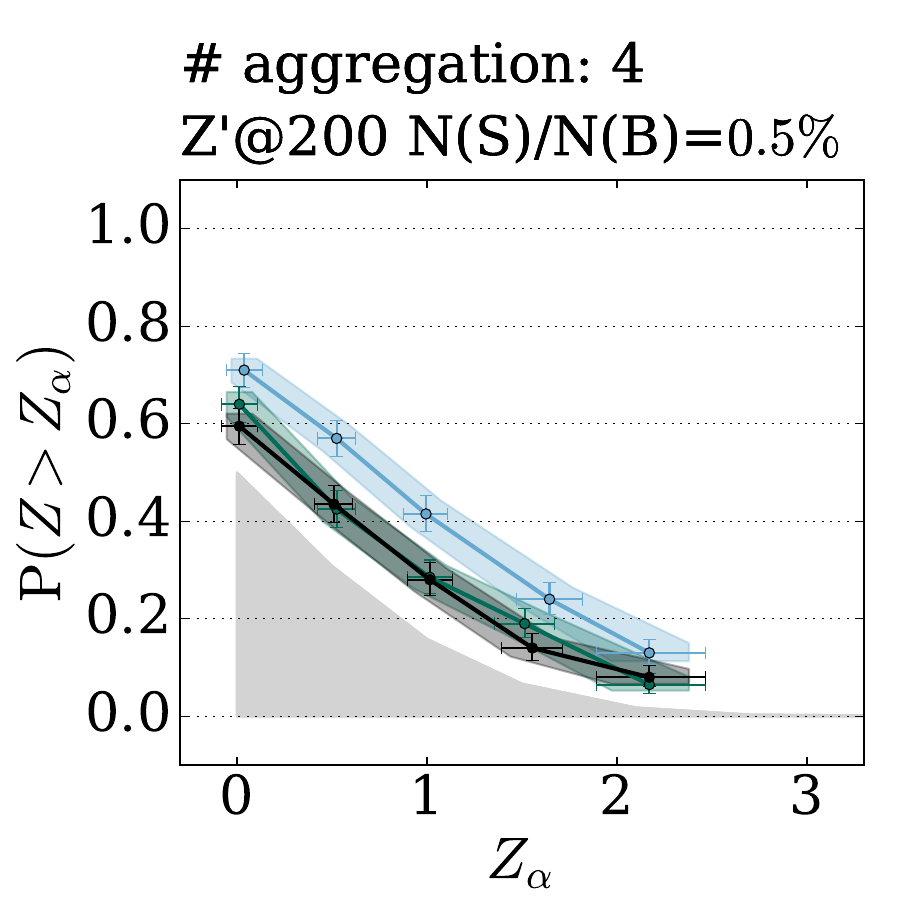}
\includegraphics[width=0.32\textwidth]{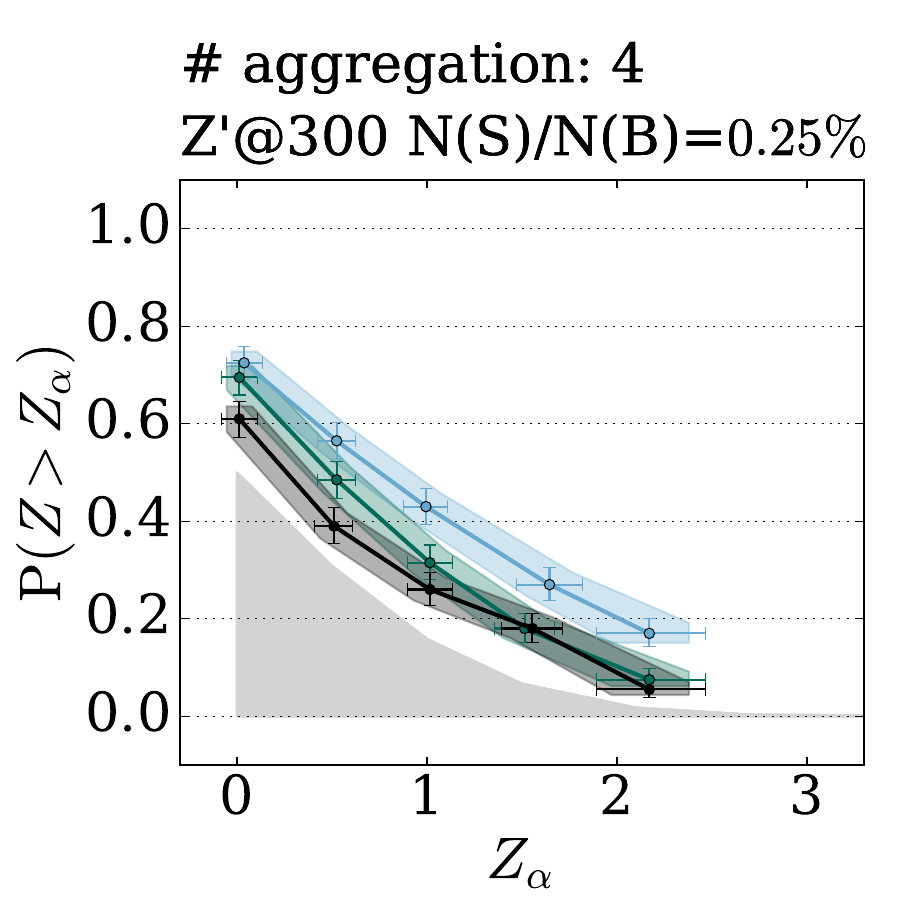}
\includegraphics[width=0.32\textwidth]{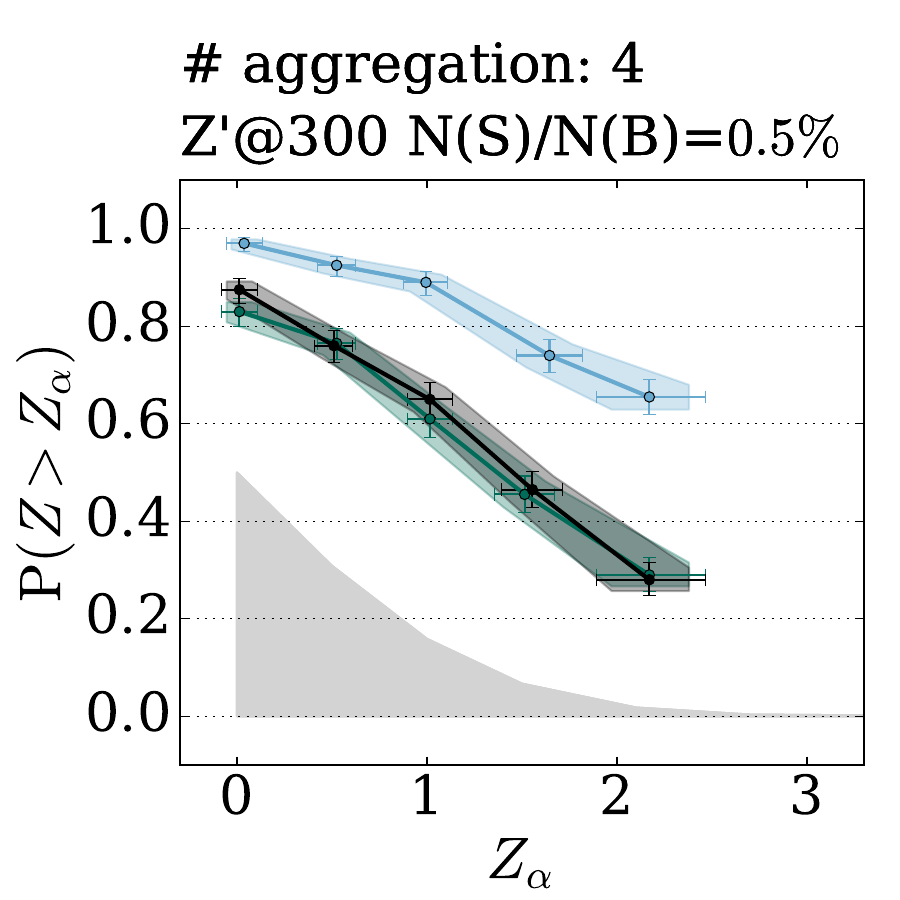}\\
\includegraphics[width=0.32\textwidth]{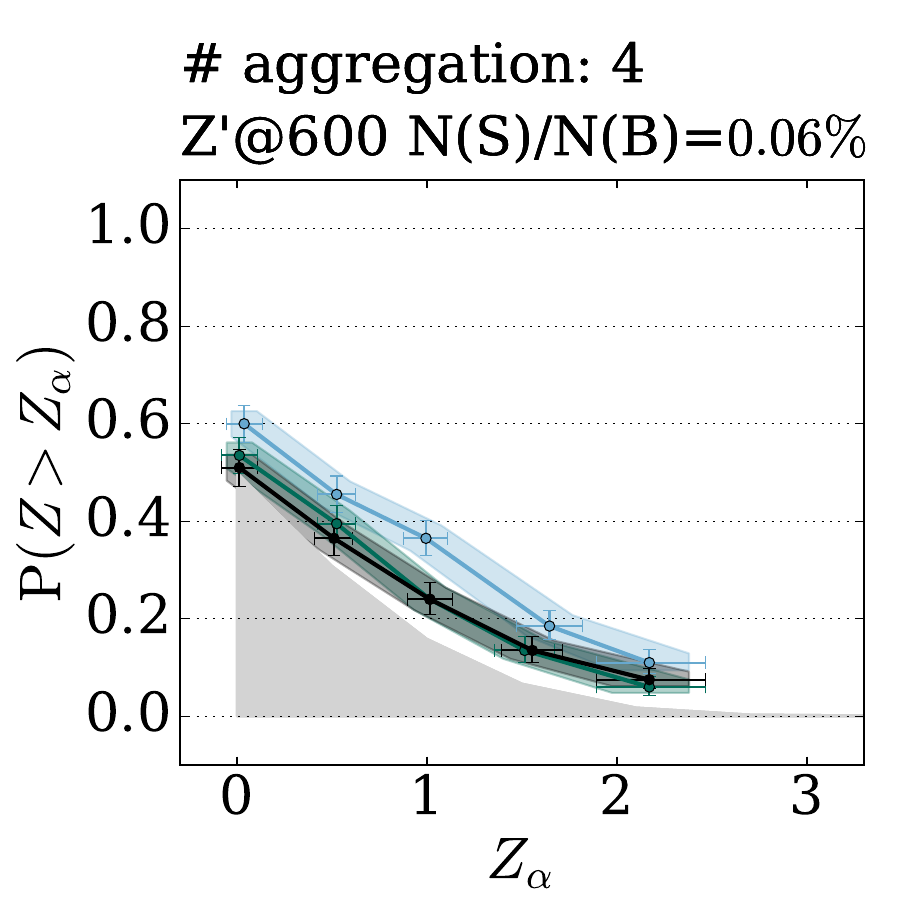}
\includegraphics[width=0.64\textwidth]{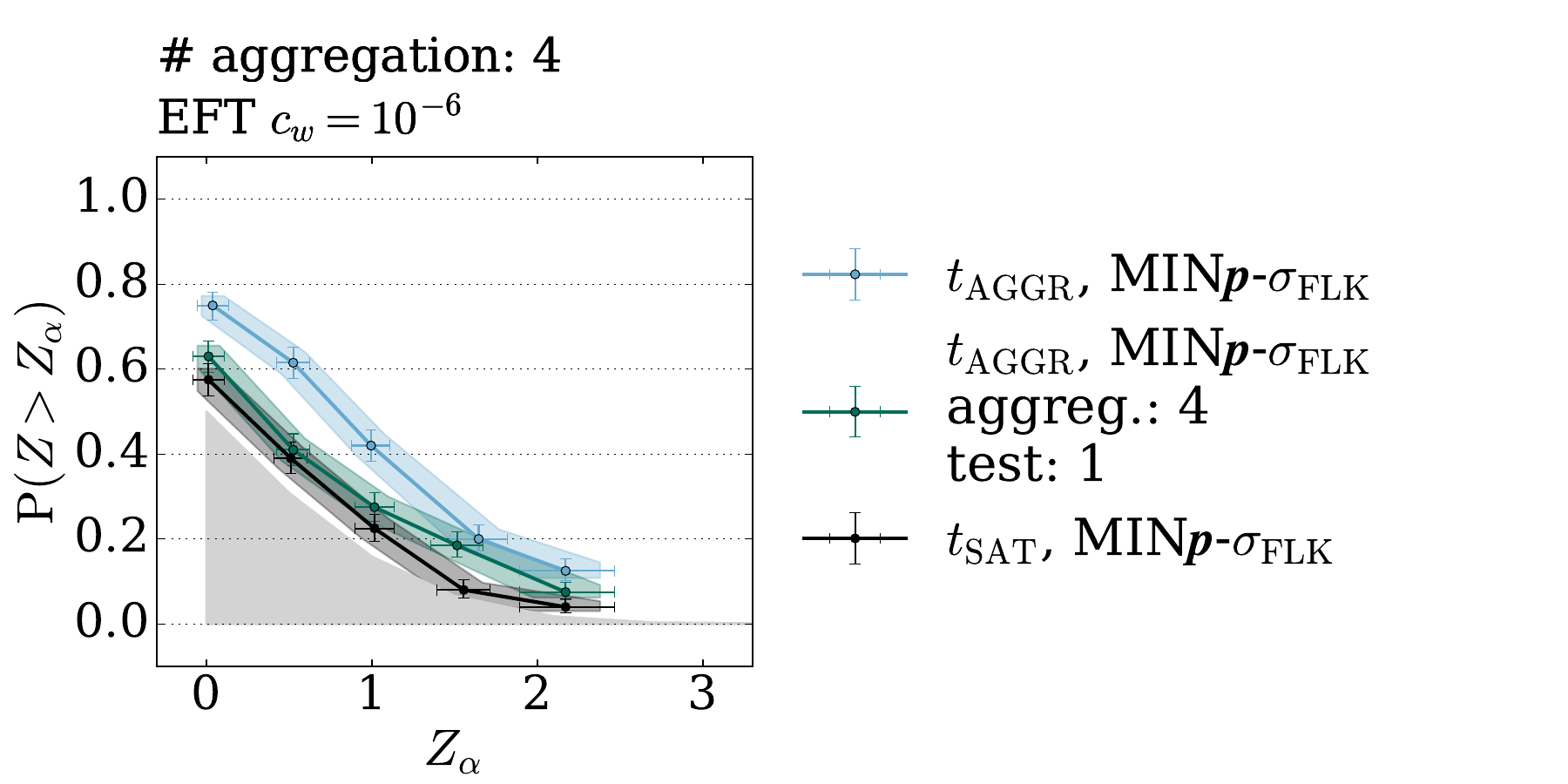}
    \caption{\textbf{NPLM-SAT. DIMUON-5D.} Comparison of the aggregated and saturated test statistic power curves for a Z' resonant signal in the bulk (left panel) and in the tail (right panel) of the mass spectrum considered in this work. The experiments have been performed splitting the dataset in 4 batches.}\label{fig:5D-nplm-sat_aggr1}
\end{figure}

In this Appendix we provide additional plots showing the power curves of the aggregation method NPLM-SAT proposed in this work (see Figures~\ref{fig:1D-nplm-sat_ap} and \ref{fig:5D-nplm-sat_aggr1}). The power of the method is compared with a simple sum of tests (details in Sections~\ref{sec:2} and~\ref{subsec:sat}).
\end{document}